\documentclass[letterpaper, 11pt]{article}

\usepackage{geometry}
\usepackage[utf8]{inputenc}
\usepackage{xcolor}
\usepackage{amsmath,amsthm,amsfonts,amssymb}
\usepackage{tikz}
\usepackage{enumitem}
\usepackage{tabularx}
\usepackage[sort, numbers, compress]{natbib}
\usepackage{mdframed}
\usepackage[ruled,vlined]{algorithm2e}
\usepackage{booktabs}
\usepackage[colorlinks, allcolors=blue]{hyperref}

\tikzstyle{dot}=[circle,fill,inner sep=1.5pt]

\newtheorem{assumption}{Assumption}
\newtheorem{definition}{Definition}
\newtheorem{theorem}{Theorem}

\newtheorem{lemma}{Lemma}

\providecommand{\iprod}[2]{\ensuremath{\left\langle #1,\,#2  \right\rangle}}
\providecommand{\norm}[1]{\ensuremath{\left\lVert#1\right\rVert }}
\providecommand{\mnorm}[1]{\ensuremath{\left\lvert#1\right\rvert}}
\providecommand{\dist}[2]{\ensuremath{\mathrm{dist}\left( #1,\,#2 \right)}}

\def\R{\mathbb{R}}
\def\Z{\mathbb{Z}}
\def\H{\mathcal{H}}
\def\B{\mathcal{B}}

\def\O{\mathcal{O}}

\def\D{\mathsf{D}}

\def\M{\mathsf{M}}
\def\W{\mathcal{W}}
\def\E{\mathbb{E}}
\def\g{\boldsymbol{g}}

\def\gf{\mathsf{GradFilter}}
\def\ga{\mathsf{GradAgg}}

\def\z{\boldsymbol{z}}
\def\bigO{\mathcal{O}}

\title{Utilizing Redundancy in Cost Functions \\ for Resilience in Distributed Optimization and Learning \thanks{This report supersedes our previous report \cite{liu2021asynchronous} as it contains the most of the results in it.}}

\author{Shuo Liu \thanks{Georgetown University. Email: {\tt sl1539@georgetown.edu}.} \hspace{0.5in} Nirupam Gupta \thanks{École Polytechnique Fédérale de Lausanne (EPFL). Email: {\tt nirupam.gupta@epfl.ch}.} \hspace{0.5in} Nitin H. Vaidya \thanks{Georgetown University. Email: {\tt nitin.vaidya@georgetown.edu}.}
}
\date{}

\begin{document}

\maketitle

\begin{abstract}
    % This paper considers the role of redundancy in cost functions for solving resilient distributed optimization problems, under a server-based system architecture. In this problem, each agent has a local cost, and the goal is to collectively find a minimum of their aggregate cost. In a synchronous fault-free scenario, distributed gradient descent (DGD) serves a standard algorithm to solve this problem. However, faulty agents and stragglers may affect the correctness and efficiency of this method; instead, more resilient solutions are needed.
    
    % In this paper, we show that the redundancy in cost functions can be utilized to provide resilient solutions to distributed optimization problems. 
    
    % %%%%%%%%%%%%%%%%%%%%%%%%%%
    This paper considers the problem of resilient distributed optimization and stochastic machine learning in a server-based architecture. The system comprises a server and multiple agents, where each agent has a local cost function. The agents collaborate with the server to find a minimum of their aggregate cost functions. We consider the case when some of the agents may be asynchronous and/or Byzantine faulty. In this case, the classical algorithm of distributed gradient descent (DGD) is rendered ineffective. Our goal is to design techniques improving the efficacy of DGD with asynchrony and Byzantine failures. To do so, we start by proposing a way to model the agents' cost functions by the generic notion of \textit{$(f, \,r; \epsilon)$-redundancy} where $f$ and $r$ are the parameters of Byzantine failures and asynchrony, respectively, and $\epsilon$ characterizes the \textit{closeness} between agents' cost functions. This allows us to quantify the level of \textit{redundancy} present amongst the agents' cost functions, for any given distributed optimization problem. We demonstrate, both theoretically and empirically, the merits of our proposed redundancy model in improving the robustness of DGD against asynchronous and Byzantine agents, and their extensions to distributed stochastic gradient descent (D-SGD) for robust distributed machine learning with asynchronous and Byzantine agents.
    % , we show that our proposed characterization can also improve the robustness of , which is more commonly used in machine learning than DGD.
    % Finally, we present some experimental results to demonstrate practical efficacy of our algorithm.
    
    % Prior works have shown that the classical implementation of the distributed gradient descent (DGD) algorithm is generally ineffective in this setting. We however show that when there is {\em redundancy in agents' costs} then a simple modified version of the classical DGD algorithm can still solve the optimization problem. In particular, we study the impact of 
    % Inspired by previous works on Byzantine fault-tolerance, we start Prior works show  know that the classical version of the distributed gradient descent (DGD) algorithm is rendered ineffective in this setting. In this context, we precisely characterize the role of {\em redundancy in cost functions} in solving this problem. In particular, we show that the classical algorithm of distributed gradient descent (DGD) 
    % serves a standard algorithm to solve this problem. However, faulty agents and stragglers may affect the correctness and efficiency of this method; instead, more resilient solutions are needed.
\end{abstract}

\newpage
\tableofcontents

\newpage
\section{Introduction}
\label{sec:intro}

With the rapid growth in the computational power of modern computer systems and the scale of optimization tasks, e.g., learning of deep neural networks~\citep{otter2020survey}, the problem of distributed optimization over a multi-agent system has gained significant attention in recent years. In the setting of multi-agent system, %there is a server and multiple agents. 
each agent has a local cost function. The goal is to design an algorithm that allows the agents to collectively minimize the aggregated cost functions of all agents~\citep{boyd2011distributed}. %, during which the server sending estimates of the model parameters to all agents, and the agents compute and send gradients to the server.
Formally, we suppose that there are $n$ agents in the system where each agent $i$ has a cost function $Q_i: \R^d \to \R$. A distributed optimization algorithm enables the agents to compute a global minimum $x^*$ such that 
\begin{equation}
     x^*\in\arg\min_{x\in\R^d}\sum_{i=1}^nQ_i(x). \label{eqn:goal}
\end{equation}
For example, amongst a group of $n$ people, the function $Q_i(x)$ may represent the cost of each person $i$ to travel to some location $x$, and thus, $x^*$ represents a location that minimizes the aggregate (or average) travel cost for all the $n$ people. Besides the obvious application to distributed machine learning~\citep{boyd2011distributed}, the above multi-agent optimization problem finds many other applications, such as distributed sensing \citep{rabbat2004distributed}, and swarm robotics \citep{raffard2004distributed}. 
% There are several algorithms known to solve this problem when the agents are reasonably fault-free \citep{boyd2011distributed, nedic2009distributed, shi2015extra, varagnolo2015newton}. 

Distributed optimization systems, however, encounter some issues in practice; such as presence of \textit{faulty agents} that send wrong information and \textit{stragglers} that operate slowly. Prior works have shown that even a single faulty agent can compromise the entire distributed optimization process \citep{chen2017distributed, su2016fault}. Stragglers, on the other hand, can lead to computational inefficiencies wasting resources~\citep{hannah2017more, leblond2018asynchronous, assran2020advances}. These issues require further attention and have sparked a plethora of research \citep{chen2018draco, blanchard2017machine, liu2021approximate, tandon2017gradient, halbawi2018improving, karakus2017encoded, niu2011hogwild}.

\textit{Resilient distributed optimization} refers to the problem of distributed optimization when there are faulty agents and/or stragglers in the system. Specifically, there are 4 types of problems listed below and summarized in Table~\ref{tab:problems}.
\begin{description}[nosep, leftmargin=\labelwidth]
    \item[Problem A] Distributed optimization in a synchronous system with no faulty agents
    \item[Problem B] Distributed optimization in a synchronous system with faulty agents
    \item[Problem C] Asynchronous distributed optimization  with no faulty agents
    \item[Problem D] Asynchronous distributed optimization  with faulty agents
\end{description}
Problem A has been well studied in the past, and can be solved using several existing distributed optimization algorithms~\citep{boyd2011distributed, nedic2009distributed, shi2015extra, varagnolo2015newton}.
Problems B, C and D on the other hand require use of techniques to irregularities due to faulty agents and/or asynchrony. Note that for faulty agents, we consider a \textit{Byzantine} fault model~\citep{lamport1982byzantine}, where no constraints are imposed on the behavior of the faulty agents. 

\begin{table}[b]
    \caption{Problems in Resilient Distributed Optimization.}
    \label{tab:problems}
    \centering
    \begin{tabular}{c|c|c}
        % \toprule
         & No faulty agents & Faulty agents \\
        \midrule
        Synchronous & A & B \\
        \midrule
        Asynchronous & C & D  
        % \bottomrule
    \end{tabular}
\end{table}

To formally study each of the above problems, we define the objective of resilient distributed optimization by the notion of {\em $(f,r;\epsilon)$-resilience} stated below. Recall that the Euclidean distance between a point $x$ and a set $Y$ in $\mathbb{R}^d$, denoted by $\dist{x}{Y}$, is defined to be
\begin{equation*}
     \dist{x}{Y}=\inf_{y\in Y}\dist{x}{y}=\inf_{y\in Y}\norm{x-y},
\end{equation*}
where $\norm{\cdot}$ represents the Euclidean norm. 

\begin{definition}[$(f,r;\epsilon)$-resilience]
    For $\epsilon \geq 0$, a distributed optimization algorithm is said to be $(f,r;\epsilon)$-resilient if its output $\widehat{x}$ satisfies
    \begin{equation*}
         \dist{\widehat{x}}{\arg\min_{x\in\R^d}\sum_{i \in\H}Q_i(x)}\leq\epsilon
    \end{equation*}
    for each set $\H$ of $n-f$ non-faulty agents, despite the presence of up to $f$ faulty agents and up to $r$ stragglers. % in the system.
\end{definition}

In this paper, we show how {\em redundancy} in agents' cost functions can be utilized to obtain $(f,r;\epsilon)$-resilience. Specifically, we consider a redundancy property of the agents' cost functions, named $(f,r;\epsilon)$-redundancy, as defined below. Recall that the Hausdorff distance between two sets $X$ and $Y$ in $\mathbb{R}^d$, denoted by $\dist{X}{Y}$, is defined to be %follows:
\begin{equation*}
    \dist{X}{Y}\triangleq\max\left\{\sup_{x\in X}\dist{x}{Y}, \sup_{y\in Y}\dist{y}{X}\right\}.
\end{equation*}

\begin{definition}[$(f,r;\epsilon)$-redundancy]
    \label{def:redundancy}
    For $\epsilon \geq 0$, the agents' cost functions are said to satisfy the $(f,r;\epsilon)$-redundancy property if and only if the distance between the minimum point sets of the aggregated cost functions of any pair of subsets of agents $S,\widehat{S}\subsetneq\{1,...,n\}$, where $\mnorm{S}=n-f$, $\mnorm{\widehat{S}}\geq n-r-2f$ and $\widehat{S}\subsetneq S$, is bounded by $\epsilon$, i.e.,
    \begin{equation*}
         \dist{\arg\min_x\sum_{i\in S}Q_i(x)}{\arg\min_x\sum_{i\in\widehat{S}}Q_i(x)}\leq\epsilon.
    \end{equation*}
\end{definition}

It should be noted that for any distributed optimization problem, for any $r< n$ and $f < (n - r)/2$, there exists $\epsilon \geq 0$ such that the agents' cost functions satisfy the $(f,r;\epsilon)$-redundancy property. Intuitively, the above redundancy property indicates the loss in accuracy due to losing some of agents in the system during the optimization process. For a fixed $f$ and $r$, smaller $\epsilon$ indicates more redundancy. On the other hand, for a fixed $\epsilon$, smaller $f$ or $r$ indicates less redundancy.

Next, we consider distributed gradient descent (DGD) based algorithms where a server maintains an estimate for the optimum. In each iteration of the algorithm, the server sends this estimate to the agents, and  the agents send the gradients of their local cost functions at that estimate to the server. The server uses these gradients to update the estimate.
% For the rest of this paper, we consider DGD-based 
Specifically, Algorithm~\ref{alg} is a framework for several instances of the distributed optimization algorithms presented later in this paper.
The main difference between these different algorithms is in the manner in which they perform the iterative update \eqref{eqn:update} at the server -- different gradient aggregators are used when solving different instances of the distributed optimization problems.
A gradient aggregator $\ga(\cdot;n,f,r):\R^{d\times (n-r)}\rightarrow\R^d$ is a function that takes $n-r$ vectors in $\R^d$ (gradients) and outputs a vector in $\R^d$ for the update, with the knowledge that there are up to $f$ Byzantine faulty agents and up to $r$ stragglers out of the $n$ agents in the system. The compact set $\W$ contains the true solution to the problem. We will explain the details when introducing our convergence analyses.

\begin{algorithm}[t]
    \SetAlgoLined
    \caption{Resilient distributed gradient-descent under $(f,r;\epsilon)$-redundancy}
    \label{alg}
    \textbf{Input:} $n$, $f$, $r$, $\epsilon$. A convex compact set $\W$. Each agent $i$ has its cost function $Q_i(x)$.
    %In the beginning of the algorithm, the server randomly 
    
    \vspace{2pt}
    The initial estimate of the optimum,
    $x^0$, is chosen by the server in $\W$. The new estimate $x^{t+1}$ is computed in iteration $t\geq 0$ as follows:
    
    % \textbf{Step S1:} The server requests from each agent the gradient of its local cost function at the current estimate $x^t$. Each agent $j$ will then be expected to send to the server the gradient $\nabla Q_j(x^t)$ with timestamp $t$.
    
    % \textbf{Step S2:} The server waits until it receives $n-r$ gradients with correct timestamp of $t$. Suppose $S^t\subsetneq\{1,...,n\}$ is the set of agents whose gradients are received by the server at step $t$ where $\mnorm{S^t}=n-r$. The server updates its estimate to
    % \begin{equation}
    %      x^{t+1}=\left[x^t-\eta_t \, \sum_{j\in S^t}\nabla Q_j(x^t) \right]_\W \label{eqn:update}
    % \end{equation}
    % where $\eta_t\geq0$ is a \textit{diminishing} step-size for iteration $t$. Specifically, 
    % \begin{equation}
    %     \sum_{i=0}^\infty\eta_t=\infty,\textrm{ and }\sum_{i=0}^\infty\eta_t^2<\infty.
    %     \label{eqn:diminishing}
    % \end{equation}
\begin{description}[nosep]
    \item[Step 1:] The server requests each agent for the gradient of its local cost function at the current estimate $x^t$. Each agent $j$ is expected to send to the server the gradient (or stochastic gradient) $g_j^t$ with timestamp $t$.  
    
    \item[Step 2:] The server waits until it receives $n-r$ gradients with the timestamp of $t$. Suppose $S^t\subseteq\{1,...,n\}$ is the set of agents whose gradients are received by the server at step $t$ where $\mnorm{S^t}=n-r$. The server updates its estimate to
    \begin{equation}
         x^{t+1}=\left[x^t-\eta_t \, \ga\left(g_j^t|\,j\in S^t;n,f,r\right) \right]_\W \label{eqn:update}
    \end{equation}
    where $\eta_t\geq0$ is the step-size for each iteration $t$,
    % where $\eta_t\geq0$ is a \textit{diminishing} step-size for iteration $t$. Specifically, 
    % \begin{equation}
    %     \sum_{i=0}^\infty\eta_t=\infty,\textrm{ and }\sum_{i=0}^\infty\eta_t^2<\infty.
    %     \label{eqn:diminishing}
    % \end{equation}
    % $\W$ is a convex compact set that contains $x^*\in\,\arg\min_{x\in\mathbb{R}^d}\sum_{i\in\H}Q_i(x)$, 
    and $[\,\cdot\,]_{\mathcal{W}}$ denotes a projection onto $\mathcal{W}$.
\end{description}
\end{algorithm}

\subsection{Distributed machine learning}
\label{sub:learning}

Distributed machine learning is a special case of distributed optimization. We consider %Each non-faulty agent $i$ in each iteration $t$ samples a batch of $k$ data points $\z_j^t=\{z_{i1}^t,...,z_{ik}^t\}$ from its data generation distribution $\mathcal{D}_i$. 
 a machine learning model $\Pi$ characterized by a real-valued parameter vector $x\in\R^d$.
Each agent $i$ has a data generation distribution $\mathcal{D}_i$ over $\R^m$. Each data point $z\in\R^m$ is a real-valued vector that incurs a \textit{cost} defined by a \textit{loss function} $\ell:(x;z)\mapsto\R$. The \textit{expected loss function} for each agent $j$ can be defined as
\begin{equation}
     Q_i(x)=\E_{z\sim\mathcal{D}_i}\ell(x;z),~\textrm{ for all }x\in\R^d. \label{eqn:learning-cost}
\end{equation}
The goal of training a machine learning model $\Pi$ is to minimize the aggregate cost over all the agents, which is of the same form as \eqref{eqn:goal}. 

Stochastic gradient descent is a more efficient method than descent-based method using full gradients for machine learning tasks \citep{bottou1998online, bottou2008tradeoffs, bottou2018optimization}, and therefore more commonly used. \textit{Resilient distributed stochastic learning} problem also has 4 types, corresponding to the four problems in Table~\ref{tab:problems}, summarized below. To solve these problems, we use stochastic optimization algorithms, with each agent's expected cost function defined in \eqref{eqn:learning-cost}.

\begin{description}[nosep, leftmargin=\labelwidth]
    \item[Problem AS] Synchronous system, no faulty agents
    \item[Problem BS] Synchronous system with faulty agents
    \item[Problem CS] Asynchronous system, no faulty agents
    \item[Problem DS] Asynchronous system with faulty agents
\end{description}

We discuss the stochastic version of Algorithm~\ref{alg} with D-SGD in detail in Section~\ref{sec:stochastic}.

\subsection{Open problems in the problem space}
\label{sub:openprob}

We briefly discuss the current status of problems listed in Table~\ref{tab:problems} and their distributed learning counterparts.

Problems A, B and AS have been solved previously. In particular,
as stated above, Problem A can be solved by DGD with standard convexity assumptions on the cost functions without redundancy. %\citep{nedic2009distributed, bertsekas2015parallel}. 
% Problem B has Byzantine faulty agents in the system, Problem C has stragglers in the system, and Problem D has both Byzantine agents and straggles at the same time. 
It was previously shown that $(f,0;\epsilon)$-redundancy is necessary and sufficient to solve Problem B \citep{gupta2020resilience, liu2021approximate}. 
For the stochastic case, Problem AS has been solved using D-SGD with convexity assumptions and without redundancy \citep{li2014communication}. 

In this paper, we show how to utilize redundancy to solve Problems C, D, BS, CS, and DS. The rest of the paper is organized as follows: We discuss other related work in Section~\ref{sec:related}. We then provide analyses on different versions of Algorithm~\ref{alg}, namely asymptotic convergence for optimization problems in Section~\ref{sec:full-grad}, and convergence rate results for learning problems in Section~\ref{sec:stochastic}. Empirical results are provided in Section~\ref{sec:experiments}. Finally, we summarize the paper and discuss limitations in our results in Section~\ref{sec:summary}. 

Proofs for all theorems and the code for experiments can be found in the supplementary materials.
\section{Related work}
\label{sec:related}

\textbf{Byzantine fault-tolerant optimization} Byzantine agents make the goal of distributed optimization \eqref{eqn:goal} generally impossible.
However, prior work has shown that the goal of minimizing the aggregate cost function of \textit{only} the non-faulty agents is achievable \citep{gupta2020resilience}, i.e., finding a point $x_\H$ such that
\begin{align}
    \textstyle x_\H\in\arg\min_x\sum_{i\in\H}Q_i(x),
    \label{eqn:goal-ft}
\end{align}
where $\H$ is a set of non-faulty agents. 
 Various methods are proposed to solve Byzantine fault-tolerant optimization or learning problems \citep{liu2021survey}, including robust gradient aggregation \citep{blanchard2017machine, chen2017distributed}, gradient coding \citep{chen2018draco}, and other methods \citep{xie2018zeno, yin2018byzantine}.

% Note that similarly, in Byzantine fault-tolerant distributed learning, the behavior of non-faulty agents follows the analysis in Section~\ref{sub:learning}, and the goal is of the same form as \eqref{eqn:goal-ft}.

\textbf{Redundancy in cost functions} Prior research showed promising results on how redundancy in cost functions can be utilized to achieve robustness in distributed optimization problems. \citet{gupta2020resilience} showed that what is equal to $(f,0;0)$-redundancy is both necessary and sufficient to solve Byzantine optimization problems \textit{exactly}, i.e., $(f,0;0)$-resilience. \citet{liu2021approximate} further showed that $(f,0;\epsilon)$-redundancy is necessary to achieve $(f,0;\epsilon)$-resilience, and sufficient to achieve $(f,0;2\epsilon)$-resilience. In practice, $(f,0;\epsilon)$-redundancy property helps DGD-based algorithms to achieve $(f,0;\bigO(\epsilon))$-resilience, where the value of $\bigO(\epsilon)$ is decided by the gradient aggregation rule \citep{liu2021approximate}. In this paper, we will further provide results on the open problems mentioned in Section~\ref{sub:openprob} with $(f,r;\epsilon)$-redundancy. To the best of our knowledge, we are the first to characterize redundancy this way, and use it to solve both Byzantine and asynchronous problems.

To rigorously analyze the utility of $(f,r;\epsilon)$-redundancy, we consider D-(S)GD in this paper. Yet one could also apply the redundancy property in other distributed optimization algorithms such as federated local SGD \citep{gupta2021byzantine-fed}.

\textbf{Asynchronous optimization} Unlike faulty agents, the existence of stragglers does not necessarily render optimization problems unsolvable, but it can make a synchronous process intolerably long. 
 In a shared-memory system, where clients and server communicate via shared memory, prior works have studied asynchronous optimization.
Specifically, prior works show that distributed optimization problems can be solved using \textit{stale} gradients with a constant delay \citep{langford2009slow} or bounded delay \citep{agarwal2012distributed, feyzmahdavian2016asynchronous}.
Furthermore, methods such as \textsc{Hogwild!} allow lock-free updates in shared memory \citep{niu2011hogwild}. Other works use variance reduction and incremental aggregation methods
to improve the convergence rate \citep{roux2012stochastic, johnson2013accelerating, defazio2014saga, shalev2013accelerated}. Although with a different architecture, these results indicate it is possible to solve asynchronous optimization problems using stale gradients. Still, a bound on allowable delay is needed, which convergence rate is related to.

\textit{Coding} has been used to mitigate the effect of stragglers or failures \citep{lee2017speeding, karakus2017encoded, karakus2017straggler, yang2017coded}.
Tandon et al. \citep{tandon2017gradient} proposed a framework using maximum-distance separable coding across gradients to tolerate failures and stragglers. Similarly, Halbawi et al. \citep{halbawi2018improving} adopted coding to construct a coding scheme with a time-efficient online decoder.
Karakus et al. \citep{karakus2019redundancy} proposed an encoding distributed optimization framework with deterministic convergence guarantee.
% Kadhe et al. \citep{kadhe2019gradient} considered stragglers in an adversarial setting, providing a way to construct approximate gradient codes lower-bounding the number of adversarial stragglers needed to inflict certain approximation error. Baharav et al. \citep{baharav2018straggler} presented a product-code-based method for distributed matrix multiplication with stragglers.
Other replication- or repetition-based techniques involve either task-rescheduling or assigning the same tasks to multiple nodes \citep{ananthanarayanan2013effective, gardner2015reducing, shah2015redundant, wang2015using, yadwadkar2016multi}. These previous methods rely on algorithm-created redundancy of data or gradients to achieve robustness. However, $(f,r;\epsilon)$-redundancy is a property of the cost functions themselves, allowing us to exploit such redundancy without extra effort.
\section{Problems C and D}
\label{sec:full-grad}

In this section, we study resilient distributed optimization Problems C and D with redundancy in cost functions. From now on, we use $[n]$ as a shorthand for the set $\{1,...,n\}$. 

\subsection{Asynchronous optimization}
\label{sub:async}

Consider Problem C in Table~\ref{tab:problems}. Specifically, the resilient optimization problem where there are up to $r$ asynchronous agents and 0 Byzantine agents in the system. The goal is to solve \eqref{eqn:goal}. We first introduce some standard assumptions on the cost functions that are necessary for our analysis.

\begin{assumption}
    \label{assum:lipschitz}
    For each (non-faulty) agent $i$, the function $Q_i(x)$ is $\mu$-Lipschitz smooth, i.e., $\forall x, x'\in\mathbb{R}^d$, 
    \begin{equation}
        \norm{\nabla Q_i(x)-\nabla Q_i(x')}\leq\mu\norm{x-x'}.
    \end{equation}
\end{assumption}

\begin{assumption}
    \label{assum:strongly-convex}
    For any set $S\subset[n]$, we define the average cost function to be $Q_S(x)=({1}/{\mnorm{S}})\sum_{j\in S}Q_j(x)$. We assume that $Q_S(x)$ is $\gamma$-strongly convex for any $S$ subject to $\mnorm{S}\geq n-r$, i.e., $\forall x, x'\in\mathbb{R}^d$, 
    \begin{equation}
        \iprod{\nabla Q_S(x)-\nabla Q_S(x')}{x-x'}\geq\gamma\norm{x-x'}^2.
    \end{equation}
\end{assumption}

Given Assumption~\ref{assum:strongly-convex}, there is only one minimum point for any $S$, $\mnorm{S}\geq n-r$. Let us define $ x^*=\arg\min_x\sum_{j\in[n]}Q_j(x)$.
Thus, $x^*$ is the unique minimum point for the aggregate cost functions of all agents. We assume that
\begin{equation}
    \label{eqn:existence-exact}
    x^*\in \W.
\end{equation}
Recall that $\W$ is used in \eqref{eqn:update}, which is an arbitrary convex compact set. For example, $\W$ can be a hypercube $[-a,a]^d$ such that each element of $x^*$ lies in $[-a,a]$. This assumption is used in our theoretical analysis.

Recall \eqref{eqn:update} in \textbf{Step 2} of our Algorithm~\ref{alg}. We define the gradient aggregation rule for asynchronous optimization to be
\begin{equation}
    \mathsf{GradAgg}\left(g_j^t|j\in S^t;n,f,r\right)=\sum_{j\in{S^t}}g_j^t%\nabla Q_j(x^t)
    \label{eqn:aggregation-rule}
\end{equation}
for every iteration $t$. That is, the algorithm updates the current estimate using the sum of the first $n-r$ gradients it receives. Note that $g_j^t=\nabla Q_j(x^t)$ is the full gradient.
% Note that although all $n$ gradients are formally mentioned in the aggregation rule notation, only those in $S^t$ are actually used, and in practice the algorithm only need to wait for $n-r$ gradients. With this notation, we state the lemma:

For this algorithm, we present below an asymptotic convergence guarantee of our proposed algorithm described in Section~\ref{sec:intro}. 
\begin{theorem}
     Suppose that Assumptions~\ref{assum:lipschitz} and \ref{assum:strongly-convex} hold true, and the agents' cost functions satisfy $(0,r;\epsilon)$-redundancy. 
     Assume that $\eta_t$ in (\ref{eqn:update}) satisfies $\sum_{t=0}^\infty\eta_t=\infty$ and $\sum_{t=0}^\infty\eta_t^2<\infty$.
     Let $\alpha$ and $D$ be defined as follows: $$\alpha\triangleq 1-\frac{r}{n}\cdot\frac{\mu}{\gamma}>0\textrm{ and }\D\triangleq\frac{2r\mu}{\alpha\gamma}\epsilon.$$ 
     Then, for the proposed Algorithm~\ref{alg}, 
    \begin{equation*}
        \lim_{t\rightarrow\infty}\norm{x^t-x^*}\leq\D, %x^t\xrightarrow[t\rightarrow\infty]{}x^* 
    \end{equation*}
    where $x^*=\arg\min_x\sum_{i\in[n]}Q_i(x)$.
    \label{thm:approx}
\end{theorem}
Intuitively, when $(0,r;\epsilon)$-redundancy is satisfied, our algorithm is guaranteed to output an approximation of the true minimum point of the aggregate cost functions of all agents, and the distance between algorithm's output and the true minimum $x^*=\arg\min_x\sum_{i\in[n]}Q_i(x)$ is bounded. The error bound $\D$ is linear to $\epsilon$.

\subsubsection{Utilizing stale gradients}
\label{sub:stale}

Let us denote $T^{t;k}$ the set of agents whose latest gradients received by the server at iteration $t$ is computed using the estimate $x^{k}$. Suppose $S^{t;k}$ is the set of agents whose gradients computed using the estimate $x^t$ is received by iteration $t$. $T^{t;k}$ can be defined in an inductive way: (i) $T^{t;t}=S^{t;t}$, and (2) $T^{t;t-i}=S^{t;t-i}\backslash\bigcup_{k=0}^{i-1}T^{t;t-k}$, $\forall i\geq1$.
% suppose $S^t$ ($\forall t$) is the set of agents sending their gradients at estimate $x^t$ received during iteration $t$. Let's denote $T^{t;t-i}$ the set of agents that at iteration $t$ send their latest gradients using the estimate $x^{t-i}$. $T^{t;t-i}$ can be defined in an inductive way:
% % \begin{align}
% %     &T^{t;t}=S^t, \textrm{ and} \\
% %     &T^{t;t-i}=S^{t-i}\backslash T^{t;t-i+1},~\forall i\geq1.
% % \end{align}
% (i) $T^{t;t}=S^t$, and (ii) $T^{t;t-i}=S^{t-i}\backslash T^{t;t-i+1},~\forall i\geq1$.
Note that the definition implies $T^{t;t-i}\cap T^{t;t-j}$ for any $i\neq j$. Let us further define $T^t=\bigcup_{i=0}^\tau T^{t;t-i}$, where $\tau\geq0$ is a predefined \textit{straggler parameter}.

To extend Algorithm~\ref{alg} to use gradients of previous iterations, in \textbf{Step 2} we replace update rule \eqref{eqn:update} to% the following one:
\begin{equation}
    \label{eqn:update-straggler}
     x^{t+1}=\left[x^t-\eta_t\sum_{i=0}^\tau\sum_{j\in T^{t;t-i}}\nabla Q_j(x^{t-i})\right]_\W,
\end{equation}
and now we only require the server wait till $T^t$ is at least $n-r$.
Intuitively, the new algorithm updates its iterative estimate $x^t$ using the latest gradients from no less than $n-r$ agents at each iteration $t$. 

\begin{theorem}
    Suppose Assumption~\ref{assum:lipschitz} and \ref{assum:strongly-convex} hold true, and the cost functions of all agents satisfies $(0,r;\epsilon)$-redundancy. 
     Assume that $\eta_t$ in (\ref{eqn:update-straggler}) satisfies $\sum_{t=0}^\infty\eta_t=\infty$, $\sum_{t=0}^\infty\eta_t^2<\infty$, and $\eta_t\geq\eta_{t+1}$ for all $t$.
     Let $\alpha$ and $D$ be defined as follows: 
     $$\alpha\triangleq1-\frac{r}{n}\cdot\frac{\mu}{\gamma}>0 \textrm{ and } D\triangleq\frac{2r\mu}{\alpha\gamma}\epsilon.$$ 
     Then, suppose there exists a $\tau\geq0$ such that $\mnorm{T^t}\geq n-r$ for all $t$, for the proposed algorithm with update rule \eqref{eqn:update-straggler}, 
    \begin{equation}
        \lim_{t\rightarrow\infty}\norm{x^t-x^*}\leq\D, %x^t\xrightarrow[t\rightarrow\infty]{}x^* 
    \end{equation}
    where $x^*=\arg\min_x\sum_{j\in[n]}Q_j(x)$.
    \label{thm:approx-generalized}
\end{theorem}
%The proof of this theorem is provided in Appendix~\ref{appdx:proof-thm-generalized}. 
This algorithm accepts gradients at most $\tau$-iteration stale. 
Theorem~\ref{thm:approx-generalized} shows exactly the same error bound $\D$ as Theorem~\ref{thm:approx}, independent from $\tau$, indicating that by using stale gradients, the accuracy of the output would not be affected, so long as the number of gradients used in each iteration is guaranteed by properly choosing $\tau$. Still, the convergence rate (the number of iterations needed to converge) will be effected by $\tau$.

\subsection{Asynchronous Byzantine optimization}
\label{sub:async-ft}

Consider Problem D in Table~\ref{tab:problems}. Specifically, the resilient optimization problem where there are up to $r$ asynchronous agents and up to $f$ Byzantine agents in the system, for which overlapping is possible. Due to the existence of Byzantine agents, the goal here is to solve \eqref{eqn:goal-ft}. Similar to what we do in Section~\ref{sub:async}, we first introduce some assumptions on the cost functions that only apply to non-faulty agents for our analysis. Suppose $\H\subset[n]$ is a subset of non-faulty agents with $\mnorm{\H}=n-f$. Apart from Assumption~\ref{assum:lipschitz}, we need to modify the strong convexity assumption.
% \begin{assumption}
%     \label{assum:lipschitz-ft}
%     For each non-faulty agent $i$, the function $Q_i(x)$ is $\mu$-Lipschitz smooth, i.e., $\forall x, x'\in\mathbb{R}^d$, 
%     \begin{equation}
%         \norm{\nabla Q_i(x)-\nabla Q_i(x')}\leq\mu\norm{x-x'}.
%     \end{equation}
% \end{assumption}

\begin{assumption}
    \label{assum:strongly-convex-ft}
    For any set $S\subset\H$, we define the average cost function to be $Q_S(x)=({1}/{\mnorm{S}})\sum_{j\in S}Q_j(x)$. We assume that $Q_S(x)$ is $\gamma$-strongly convex for any $S$ subject to $\mnorm{S}\geq n-f$, i.e., $\forall x, x'\in\mathbb{R}^d$, 
    \begin{equation}
        \iprod{\nabla Q_S(x)-\nabla Q_S(x')}{x-x'}\geq\gamma\norm{x-x'}^2.
    \end{equation}
\end{assumption}
Similar to the conclusion in Section~\ref{sub:async} regarding the existence of a minimum point \eqref{eqn:existence-exact}, we also require the existence of a solution to the fault-tolerant optimization problem: 
% \begin{assumption}[Existence]
%     \label{assum:existence-ft}
    For each subset of non-faulty agents $S$ with $\mnorm{S}=n-f$, we assume that there exists a point $x_S\in\arg\min_{x\in\R^d}\sum_{j\in S}Q_j(x)$ such that $x_S\in\W$.
% \end{assumption}
% Suppose $\H$ is an arbitrary set of non-faulty agents with $\mnorm{\H}=n-f$. 
By Assumptions~\ref{assum:strongly-convex-ft}, %and \ref{assum:existence-ft}
there exists a unique minimum point $x_\H$ in $\W$ that minimize the aggregate cost functions of agents in $\H$. Specifically,
\begin{equation}
     \{x_\H\}=\W\cap\arg\min_{x\in\R^d}\sum_{j\in\H}Q_j(x).
    \label{eqn:existence-ft}
\end{equation}

Recall \eqref{eqn:update} in \textbf{Step 2} of our Algorithm~\ref{alg}. We define the gradient aggregation rule for Problem D to be
\begin{align}
    &\mathsf{GradAgg}\left(g_j^t|j\in S^t;n,f,r\right) 
    =\mathsf{GradFilter}\left(g_j^t|j\in S^t;n-r,f\right).
    \label{eqn:aggregation-rule-async-ft}
\end{align}
for every iteration $t$, where $\mathsf{GradFilter}$ is a \textit{robust aggregation rule}, or a \textit{gradient filter}, that allows fault-tolerance \citep{liu2021approximate, damaskinos2019aggregathor, karimireddy2021learning}. Specifically, a gradient filter $\mathsf{GradFilter}(\cdot;m,f):\R^{d\times m}\rightarrow\R^d$ is a function that takes $m$ vectors of $d$-dimension and outputs a $d$-dimension vector given that there are up to $f$ Byzantine agents. Generally, $m>f\geq0$. Each agent $j$ sends a vector
\begin{equation}
    g_j^t=\left\{\begin{array}{cl}
        \nabla Q_j(x^t), & \textrm{ if the agent is non-faulty,} \\
        \textrm{arbitrary vector}, & \textrm{ if the agent is faulty} 
    \end{array}\right.
\end{equation}
to the server at iteration $t$. 

Following the above aggregation rule, the server receives the first $n-r$ vectors from the agents in the set $S^t$, and send the vectors through a gradient filter. 

\begin{theorem}
    \label{thm:async-fault-toler}
    Suppose that Assumptions~\ref{assum:lipschitz} and \ref{assum:strongly-convex-ft} hold true, and the cost functions of all agents satisfies $(f,r;\epsilon)$-redundancy. Assume that $\eta_t$ %in \eqref{eqn:step-size} 
    satisfies $\sum_{t=0}^\infty=\infty$ and $\sum_{t=0}^\infty\eta_t^2<\infty$. Suppose that $\norm{\mathsf{GradFilter}\left(n-r,f;\,\left\{g_i^t\right\}_{i\in S^t}\right)}<\infty$ for all $t$. The proposed algorithm with aggregation rule \eqref{eqn:aggregation-rule-async-ft} satisfies the following: For some point $x_\H\in\W$, if there exists a real-valued constant $\D^*\in\left[0,\max_{x\in\W}\norm{x-x_\H}\right)$ and $\xi>0$ such that for each iteration $t$,
    \begin{equation*}
        \phi_t\triangleq\iprod{x^t-x_\H}{\mathsf{GradFilter}\left(n-r,f;\,\left\{g_i^t\right\}_{i\in S^t}\right)}\geq\xi\textrm{ when }\norm{x^t-x_\H}\geq\D^*,
    \end{equation*}
    we have $\lim_{t\rightarrow\infty}\norm{x^t-x_\H}\leq\D^*$.
\end{theorem}
% Theorem~\ref{thm:async-fault-toler} already stands combining \eqref{eqn:aggregation-rule-ft} and Lemma~\ref{lemma:bound}. 
Intuitively, so long as the gradient filter in use satisfies the desired properties in Theorem~\ref{thm:async-fault-toler}, our adapted algorithm can tolerate up to $f$ Byzantine faulty agents and up to $r$ stragglers in distributed optimization. Such kind of gradient filters includes CGE \citep{liu2021approximate} and coordinate-wise trimmed mean \citep{yin2018byzantine}. %, using a similar proof as in \cite[Section 4]{liu2021approximate}. 
For the gradient filter CGE, we obtain the following parameters in Theorem~\ref{thm:async-fault-toler}:
\begin{gather*}
    \alpha = 1 - \frac{f-r}{n-r} + \frac{2\mu}{\gamma}\cdot\frac{f+r}{n-r}>0, \\
    \xi = \alpha m \gamma \delta \left( \frac{4\mu (f+r)\epsilon}{\alpha\gamma} + \delta \right) > 0, \\
    \D^* = \frac{4\mu (f+r)\epsilon}{\alpha\gamma}+\delta,
\end{gather*}
where $\delta>0$ is an arbitrary positive number.

Intuitively, by applying CGE gradient filter, the output of our new algorithm can converge to a $\epsilon$-related area centered by the true minimum point $x_\H$ of aggregate cost functions of non-faulty agents, with up to $f$ Byzantine agents and up to $r$ stragglers. 

Results in Theorems~\ref{thm:approx}, \ref{thm:approx-generalized}, and the specific result for CGE shows that Algorithm~\ref{alg} is $(f,r;\O(\epsilon))$-resilient for Problems C and D.
Note that when $f=0$, Problem D becomes Problem C with $(0,r;\epsilon)$-redundancy, while when $r=0$, Problem D becomes Problem B with $(f,0;\epsilon)$-redundancy solved in \citep{liu2021approximate}. That is, results in Theorem~\ref{thm:async-fault-toler} using CGE with $(f,r;\epsilon)$-redundancy generalizes Problems B and C.\footnote{Theorem~\ref{thm:async-fault-toler} using CGE does not match exactly the special case Theorem~\ref{thm:approx} when $f=0$, for some inequalities in the proofs are not strict bounds.}

\section{Resilient distributed stochastic machine learning problems}
\label{sec:stochastic}

As discussed in Section~\ref{sub:learning}, resilient distributed machine learning problems can be formulated in the same forms of resilient distributed optimization problems, but the stochastic method D-SGD is more commonly used than DGD. In this section, we show how to utilize $(f,r;\epsilon)$-redundancy in solving resilient distributed machine learning problems: Problems BS, CS, and DS.

We first briefly revisit the computation of stochastic gradients in a distributed machine learning system. To compute a stochastic gradient in iteration $t$, a (non-faulty) agent $i$ samples $k$ i.i.d. data points $z_{i_1}^t,...,z_{i_k}^t$ from its distribution $\mathcal{D}_i$ and computes
\begin{equation}
     g_i^t=\frac{1}{k}\sum_{i=1}^k\nabla\ell(x^t, z_{i_j}^t),
    \label{eqn:def-g-i-t}
\end{equation}
where $k$ is referred to as the \textit{batch size}.

We will also use the following notations in this section. Suppose faulty agents (if any) in the system are \textit{fixed} during a certain execution. For each non-faulty agent $i$, let $\z_i^t=\left\{z_{i_1}^t,...,z_{i_k}^t\right\}$ denote the collection of $k$ i.i.d. data points sampled by agent $i$ at iteration $t$. For each agent $i$ and iteration $t$, we define a random variable 
\begin{equation}
    \zeta_i^t=\begin{cases}
        \z_i^t, & \textrm{ agent $i$ is non-faulty}, \\
        g_i^t, & \textrm{ agent $i$ is faulty}.
    \end{cases}
    \label{eqn:zeta-i-t-cge}
\end{equation}
Recall that $g_i^t$ can be an arbitrary $d$-dimensional random variable for each Byzantine faulty agent $i$. For each iteration $t$, let $\zeta^t=\left\{\zeta_i^t,\,i=1,...,n\right\}$, and let $\E_t$ denote the expectation with respect to the collective random variables $\zeta^0,...,\zeta^t$, given the initial estimate $x^0$. Specifically, 
\begin{equation}
    \E_t(\cdot)=\E_{\zeta^0,...,\zeta^t}(\cdot), ~\forall t\geq0.
\end{equation}

Similar to Section~\ref{sec:full-grad}, we make standard Assumptions~\ref{assum:lipschitz} and \ref{assum:strongly-convex} (or Assumption~\ref{assum:strongly-convex-ft} depending on the existence of Byzantine agents). We also need an extra assumption to bound the variance of stochastic gradients from all (non-faulty) agents. 
\begin{assumption}
     \label{assum:bound-grad}
     For each (non-faulty) agent $i$, assume that the variance of $g_i^t$ is bounded. Specifically, there exists a finite real value $\sigma$ such that for all non-faulty agent $i$,
     \begin{equation}
          \E_{\zeta_i^t}\norm{g_i^t-\E_{\zeta_i^t}\left(g_i^t\right)}^2\leq\sigma^2.
          \label{eqn:assum-bound-var-ft}
     \end{equation}
\end{assumption}
We make no assumption over Byzantine agents. 

Suppose $\H\subset[n]$ is a subset of non-faulty agents with $\mnorm{\H}=n-f$, and a solution $x_\H$ exists in $\W$. Note that when $f=0$, $\H=[n]$. The following theorem shows the general results for Problems BS, CS, and DS.
\begin{theorem}
    \label{thm:cge}
    Consider Algorithm~\ref{alg} with stochastic updates and a certain gradient aggregator (\emph{specified later}). Suppose Assumptions~\ref{assum:lipschitz}, \ref{assum:strongly-convex} (or \ref{assum:strongly-convex-ft}), and \ref{assum:bound-grad} hold true, the expected cost functions of non-faulty functions satisfy $(f,r;\epsilon)$-redundancy, $\alpha>0$ and step size in \eqref{eqn:update} $\eta_t=\eta>0$ for all $t$. Let $\M$ be an error-related margin.
    If $\eta<\overline{\eta}$, the following holds true:
    \begin{itemize}[nosep]
        \item The value of a convergence rate parameter $\rho$
        satisfies $0\leq\rho<1$, and 
        \item Given the initial estimate $x^0$ arbitrarily chosen from $\W$, for all $t\geq0$,
            \begin{align}
                \E_{t}\norm{x^{t+1}-x_\H}^2&\leq\rho^{t+1}\norm{x^0-x_\H}^2 + \frac{1-\rho^{t+1}}{1-\rho}\M.
                \label{eqn:expectation-bound-1}
            \end{align}
    \end{itemize}
\end{theorem}
Where the gradient aggergator in use and values of $\alpha$, $\overline{\eta}$, $\rho$, and $\M$ depend on the problems. Specifically, let $\Gamma=\max_{x\in\W}\norm{x-x^*}$, we have

\textbf{Problem BS } For \textit{Byzantine learning}, we have $r=0$. We use the following gradient aggregator
\begin{align*}
    \mathsf{GradAgg}\left(g_j^t|j\in S^t;n,f,0\right)
    =\mathsf{GradFil}\left(g_j^t|j\in[n];n,f\right)
    \label{eqn:aggregation-rule-ft}
\end{align*}
Specifically, we show the result of using CGE gradient filter, where the parameters are defined as follows:

The \textit{resilience margin}
\begin{align*}
     \alpha = 1-\frac{f}{n}\cdot\frac{\gamma+2\mu}{\gamma}.
    % \label{eqn:def-alpha-ft}
\end{align*}
The parameter determines step size
\begin{align*}
     \overline{\eta} = \frac{2n\gamma\alpha}{(n-f)^2\mu^2}.
    % \label{eqn:def-eta-bar-ft}
\end{align*}
The convergence rate parameter
\begin{align*}
    \rho &= 1-2(n-f)\eta\gamma+4f\eta\mu+(n-f)^2\eta^2\mu^2.
    % \label{eqn:def-rho-ft}
\end{align*}
And the error-related margin
\begin{align*}
    \M &= 4n\eta\mu\epsilon\left(2f+(n-f)^2\eta\mu\right)\Gamma + 4n^2(n-f)^2\eta^2\mu^2\epsilon^2  + 2f\eta\sigma\Gamma + (n-f)^2\eta^2\sigma^2.
    % \label{eqn:def-m-ft}
\end{align*}

\textbf{Problem CS } For \textit{asynchronous learning}, we have $f=0$. Note that since $\H=[n]$, $x_\H=x^*$ in \eqref{eqn:expectation-bound-1}. We use the gradient aggregator \eqref{eqn:aggregation-rule}.
The said parameters are defined as follows:
% The \textit{resilience margin}
\begin{align*}
     \alpha = 1-\frac{r}{n}\cdot\frac{\mu}{\gamma},
    % \label{eqn:def-alpha} \\
\end{align*}
% % The parameter determines step size
\begin{align*}
     \overline{\eta} = \frac{2n\gamma\alpha}{(n-r)^2\mu^2},
    % \label{eqn:def-eta-bar} \\
\end{align*}
% % The convergence rate parameter
\begin{align*}
    \rho = 1-2(n\gamma-r\mu)\eta + (n-r)^2\eta^2\mu^2,
    % \label{eqn:def-rho}
\end{align*}
\begin{align*}
    \M &=  4n\eta\mu\epsilon\left(r + (n-r)^2\eta\mu\right)\Gamma + 4n^2(n-r)^2\eta^2\mu^2\epsilon^2 + (n-r)^2\eta^2\sigma^2.
    % \label{eqn:def-m}
\end{align*}

\textbf{Problem DS } For \textit{asynchronous Byzantine learning}, we have $f,r\geq0$. We use the gradient aggregator \eqref{eqn:aggregation-rule-async-ft}.

Specifically, we show the result of using CGE gradient filter, where the parameters are defined as follows:
% The \textit{resilience margin}
\begin{align*}
     \alpha = 1-\frac{f-r}{n-r}+\frac{f+r}{n-r}\cdot\frac{2\mu}{\gamma}, 
    % \label{eqn:def-alpha-cge} \\
\end{align*}
% % The parameter determines step size
\begin{align*}
     \overline{\eta} = \frac{2(n-r)\gamma\alpha}{(n-r-f)^2\mu^2}, 
    % \label{eqn:def-eta-bar-cge} \\
\end{align*}
% % The convergence rate parameter
\begin{align*}
    \rho = 1-2(n-f)\eta\gamma+4(f+r)\eta\mu+(n-r-f)^2\eta^2\mu^2,
    % \label{eqn:def-rho-cge}
\end{align*}
% And the error-bound parameter
\begin{align*}
    \M =& 4(n-r)\eta\mu\epsilon\left(2(f+r)+(n-r-f)^2\eta\mu\right)\Gamma + 4(n-r)^2(n-r-f)^2\eta^2\mu^2\epsilon^2  \nonumber \\
    &+2(f+r)\eta\sigma\Gamma + (n-r-f)^2\eta^2\sigma^2.
    % \label{eqn:def-m-cge}
\end{align*}
For Problem DS, we also need an extra assumption that $n\geq 2f+r/2$ to guarantee that $\rho\geq0$. 

\subsection{Discussions}
\label{sub:stochastic-discuss}

The three results provided above, each for a type of resilient distributed learning problems, indicates that with $(f,r;\epsilon)$-redundancy, there exist algorithms to approximate the true solution to \eqref{eqn:goal} or \eqref{eqn:goal-ft} with D-SGD, where \textit{linear} convergence is achievable, and the error range of that approximation is expected to be proportional to $\epsilon$ and $\sigma$. Specifically, in \eqref{eqn:expectation-bound-1} when $t\rightarrow\infty$, %we have
\begin{equation}
     \lim_{t\rightarrow\infty}\E_t\norm{x^{t+1}-x_H}^2\leq\frac{1}{1-\rho}\M,
\end{equation}
where $\M$ is changes monotonically as $\epsilon$ and $\sigma$. 

Note that the result for Problem DS generalized the two results for Problems BS and CS. Also, \citet{gupta2021byzantine} showed a special case of Problem BS, where all agents have the same data distribution. Since the distribution of each agent can be different, our results can be applied to a broader range of distributed machine learning problems, including heterogeneous problems like \textit{federated learning} \citep{konevcny2015federated}.  

\begin{figure*}[t]
    \centering
    \includegraphics[width=\linewidth]{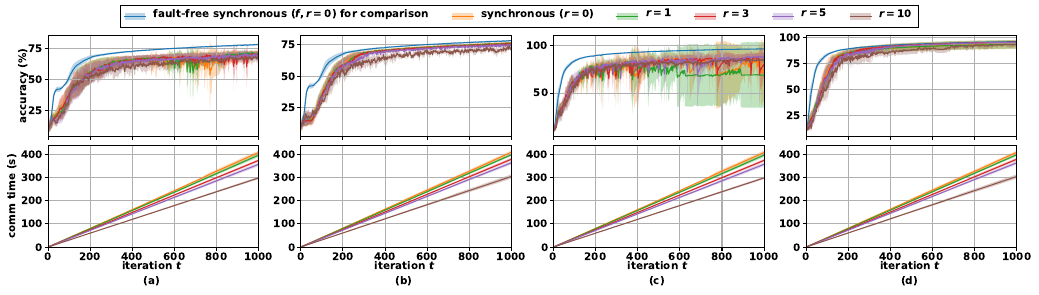}
    \caption{Empirical results for resilient distributed learning tasks with $n=20$, $f=3$, and various $r$'s with Algorithm~\ref{alg} for Problem DS with D-SGD and CGE gradient filter. Datasets: (ab) Fashion-MNIST, (cd) MNIST. Byzantine faults: (ac) \textit{reverse-gradient}, (bd) \textit{label-flipping}. Solid line for average; shade for standard deviation.}
    \label{fig:ds}
\end{figure*}

\section{Empirical studies}
\label{sec:experiments}

In this section, we empirically show the effectiveness of our scheme in Algorithm~\ref{alg} on some benchmark distributed machine learning tasks. It is worth noting that even though the actual values of redundancy parameter $\epsilon$ are difficult to compute, through the following results we can still see that the said redundancy property exists in real-world scenarios and it supports our algorithm to realize its applicability.

We simulate a server-based distributed learning system using multiple threads, one for the server, the rest for agents, with inter-thread communications handled by message passing interface. The simulator is built in Python with PyTorch \citep{paszke2019pytorch} and MPI4py \citep{dalcin2011parallel}, and deployed on a virtual machine with 14 vCPUs and 16 GB memory.

The experiments are conducted on two benchmark image-classification datasets: MNIST \citep{bottou1998online} of monochrome handwritten digits, and Fashion-MNIST \citep{xiao2017fashion} of grayscale images of clothes. Each dataset comprises of 60,000 training and 10,000 testing data points in 10 non-overlapping classes, with each image of size $28\times28$. For each dataset, we train a benchmark neural network LeNet \citep{lecun1998gradient} with 431,080 learnable parameters. 
Data points are divided among agents such that each agent gets 2 out of 10 classes, and each class appears in 4 agents; $\mathcal{D}_i$ for each agent $i$ is unique.
We show the performance of Algorithm~\ref{alg} for \textit{Problem DS}.

In each of our experiments, we simulate a distributed system of $n=20$ agents with $f=3$ and different values of $r=0,1,3,5,10$. Note that when $r=0$, our algorithm becomes the synchronous \textit{Problem BS}. We also compare these results with the fault-free synchronous \textit{Problem AS} ($f,r=0$). We choose batch size $b=128$ for D-SGD, and fixed step size $\eta=0.01$. Performance of algorithms is measured by model accuracy at each step. We also document the cumulative communication time of each setting. Experiments of each setting are run 4 times with different random seeds\footnote{Randomnesses exist in drawing of data points and stragglers in each iteration of each execution.}, and the averaged performance is reported. The results are shown in Figure~\ref{fig:ds}. We show the first 1,000 iterations as there is a clear trend of converging by the end of 1,000 iterations for both tasks in all four settings.

For Byzantine agents, we evaluate two types of faults: \textit{reverse-gradient} where faulty agents reverse the direction of its true gradients, and \textit{label-flipping} where faulty agents label data points of class $i$ as class $9-i$.

As is shown in the first row of Figure~\ref{fig:ds}, Algorithm~\ref{alg} converges in a similar rate, and the learned model reaches comparable accuracy to the one learned by synchronous algorithm at the same iteration; there is a gap between them and the fault-free case, echoing the error bound $\M$. In the second row of Figure~\ref{fig:ds}, we see that by dropping out $r$ stragglers, the communication overhead is gradually reduced with increasing value of $r$. It is worth noting that 
reduction in communication time over $r$ when $r$ is small is more significant than that when $r$ is large, indicating a small number of stragglers work very slow, from which our algorithm improves more in communication overhead.
\section{Summary}
\label{sec:summary}

We studied the impact of $(f,r;\epsilon)$-redundancy in cost functions on resilient distributed optimization and machine learning. Specifically, we presented an algorithm for resilient distributed optimization and learning, and analyzed its convergence when agents' cost functions have $(f,r;\epsilon)$-redundancy - a generic characterization of redundancy in costs functions. We examined the resilient optimization and learning problem space. We showed that, under $(f,r;\epsilon)$-redundancy, Algorithm~\ref{alg} with DGD achieves $(f,r;\O(\epsilon))$-resilience for optimization Problems C and D, and Algorithm~\ref{alg} with D-SGD can solve resilient distributed learning Problems BS, CS, DS with error margins proportional to $\epsilon$.
We presented empirical results showing efficacy of Algorithm~\ref{alg} solving the most generalized Problem DS for resilient distributed learning.

\textbf{Discussion on limitations } 
Note that our results in this paper are proved under strongly-convex assumptions. One may argue that such kind of assumptions are too strong to be realistic. However, previous research has pointed out that although not a global property, cost functions of many machine learning problems are strongly-convex in the neighborhood of local minimizers \citep{bottou2018optimization}. Also, there is a research showing that the results on non-convex cost functions can be derived from those on strongly-convex cost functions \citep{allen2016optimal}, and therefore our results can be applied to a broader range of real-world problems. 
Our empirical results showing efficacy of our algorithm also concur with this argument.

It is also worth noting that the approximation bounds in optimization problems are linearly associated with the number of agents $n$, and the error margins in learning problems are related to $\Gamma$, the size of $\W$. These bounds can be loose when $n$ or $\Gamma$ is large. We do note that in practice $\W$ can be arbitrary, for example, a neighborhood of local minimizers mentioned above, making $\Gamma$ acceptably small. The value of $\epsilon$ can also be small in practice, as indicated by results in Section~\ref{sec:experiments}.

% \input{related}

% \input{asynchrony_v2}
% %\input{learning-discussion}
% \input{extensions}

% % \input{experiment_v2}
% \input{experiment-learning}
% \input{summary}

\subsection*{Acknowledgements}
Research reported in this paper was supported in part by the Army Research Laboratory under Cooperative Agreement W911NF- 17-2-0196, and by the National Science Foundation award 1842198. The views and conclusions contained in this document are those of the authors and should not be interpreted as representing the official policies, either expressed or implied, of the Army Research Laboratory, National Science Foundation, or the U.S. Government. Research reported in this paper is also supported in part by a Fritz Fellowship from Georgetown University.

% \subsubsection*{References}
\bibliographystyle{plainnat}
\bibliography{bib}

\appendix
\vfill \newpage
\vfill \newpage
\section{Proofs of theorems in Section 3}
\label{part:2}

In this section, we present the detailed proof of the asymptotic convergence results presented in Section 3, namely Theorems~\ref{thm:approx}, \ref{thm:approx-generalized}, and \ref{thm:async-fault-toler}. For each of them, we restate the assumptions, then proceed with the proofs of the theorems.

\subsection{Lemma~\ref{lemma:bound}}
\label{appdx:proof-lemma-bound}

First of all, the proofs of the results presented in Section~3 utilize the following lemma:

\begin{mdframed}[default]
\begin{lemma}
\label{lemma:bound}
    Consider the general iterative update rule (\ref{eqn:update}). Let $\eta_t$ satisfy $$\sum_{t=0}^\infty\eta_t=\infty~\textrm{ and }~\sum_{t=0}^\infty\eta_t^2<\infty.$$
    For any given gradient aggregation rule $\ga\left(g_j^t|\,j\in S^t;n,f,r\right)$, if there exists $\M<\infty$ such that
    \begin{equation}
        \norm{\ga\left(g_j^t|\,j\in S^t;n,f,r\right)}\leq\M
        \label{eqn:bounded-aggregator-norm}
    \end{equation} for all $t$, and there exists $\D^*\in\left[0,\max_{x\in\W}\norm{x-x^*}\right)$ and $\xi>0$ such that when $\norm{x^t-x^*}\geq\D^*$,
    \begin{equation}
        \phi_t\triangleq\iprod{x^t-x^*}{\ga\left(g_j^t|\,j\in S^t;n,f,r\right)}\geq\xi,~
        \label{eqn:phi-def}
    \end{equation}
    we have $\lim_{t\rightarrow\infty}\norm{x^t-x^*}\leq\D^*$.
\end{lemma}
\end{mdframed}

Consider the iterative process \eqref{eqn:update}. Assume that $x^*\in\W$. From now on, we use $\mathsf{GradAgg}[t]$ as a short hand for the output of the gradient aggregation rule at iteration $t$, i.e. 
\begin{equation}
    \label{eqn:def-gradagg-t}
    \mathsf{GradAgg}[t]\triangleq\ga\left(g_j^t|\,j\in S^t;n,f,r\right).
\end{equation}

The proof of this lemma uses the following sufficient criterion for the convergence of non-negative sequences:
\begin{lemma}[Ref. \citep{bottou1998online}]
    \label{lemma:converge}
    Consider a sequence of real values $\{u_t\}$, $t=0,1,\dots$. If $u_t\geq0$, $\forall t$, 
    \begin{align*}
        &\sum_{t=0}^\infty(u_{t+1}-u_t)_+<\infty ~\textrm{ implies }~\left\{\begin{array}{l}
            u_t\xrightarrow[t\rightarrow\infty]{}u_\infty<\infty, \\
            \sum_{t=0}^\infty(u_{t+1}-u_t)_->-\infty,
        \end{array}\right.
    \end{align*}
    where the operators $(\cdot)_+$ and $(\cdot)_-$ are defined as follows for a real value scalar $x$:
    \begin{align*}
        (x)_+=\left\{\begin{array}{ll}
            x, & x>0, \\
            0, & \textrm{otherwise},
        \end{array}\right.~\textrm{ and }~
        (x)_-=\left\{\begin{array}{ll}
            0, & x>0, \\
            x, & \textrm{otherwise}.
        \end{array}\right.
    \end{align*}
\end{lemma}

\subsubsection{Proof of Lemma~\ref{lemma:bound}}
Let $e_t$ denote $\norm{x^t-x^*}$. Define a scalar function $\psi$,
\begin{equation}
    \psi(y)=\left\{\begin{array}{ll}
        0, & y<\left(\D^*\right)^2, \\
        \left(y-\left(\D^*\right)^2\right)^2, & \textrm{otherwise}.
    \end{array}\right.
    \label{eqn:psi_def}
\end{equation}
Let $\psi'(y)$ denote the derivative of $\psi$ at $y$. Then (cf. \cite{bottou1998online})
\begin{equation}
    \label{eqn:psi_bnd}
    \psi(z)-\psi(y)\leq(z-y)\psi'(y)+(z-y)^2,~\forall y,z\in\mathbb{R}_{\geq0}.
\end{equation}
Note,
\begin{equation}
    \label{eqn:psi_prime}
    \psi'(y)=\max\left\{0,2\left(y-\left(\D^*\right)^2\right)\right\}.
\end{equation}

Now, define
\begin{equation}
    \label{eqn:ht_def}
    h_t\triangleq\psi(e_t^2).
\end{equation}
From \eqref{eqn:psi_bnd} and \eqref{eqn:ht_def},
\begin{equation}
    h_{t+1}-h_t=\psi\left(e_{t+1}^2\right)-\psi\left(e_t^2\right)\leq\left(e_{t+1}^2-e_t^2\right)\psi'\left(e_t^2\right)+\left(e_{t+1}^2-e_t^2\right)^2,~\forall t\in\mathbb{Z}_{\geq0}.
\end{equation}
From now on, we use $\psi_t'$ as the shorthand for $\psi'\left(e_t^2\right)$, i.e.,
\begin{equation}
    \label{eqn:def-psi-t}
    \psi_t'\triangleq\psi'\left(e_t^2\right).
\end{equation}
From above, for all $t\geq0$,
\begin{equation}
    \label{eqn:ht_1}
    h_{t+1}-h_t\leq\left(e_{t+1}^2-e_t^2\right)\psi'_t+\left(e_{t+1}^2-e_t^2\right)^2.%~\forall t\in\mathbb{Z}_{\geq0}.
\end{equation}
Recall the iterative process \eqref{eqn:update}. Using the non-expansion property of Euclidean projection onto a closed convex set\footnote{$\norm{x-x^*}\geq\norm{[x]_\W-x^*},~\forall w\in\mathbb{R}^d$.}, 
\begin{equation}
    \norm{x^{t+1}-x^*}\leq\norm{x^t-\eta_t\mathsf{GradAgg}[t]-x^*}.%~\forall t\in\mathbb{Z}_{\geq0}.
\end{equation}
Taking square on both sides, and recalling that $e_t\triangleq\norm{x^t-x^*}$,
\begin{equation*}
    e_{t+1}^2\leq e_t^2-2\eta_t\iprod{x_t-x^*}{\mathsf{GradAgg}[t]}+\eta_t^2\norm{\mathsf{GradAgg}[t]}^2.%~\forall t\in\mathbb{Z}_{\geq0}.
\end{equation*}
Recall from \eqref{eqn:phi-def} that $\iprod{x_t-x^*}{\mathsf{GradAgg}[t]}=\phi_t$, therefore,
\begin{equation}
    \label{eqn:proj_bound}
    e_{t+1}^2\leq e_t^2-2\eta_t\phi_t+\eta_t^2\norm{\mathsf{GradAgg}[t]}^2,~\forall t\geq0.
\end{equation}

As $\psi'_t\geq0,~\forall t\in\mathbb{Z}_{\geq0}$, combining \eqref{eqn:ht_1} and \eqref{eqn:proj_bound},
\begin{equation}
    \label{eqn:ht_2}
    h_{t+1}-h_t\leq\left(-2\eta_t\phi_t+\eta_t^2\norm{\mathsf{GradAgg}[t]}^2\right)\psi'_t+\left(e_{t+1}^2-e_t^2\right)^2,~\forall t\geq0.
\end{equation}
Note that 
\begin{equation}
    \mnorm{e_{t+1}^2-e_t^2}=(e_{t+1}+e_t)\mnorm{e_{t+1}-e_t}.
\end{equation}
As $\W$ is assumed compact, there exists
\begin{equation}
    \Gamma=\max_{x\in\W}\norm{x-x^*}\leq\infty.
\end{equation}
Let $\Gamma>0$, since otherwise $\W=\{x^*\}$ only contains one point, and the problem becomes trivial. As $x^t\in\W$, $\forall t\geq0$,
\begin{equation}
    \label{eqn:e_t_bound}
    e_t\leq\Gamma,
\end{equation}
which implies
\begin{equation}
    e_{t+1}+e_t\leq2\Gamma.
\end{equation}
Therefore,
\begin{equation}
    \mnorm{e_{t+1}^2-e_t^2}\leq2\Gamma\mnorm{e_{t+1}-e_t},~\forall t\geq0.
    \label{eqn:e2t-bound-1}
\end{equation}
By triangle inequality,
\begin{equation}
    \mnorm{e_{t+1}-e_t}=\mnorm{\norm{x^{t+1}-x^*}-\norm{x^t-x^*}}\leq\norm{x^{t+1}-x^t}.
    \label{eqn:e2t-bound-2}
\end{equation}
From \eqref{eqn:update} and the non-expansion property of Euclidean projection onto a closed convex set,
\begin{equation}
    \norm{x^{t+1}-x^t}=\norm{\left[x^t-\eta_t\mathsf{GradAgg}[t]\right]_\W-x^t}\leq\eta_t\norm{\mathsf{GradAgg}[t]}.
    \label{eqn:e2t-bound-3}
\end{equation}
So from \eqref{eqn:e2t-bound-1}, \eqref{eqn:e2t-bound-2}, and \eqref{eqn:e2t-bound-3},
\begin{align}
    &\mnorm{e_{t+1}^2-e_t^2}\leq2\eta_t\Gamma\norm{\mathsf{GradAgg}[t]}, \nonumber \\
    \Longrightarrow&\left(e_{t+1}^2-e_t^2\right)^2\leq4\eta_t^2\Gamma^2\norm{\mathsf{GradAgg}[t]}^2.
\end{align}
Substituting above in \eqref{eqn:ht_2},
\begin{align}
    \label{eqn:ht_3}
    h_{t+1}-h_t\leq&\left(-2\eta_t\phi_t+\eta_t^2\norm{\mathsf{GradAgg}[t]}^2\right)\psi'_t+4\eta_t^2\Gamma^2\norm{\mathsf{GradAgg}[t]}^2, \nonumber \\
    =&-2\eta_t\phi_t\psi'_t+(\psi'_t+4\Gamma^2)\eta_t^2\norm{\mathsf{GradAgg}[t]}^2,~\forall t\geq0.
\end{align}

Recall that the statement of Lemma~\ref{lemma:bound} assumes that $\D^*\in[0,\max_{x\in\W}\norm{x-x^*})$, which indicates $\D^*<\Gamma$. Using \eqref{eqn:psi_prime} and \eqref{eqn:e_t_bound}, we have
\begin{equation}
    0\leq\psi'_t\leq2\left(e_t^2-\left(\D^*\right)^2\right)\leq2\left(\Gamma^2-\left(\D^*\right)^2\right)\leq2\Gamma^2.
    \label{eqn:psi_prime_t_bnd}
\end{equation}
Recall from \eqref{eqn:bounded-aggregator-norm} that  $\norm{\mathsf{GradAgg}[t]}\leq\M<\infty$ for all $t$. Substituting \eqref{eqn:psi_prime_t_bnd} in \eqref{eqn:ht_3},
\begin{align}
    \label{eqn:ht_4}
    h_{t+1}-h_t\leq&-2\eta_t\phi_t\psi'_t+(2\Gamma^2+4\Gamma^2)\eta_t^2\M^2=-2\eta_t\phi_t\psi'_t+6\Gamma^2\eta_t^2\M^2,~\forall t\geq0.
\end{align}

Now we use Lemma~\ref{lemma:converge} to show that $h_\infty=0$ as follows. For each iteration $t$, consider the following two cases:
\begin{description}
    \item[Case 1)] Suppose $e_t<\D^*$. In this case, $\psi'_t=0$. By Cauchy-Schwartz inequality,
        \begin{equation}
            \mnorm{\phi_t}=\mnorm{\iprod{x^t-x^*}{\mathsf{GradAgg}[t]}}\leq e_t\norm{\mathsf{GradAgg}[t]}.
        \end{equation}
        By \eqref{eqn:bounded-aggregator-norm} and \eqref{eqn:e_t_bound}, this implies that
        \begin{equation}
            \mnorm{\phi_t}\leq\Gamma\M<\infty.
        \end{equation}
        Thus,
        \begin{equation}
            \label{eqn:phitpsit_1}
            \phi_t\psi'_t=0.
        \end{equation}
    \item[Case 2)] Suppose $e_t\geq\D^*$. Therefore, there exists $\delta\geq0$, $e_t=\D^*+\delta$. From \eqref{eqn:psi_prime}, we obtain that
    \begin{equation}
        \psi_t'=2\left(\left(\D^*+\delta\right)^2-\left(\D^*\right)^2\right)=2\delta\left(2\D^*+\delta\right).
    \end{equation}
    The statement of Lemma~\ref{lemma:bound} assumes that $\phi_t\geq\xi>0$ when $e_t\geq\D^*$, thus, 
        \begin{equation}
            \label{eqn:phitpsit_2}
            \phi_t\psi'_t\geq2\delta\xi\left(2\D^*+\delta\right)>0.%,~\textrm{if}~e_t\geq\D^*+\delta.
        \end{equation}
\end{description}
From \eqref{eqn:phitpsit_1} and \eqref{eqn:phitpsit_2}, for both cases,
\begin{equation}
    \phi_t\psi'_t\geq0, ~\forall t\geq0.
    \label{eqn:phi_psi_bound}
\end{equation}
Combining this with \eqref{eqn:ht_4},
\begin{equation}
    h_{t+1}-h_t\leq6\Gamma^2\eta_t^2\M^2.
\end{equation}
%Recall Lemma~\ref{lemma:converge}. 
From above we have
\begin{equation}
    \left(h_{t+1}-h_t\right)_+\leq6\Gamma^2\eta_t^2\M^2.%,~\forall t\in\mathbb{Z}_{\geq0}.
\end{equation}
Since $\sum_{t=0}^\infty\eta_t^2<\infty$, $\Gamma,\M<\infty$,
\begin{equation}
    \sum_{t=0}^\infty\left(h_{t+1}-h_t\right)_+\leq6\Gamma^2\M^2\sum_{t=0}^\infty\eta_t^2<\infty.
\end{equation}
Then Lemma~\ref{lemma:converge} implies that by the definition of $h_t$, we have $h_t\geq0,~\forall t$, 
\begin{align}
    h_t\xrightarrow[t\rightarrow\infty]{}h_\infty<\infty,~\textrm{and} \label{eqn:upper_bound_h_infty}\\
    \sum_{t=0}^\infty\left(h_{t+1}-h_t\right)_->-\infty.
\end{align}
Note that $h_\infty-h_0=\sum_{t=0}^\infty(h_{t+1}-h_t)$. Thus, from \eqref{eqn:ht_4} we have 
\begin{equation}
    h_\infty-h_0\leq-2\sum_{t=0}^\infty\eta_t\phi_t\psi_t'+6\Gamma^2\M^2\sum_{t=0}^\infty\eta_t^2.
\end{equation}
By \eqref{eqn:ht_def} the definition of $h_t$, $h_t\geq0$ for all $t$. Therefore, from above we obtain
\begin{equation}
    2\sum_{t=0}^\infty\eta_t\phi_t\psi_t'\leq h_0-h_\infty+6\Gamma^2\M^2\sum_{t=0}^\infty\eta_t^2.
    \label{eqn:bound_2_sum}
\end{equation}
% Thus,
% \begin{equation}
%     2\mnorm{\sum_{t=0}^\infty\eta_t\phi_t\psi_t'}\leq h_0+h_\infty+6\Gamma^2\M^2\sum_{t=0}^\infty\eta_t^2.
%     \label{eqn:bound_abs_sum}
% \end{equation}
% By assumption, $\sum_{t=0}^\infty\eta_t^2<\infty$. From \eqref{eqn:upper_bound_h_infty}, $h_\infty<\infty$. Substituting from \eqref{eqn:e_t_bound} that $e_t<\infty$ in \eqref{eqn:ht_def}, we obtain that $h_0=\psi\left(e_0^2\right)<\infty$. Also from \eqref{eqn:ht_def}, $h_t\geq0~(\forall t)$. Therefore, \eqref{eqn:bound_abs_sum} implies that
% \begin{equation}
%     2\mnorm{\sum_{t=0}^\infty\eta_t\phi_t\psi_t'}<\infty.
% \end{equation}
By assumption, $\sum_{t=0}^\infty\eta_t^2<\infty$. Substituting from \eqref{eqn:e_t_bound} that $e_t<\infty$ in \eqref{eqn:ht_def}, we obtain that 
\begin{equation}
    h_0=\psi\left(e_0^2\right)<\infty.
\end{equation} 
Therefore, \eqref{eqn:bound_2_sum} implies that
\begin{equation}
    2\sum_{t=0}^\infty\eta_t\phi_t\psi_t'\leq h_0+6\Gamma^2\M^2\sum_{t=0}^\infty\eta_t^2<\infty.
\end{equation}
Or simply,
\begin{equation}
    \sum_{t=0}^\infty\eta_t\phi_t\psi_t'<\infty.
    \label{eqn:upper_bound_etatphitpsit}
\end{equation}

Finally, we reason below by contradiction that $h_\infty=0$. Note that for any $\zeta>0$, there exists a unique positive value $\beta$ such that $\zeta=2\beta\left(2\D^*+\sqrt{\beta}\right)^2$. Suppose that $h_\infty=2\beta\left(2\D^*+\sqrt{\beta}\right)^2$ for some positive value $\beta$. As the sequence $\{h_t\}_{t=0}^\infty$ converges to $h_\infty$ (see \eqref{eqn:upper_bound_h_infty}), there exists some finite $\tau\in\Z_{\geq0}$ such that for all $t\geq\tau$, 
\begin{align}
    &\mnorm{h_t-h_\infty}\leq\beta\left(2\D^*+\sqrt{\beta}\right)^2 \\
    \Longrightarrow & h_t\geq h_\infty-\beta\left(2\D^*+\sqrt{\beta}\right)^2.
\end{align}
As $h_\infty=2\beta\left(2\D^*+\sqrt{\beta}\right)^2$, the above implies that
\begin{equation}
    h_t\geq \beta\left(2\D^*+\sqrt{\beta}\right)^2, \forall t\geq\tau.
    \label{eqn:ht_lower_bound}
\end{equation}
Therefore (cf. \eqref{eqn:psi_def} and \eqref{eqn:ht_def}), for all $t\geq\tau$,
\begin{eqnarray*}
    \left(e_t^2-\left(\D^*\right)^2\right)^2\geq\beta\left(2\D^*+\sqrt{\beta}\right)^2, \textrm{ or} \\
    \mnorm{e_t^2-\left(\D^*\right)^2}\geq\sqrt{\beta}\left(2\D^*+\sqrt{\beta}\right).
\end{eqnarray*}
Thus, for each $t\geq\tau$, either
\begin{equation}
    e^2_t\geq\left(\D^*\right)^2+\sqrt{\beta}\left(2\D^*+\sqrt{\beta}\right)=\left(\D^*+\sqrt{\beta}\right)^2,
    \label{eqn:et_case_1}
\end{equation}
or
\begin{equation}
    e^2_t\leq\left(\D^*\right)^2-\sqrt{\beta}\left(2\D^*+\sqrt{\beta}\right)<\left(\D^*\right)^2.
    \label{eqn:et_case_2}
\end{equation}
If the latter, i.e., \eqref{eqn:et_case_2} holds true for some $t'\geq\tau$, 
\begin{equation}
    h_{t'}=\psi\left(e_{t'}^2\right)=0,
\end{equation}
which contradicts \eqref{eqn:ht_lower_bound}. Therefore, \eqref{eqn:ht_lower_bound} implies \eqref{eqn:et_case_1}.

From above we obtain that if $h_\infty=2\beta\left(2\D^*+\sqrt{\beta}\right)^2$, there exists $\tau<\infty$ such that for all $t\geq\tau$, 
\begin{equation}
    e_t\geq\D^*+\sqrt{\beta}.
\end{equation}
Thus, from \eqref{eqn:phitpsit_2}, with $\delta=\sqrt{\beta}$, we obtain that %there exists a $\xi>0$, for $\epsilon=2\xi\sqrt{\beta}\left(2\D^*+\sqrt{\beta}\right)>0$ and $\delta=\sqrt{\beta}>0$ in \eqref{eqn:phitpsit_2},
\begin{equation}
    \phi_t\psi_t'\geq2\xi\sqrt{\beta}\left(2\D^*+\sqrt{\beta}\right), \forall t\geq\tau.
\end{equation}
Therefore,
\begin{equation}
    \sum_{t=\tau}^\infty\eta_t\phi_t\psi_t'\geq2\xi\sqrt{\beta}\left(2\D^*+\sqrt{\beta}\right)\sum_{t=\tau}^\infty\eta_t=\infty.
\end{equation}
This is a contradiction to \eqref{eqn:upper_bound_etatphitpsit}. Therefore, $h_\infty=0$, and by \eqref{eqn:ht_def}, the definition of $h_t$, 
\begin{equation}
    h_\infty=\lim_{t\rightarrow\infty}\psi\left(e_t^2\right)=0.
\end{equation}
Hence, by \eqref{eqn:psi_def}, the definition of $\psi(\cdot)$, 
\begin{equation}
    \lim_{t\rightarrow\infty}\norm{x^t-x^*}\leq\D^*.
\end{equation}

\vfill \newpage
\subsection{Proof of Theorem~\ref{thm:approx}}

Theorem~\ref{thm:approx} states the asymptotic convergence of Algorithm~\ref{alg} when solving \textbf{Problem C}, asynchronous optimization with no Byzantine agents ($f=0, r\geq0$), using the following gradient aggregator:
\begin{equation}
    \textstyle\mathsf{GradAgg}\left(g_j^t|j\in S^t;n,f,r\right)=\sum_{j\in{S^t}}\nabla Q_j(x^t)
    \label{eqn:aggregation-rule}
\end{equation}

The result is established upon the following assumptions:

% \begin{assumption}
    % \label{assum:lipschitz}
    \textbf{Assumption \ref{assum:lipschitz}.} 
    \textit{For each (non-faulty) agent $i$, the function $Q_i(x)$ is $\mu$-Lipschitz smooth, i.e., $\forall x, x'\in\mathbb{R}^d$, 
    \begin{equation}
        \norm{\nabla Q_i(x)-\nabla Q_i(x')}\leq\mu\norm{x-x'}.
    \end{equation}}
% \end{assumption}

% \begin{assumption}
    \textbf{Assumption \ref{assum:strongly-convex}.}
    For any set $S\subset[n]$, we define the average cost function to be $Q_S(x)=({1}/{\mnorm{S}})\sum_{j\in S}Q_j(x)$. We assume that $Q_S(x)$ is $\gamma$-strongly convex for any $S$ subject to $\mnorm{S}\geq n-r$, i.e., $\forall x, x'\in\mathbb{R}^d$, 
    \begin{equation}
        \iprod{\nabla Q_S(x)-\nabla Q_S(x')}{x-x'}\geq\gamma\norm{x-x'}^2.
    \end{equation}
% \end{assumption}

Now we prove Theorem~\ref{thm:approx}. Note that $x^*=\arg\min_{x\in\W}\sum_{j\in[n]}Q_j(x)$ is the unique minimum of the aggregate cost functions of all agents, and the vector $g_j^t=\nabla Q_j(x^t)$ is the gradient of $Q_j$ at $x^t$. 

\begin{mdframed}[everyline=true]
% \begin{theorem}
    \textbf{Theorem \ref{thm:approx}.}
    \textit{Suppose that Assumptions~\ref{assum:lipschitz} and \ref{assum:strongly-convex} hold true, and the agents' cost functions satisfy $(0,r;\epsilon)$-redundancy. 
    Assume that $\eta_t$ in (\ref{eqn:update}) satisfies $\sum_{t=0}^\infty\eta_t=\infty$ and $\sum_{t=0}^\infty\eta_t^2<\infty$.
    Let $\alpha$ and $D$ be defined as follows: $$\alpha\triangleq 1-\frac{r}{n}\cdot\frac{\mu}{\gamma}>0 \textrm{ and } D\triangleq\frac{2r\mu}{\alpha\gamma}\epsilon.$$ Then, for the proposed Algorithm~\ref{alg}, 
    \begin{equation}
        \lim_{t\rightarrow\infty}\norm{x^t-x^*}\leq\D, %x^t\xrightarrow[t\rightarrow\infty]{}x^* 
    \end{equation}
    where $x^*=\arg\min_x\sum_{i\in[n]}Q_i(x)$.}
    % \label{thm:approx}
% \end{theorem}
\end{mdframed}

\begin{proof}
\textbf{First}, we need to show that $\norm{\sum_{j\in S^t}\nabla Q_j(x^t)}$ is bounded for all $t$. By Assumption~\ref{assum:lipschitz}, for all $j\in[n]$, 
\begin{equation}
    \label{eqn:thm2-lip}
    \norm{\nabla Q_j(x)-\nabla Q_j(x^*)}\leq\mu\norm{x-x^*}.
\end{equation}

Let $x_S=\arg\min_x\sum_{j\in S}Q_j(x)$ be the minimum point of the aggregated cost functions of a set $S$ of $n-r$ agents, i.e., $\mnorm{S}=n-r$. Note that $\sum_{j\in S}\nabla Q_j(x_S)=0$.
By triangle inequality,
\begin{equation}
    \label{eqn:thm2-triangle}
    \norm{\sum_{j\in S}\nabla Q_j(x_S)-\sum_{j\in S}\nabla Q_j(x^*)}\leq\sum_{j\in S}\norm{\nabla Q_j(x_S)-\nabla Q_j(x^*)}.
\end{equation}
Combining~\eqref{eqn:thm2-lip} through \eqref{eqn:thm2-triangle}, %also recall $Q_S(x)=\frac{1}{\mnorm{S}}\sum_{j\in S}Q_j(x)$,
\begin{equation}
    \norm{\sum_{j\in S}\nabla Q_j(x^*)}\leq\mnorm{S}\mu\norm{x_S-x^*}.
\end{equation}
By Definition~\ref{def:redundancy} of $(0,r;\epsilon)$-redundancy, $\norm{x_S-x^*}\leq\epsilon$.
Therefore,
\begin{equation}
    \norm{\sum_{j\in S}\nabla Q_j(x^*)}\leq\mu\epsilon (n-r).
\end{equation}

Now, consider an arbitrary agent $i\in[n]\backslash S$. Let $T=S\cup\{i\}$. Using a similar argument as above, we obtain %that with Assumption~\ref{assum:strongly-convex}, for all $x_T=\arg\min_x\sum_{j\in T}Q_j(x)$,
\begin{equation}
    \norm{\sum_{j\in T}\nabla Q_j(x^*)}\leq\mu\epsilon (n-r+1).
\end{equation}
Therefore, %by the arbitrariness of $S$ and $i$, for each agent $i\in[n]$,
\begin{align}
    \norm{\nabla Q_i(x^*)}=&\norm{\sum_{j\in T}\nabla Q_j(x^*)-\sum_{j\in S}\nabla Q_j(x^*)} \leq\norm{\sum_{j\in T}\nabla Q_j(x^*)}+\norm{\sum_{j\in S}\nabla Q_j(x^*)} \nonumber \\
        \leq&(n-r)\mu\epsilon+(n-r+1)\mu\epsilon=(2n-2r+1)\mu\epsilon. \label{eqn:apprx_gradient_bnd}
\end{align}
Note that the inequality \eqref{eqn:apprx_gradient_bnd} can be applied to any $i\in[n]$ with a suitable choice of $S$ above.\\

On the other hand, by Assumption~\ref{assum:lipschitz}, for any $x\in\mathbb{R}^d$,
\begin{equation}
    \norm{\nabla Q_i(x)-\nabla Q_i(x^*)}\leq\mu\norm{x-x^*}.
\end{equation}
By triangle inequality,
\begin{equation}
    \norm{\nabla Q_i(x)}\leq\norm{\nabla Q_i(x^*)}+\mu\norm{x-x^*}.
\end{equation}
Combining above and \eqref{eqn:apprx_gradient_bnd},
\begin{equation}
    \label{eqn:apprx_gradient_bnd_2}
    \norm{\nabla Q_i(x)}\leq(2n-2r+1)\mu\epsilon+\mu\norm{x-x^*}\leq2n\mu\epsilon+\mu\norm{x-x^*}.
\end{equation}
Recall that $x^t\in\W$ for all $t$, where $\W$ is a compact set. There exists a $\Gamma=\max_{x\in\W}\norm{x-x^*}<\infty$, such that $\norm{x^t-x^*}\leq\Gamma$ for all $t$. 
Therefore, for all $t$, 
\begin{align}
    \label{eqn:apprx_phi_t_bnd}
    \norm{\sum_{j\in S^t}\nabla Q_j(x^t)}&=\norm{\sum_{j\in S^t}\nabla Q_j(x^t)}\leq\sum_{j\in S^t}\norm{\nabla Q_j(x^t)} \nonumber \\
        &\leq\mnorm{S^t}\cdot\left(2n\mu\epsilon+\mu\norm{x^t-x^*}\right)\leq(n-r)\left(2n\mu\epsilon+\mu\Gamma\right)<\infty.
\end{align}

\textbf{Second}, consider the following term $\displaystyle\phi_t=\iprod{x^t-x^*}{\sum_{j\in S^t}\nabla Q_j(x^t)}$.
We have
\begin{align}
    \label{eqn:apprx_phi_t_1}
    \phi_t=&\iprod{x^t-x^*}{\sum_{j\in S^t}\nabla Q_j(x^t)} \nonumber \\
        =&\iprod{x^t-x^*}{\sum_{j\in S^t}\nabla Q_j(x^t)+\sum_{k\in [n]\backslash S^t}\nabla Q_k(x^t)-\sum_{k\in [n]\backslash S^t}\nabla Q_k(x^t)} \nonumber \\
        =&\iprod{x^t-x^*}{\sum_{j\in [n]}\nabla Q_j(x^t)}-\iprod{x^t-x^*}{\sum_{k\in [n]\backslash S^t}\nabla Q_k(x^t)}.
\end{align}

For the first term in \eqref{eqn:apprx_phi_t_1}, recall that $Q_S(x)=\frac{1}{\mnorm{S}}\sum_{j\in S}Q_j(x)$ for any set $S$ that satisfies $\mnorm{S}\geq n-r$. With Assumption~\ref{assum:strongly-convex}, 
\begin{align}
    \label{eqn:apprx_phi_t_1_1}
    &\iprod{x^t-x^*}{\sum_{j\in [n]}\nabla Q_j(x^t)}=\iprod{x^t-x^*}{\sum_{j\in [n]}\nabla Q_j(x^t)-\sum_{j\in [n]}\nabla Q_j(x^*)} \nonumber \\
        =&n\cdot\iprod{x^t-x^*}{\nabla Q_{[n]}(x^t)-\nabla Q_{[n]}(x^*)}\geq n\gamma\norm{x^t-x^*}^2.
\end{align}

For the second term in \eqref{eqn:apprx_phi_t_1}, by Cauchy-Schwartz inequality,
\begin{align}
    \label{eqn:apprx_phi_t_1_2}
    \iprod{x^t-x^*}{\sum_{k\in [n]\backslash S^t}\nabla Q_k(x^t)}&=\sum_{k\in [n]\backslash S^t}\iprod{x^t-x^*}{\nabla Q_k(x^t)}\nonumber \\
        &\leq\sum_{k\in [n]\backslash S^t}\norm{x^t-x^*}\cdot\norm{\nabla Q_k(x^t)}.
\end{align}

Combining \eqref{eqn:apprx_phi_t_1}, \eqref{eqn:apprx_phi_t_1_1}, and \eqref{eqn:apprx_phi_t_1_2},
\begin{equation}
    \label{eqn:apprx_phi_t_2}
    \phi_t\geq n\gamma\norm{x^t-x^*}^2-\sum_{k\in[n]\backslash S^t}\norm{x^t-x^*}\norm{\nabla Q_k(x^t)}.
\end{equation}
Substituting from \eqref{eqn:apprx_gradient_bnd_2} in above, note that $\mnorm{[n]\backslash S^t}=r$,
\begin{align}
    \label{eqn:apprx_phi_t_3}
    \phi_t&\geq n\gamma\norm{x^t-x^*}^2-\sum_{k\in[n]\backslash S^t}\norm{x^t-x^*}(2n\mu\epsilon+\mu\norm{x^t-x^*}) \nonumber \\
        &=n\gamma\norm{x^t-x^*}^2-r\norm{x^t-x^*}(2n\mu\epsilon+\mu\norm{x^t-x^*}) \nonumber \\
        &=n\gamma\left(1-\dfrac{r}{n}\cdot\dfrac{\mu}{\gamma}\right)\norm{x^t-x^*}\left(\norm{x^t-x^*}-\dfrac{2r\mu\epsilon}{\gamma\left(1-\dfrac{r}{n}\cdot\dfrac{\mu}{\gamma}\right)}\right).
\end{align}
Recall that we assume $\displaystyle\alpha=1-\dfrac{r}{n}\cdot\dfrac{\mu}{\gamma}>0$.
We have
\begin{equation}
    \phi_t\geq\alpha n\gamma\norm{x^t-x^*}\left(\norm{x^t-x^*}-\dfrac{2r\mu}{\alpha\gamma}\epsilon\right).
    \label{eqn:phi_t_final}
\end{equation}
Let $\D\triangleq\dfrac{2r\mu}{\alpha\gamma}\epsilon$. \eqref{eqn:phi_t_final} implies for an arbitrary $\delta>0$, 
\begin{equation}
    \phi_t\geq\alpha n\gamma\delta\left(D+\delta\right)>0~\textrm{ when }~\norm{x^t-x^*}\geq\D+\delta. 
\end{equation}
Therefore, by Lemma~\ref{lemma:bound}, 
\begin{equation}
    \lim_{t\rightarrow\infty}\norm{x^t-x^*}\leq\D.
\end{equation}
\end{proof}

\vfill \newpage
\subsection{Proof of Theorem~\ref{thm:approx-generalized}}

Theorem~\ref{thm:approx-generalized} states the asymptotic convergence of Algorithm~\ref{alg} when solving Problem C while using stale gradients. Specifically, we replace \eqref{eqn:update} in \textbf{Step 2} of Algorithm~\ref{alg} with
\begin{equation}
    \label{eqn:update-straggler}
    \textstyle x^{t+1}=\left[x^t-\eta_t\sum_{i=0}^\tau\sum_{j\in T^{t;t-i}}\nabla Q_j(x^{t-i})\right]_\W,
\end{equation}
where the sets $T^{t;t-i}$'s are defined as follows: (1) $T^{t;t}=S^{t;t}$, and (2) $T^{t;t-i}=S^{t;t-i}\backslash\bigcup_{k=0}^{i-1}T^{t;t-k}$, $\forall i\geq1$. We further define $T^t=\bigcup_{i=0}^\tau T^{t;t-i}$. The \textit{straggler parameter} $\tau\geq0$ is a predefined integer.

\begin{mdframed}
% \begin{theorem}
    \textbf{Theorem \ref{thm:approx-generalized}.}
    \textit{Suppose Assumption~\ref{assum:lipschitz} and \ref{assum:strongly-convex} hold true, and the cost functions of all agents satisfies $(0,r;\epsilon)$-redundancy. 
     Assume that $\eta_t$ in (\ref{eqn:update-straggler}) satisfies $\sum_{t=0}^\infty\eta_t=\infty$, $\sum_{t=0}^\infty\eta_t^2<\infty$, and $\eta_t\geq\eta_{t+1}$ for all $t$.
     Let $\alpha$ and $D$ be defined as follows: 
     $$\alpha\triangleq1-\frac{r}{n}\cdot\frac{\mu}{\gamma}>0 \textrm{ and } D\triangleq\frac{2r\mu}{\alpha\gamma}\epsilon.$$
     Then, suppose there exists a $\tau\geq0$ such that $\mnorm{T^t}\geq n-r$ for all $t$, for the proposed algorithm with update rule \eqref{eqn:update-straggler}, 
    \begin{equation}
        \lim_{t\rightarrow\infty}\norm{x^t-x^*}\leq\D, %x^t\xrightarrow[t\rightarrow\infty]{}x^* 
    \end{equation}
    where $x^*=\arg\min_x\sum_{j\in[n]}Q_j(x)$.}
%     \label{thm:approx-generalized}
% \end{theorem}
\end{mdframed}

The proof consists of three parts. In the first part, we show the norm of the update is bounded for all $t$. In the second part, we consider the inner product term $\phi_{t;1}$ (defined later) is lower-bounded. And in the third part, we show that with a lower-bounded $\phi_{t;1}$ term, the iterative estimate converges. 

\begin{proof}
\textbf{First}, we show that $\norm{\sum_{i=0}^\tau\sum_{j\in T^{t;t-i}}\nabla Q_j(x^{t-i})}$ is bounded for all $t$. By Assumption~\ref{assum:lipschitz} and $(0,r;\epsilon)$-redundancy, following the same argument in the proof of Theorem~\ref{thm:approx}, we obtain that for all $j\in[n]$,
\begin{equation}
    \norm{\nabla Q_j(x^*)}\leq(2n-2r+1)\mu\epsilon. \tag{\ref{eqn:apprx_gradient_bnd}}
\end{equation}
Furthermore, by Assumption~\ref{assum:lipschitz}, for all $x\in\R^d$,
\begin{equation}
    \norm{\nabla Q_j(x)}\leq 2n\mu\epsilon+\mu\norm{x-x^*}. \tag{\ref{eqn:apprx_gradient_bnd_2}}
\end{equation}
Recall that $x^t\in\W$ for all $t$, where $\W$ is a compact set. There exists a $\Gamma=\max_{x\in\W}\norm{x-x^*}<\infty$, such that $\norm{x^t-x^*}\leq\Gamma$ for all $t$. Therefore, for all $t$,
\begin{align}
    &\norm{\sum_{i=0}^\tau\sum_{j\in T^{t;t-i}}\nabla Q_j(x^{t-i})}\leq \sum_{i=0}^\tau\sum_{j\in T^{t;t-i}}\norm{\nabla Q_j(x^{t-i})} \nonumber \\
    \leq& \sum_{i=0}^\tau\mnorm{T^{t;t-i}}\cdot(2n\mu\epsilon+\mu\norm{x^{t-i}-x^*})\leq \sum_{i=0}^\tau\mnorm{T^{t;t-i}}\cdot(2n\mu\epsilon+\mu\Gamma)
\end{align}
Recall the definition of $T^{t;t-i}$, that $T^{t;t-i_1}\cap T^{t;t-i_2}=\varnothing$ for all $0\leq i_1, i_2\leq\tau$, $i_1\neq i_2$. Therefore, $\sum_{i=0}^\tau\mnorm{T^{t;t-i}}\leq n$. Thus,
\begin{align}
    \label{eqn:update-bnd-generalized}
    &\norm{\sum_{i=0}^\tau\sum_{j\in T^{t;t-i}}\nabla Q_j(x^{t-i})}\leq n(2n\mu\epsilon+\mu\Gamma)<\infty.
\end{align}

\textbf{Second}, consider the term 
\begin{equation}
    \label{eqn:phi_t-gen}
    \phi_t=\iprod{x^t-x^*}{\sum_{i=0}^\tau\sum_{j\in T^{t;t-i}}\nabla Q_j(x^{t-i})}.
\end{equation}
Note that with the non-expansion property of Euclidean projection onto a closed convex set\footnote{$\norm{x-x^*}\geq\norm{[x]_\W-x^*},~\forall w\in\mathbb{R}^d$.} we have
\begin{align}
    \label{eqn:origin-of-phi-t}
    \norm{x^{t+1}-x^*}&=\norm{x^t-\eta_t\sum_{i=0}^\tau\sum_{j\in T^{t;t-i}}\nabla Q_j(x^{t-i})-x^*}^2 \nonumber \\
    &\leq\norm{x^t-x^*}^2-2\eta_t\phi_t+\eta_t^2\norm{\sum_{i=0}^\tau\sum_{j\in T^{t;t-i}}\nabla Q_j(x^{t-i})}^2.
\end{align}

Let $T^t=\bigcup_{i=0}^\tau T^{t;t-i}$ be the set of agents whose gradients are used for the update at iteration $t$.
\begin{align}
    \sum_{i=0}^\tau\sum_{j\in T^{t;t-i}}\nabla Q_j(x^{t-i})=&\sum_{i=0}^\tau\sum_{j\in T^{t;t-i}}\left(\nabla Q_j(x^t) - \nabla Q_j(x^t) + \nabla Q_j(x^{t-i}) \right) \nonumber \\
    =&\sum_{i=0}^\tau\sum_{j\in T^{t;t-i}}\nabla Q_j(x^t) - \sum_{i=0}^\tau\sum_{j\in T^{t;t-i}}\left(\nabla Q_j(x^{t}) - \nabla Q_j(x^{t-i})\right) \nonumber \\
    =&\sum_{j\in T^t}\nabla Q_j(x^t) - \sum_{i=0}^\tau\sum_{j\in T^{t;t-i}}\left(\nabla Q_j(x^{t}) - \nabla Q_j(x^{t-i})\right).
\end{align}
Therefore, we have
\begin{align}
    \label{eqn:phi_t-generalized}
    \phi_t=&\iprod{x^t-x^*}{\sum_{i=0}^\tau\sum_{j\in T^{t;t-i}}\nabla Q_j(x^{t-i})} \nonumber \\
    =&\iprod{x^t-x^*}{\sum_{j\in T^{t}}\nabla Q_j(x^{t})} - \iprod{x^t-x^*}{\sum_{i=0}^\tau\sum_{j\in T^{t;t-i}}\left(\nabla Q_j(x^t)-\nabla Q_j(x^{t-i})\right)}.
\end{align}

For the first term in \eqref{eqn:phi_t-generalized}, denote it $\phi_{t;1}$, we have
\begin{align}
    \label{eqn:phi_t1}
    \phi_{t;1}=&\iprod{x^t-x^*}{\sum_{j\in T^{t}}\nabla Q_j(x^{t})} \nonumber \\
    =&\iprod{x^t-x^*}{\sum_{j\in T^{t}}\nabla Q_j(x^{t})+\sum_{k\in[n]\backslash T^t}\nabla Q_k(x^t)-\sum_{k\in[n]\backslash T^t}\nabla Q_k(x^t)} \nonumber \\
    =&\iprod{x^t-x^*}{\sum_{j\in[n]}\nabla Q_j(x^{t})}-\iprod{x^t-x^*}{\sum_{k\in[n]\backslash T^t}\nabla Q_k(x^t)}.
\end{align}
Recall \eqref{eqn:apprx_phi_t_1_1} and \eqref{eqn:apprx_phi_t_1_2}, we have the same result as we have in \eqref{eqn:apprx_phi_t_2}:
\begin{equation}
    \phi_{t;1}\geq n\gamma\norm{x^t-x^*}^2-\sum_{k\in[n]\backslash T^t}\norm{x^t-x^*}\norm{\nabla Q_k(x^t)}.
\end{equation}
Note that $\mnorm{T^t}\geq n-r$, or $\mnorm{[n]\backslash T^t}\leq r$. Combining above and \eqref{eqn:apprx_gradient_bnd_2},
\begin{align}
    \label{eqn:phi_t1_bnd}
    \phi_{t;1}\geq& n\gamma\norm{x^t-x^*}^2-r\norm{x^t-x^*}\left(2n\mu\epsilon+\mu\norm{x-x^*}\right) \nonumber \\
        =&n\gamma\left(1-\dfrac{r}{n}\cdot\dfrac{\mu}{\gamma}\right)\norm{x^t-x^*}\left(\norm{x^t-x^*}-\dfrac{2r\mu\epsilon}{\gamma\left(1-\dfrac{r}{n}\cdot\dfrac{\mu}{\gamma}\right)}\right).
\end{align}
Recall that we assume $\displaystyle\alpha=1-\dfrac{r}{n}\cdot\dfrac{\mu}{\gamma}>0$.
We have
\begin{equation}
    \phi_{t;1}\geq\alpha n\gamma\norm{x^t-x^*}\left(\norm{x^t-x^*}-\dfrac{2r\mu}{\alpha\gamma}\epsilon\right).
    \label{eqn:phi_t1_final}
\end{equation}
Let $\D\triangleq\dfrac{2r\mu}{\alpha\gamma}\epsilon$. \eqref{eqn:phi_t1_final} implies for an arbitrary $\delta>0$, 
\begin{equation}
    \label{eqn:phi_t1_final_final}
    \phi_{t;1}\geq\alpha n\gamma\delta\left(D+\delta\right)>0~\textrm{ when }~\norm{x^t-x^*}\geq\D+\delta. 
\end{equation}
%Therefore, by Lemma~\ref{lemma:bound}, $\lim_{t\rightarrow\infty}\norm{x^t-x^*}\leq\D$.

For the second term in \eqref{eqn:phi_t-generalized}, denote it $\phi_{t;2}$, by Cauchy-Schwartz inequality, we have
\begin{align}
    \label{eqn:phi_t2}
    \phi_{t;2}=&\iprod{x^t-x^*}{\sum_{i=0}^\tau\sum_{j\in T^{t;t-i}}\left(\nabla Q_j(x^{t}) - \nabla Q_j(x^{t-i})\right)} \nonumber \\
    \leq& \norm{x^t-x^*}\cdot\norm{\sum_{i=0}^\tau\sum_{j\in T^{t;t-i}}\left(\nabla Q_j(x^{t}) - \nabla Q_j(x^{t-i})\right)}.
\end{align}
Consider the factor $\norm{\sum_{i=0}^t\sum_{j\in T^{t;t-i}}\left(\nabla Q_j(x^{t}) - \nabla Q_j(x^{t-i})\right)}$. By triangle inequality,
\begin{align}
    \label{eqn:phi_t2_step_a}
    \norm{\sum_{i=0}^\tau\sum_{j\in T^{t;t-i}}\left(\nabla Q_j(x^{t}) - \nabla Q_j(x^{t-i})\right)}\leq&\sum_{i=0}^\tau\sum_{j\in T^{t;t-i}}\norm{\nabla Q_j(x^{t}) - \nabla Q_j(x^{t-i})}.
\end{align}
By Assumption~\ref{assum:lipschitz} and $x^t\in\W$ for all $t$, 
\begin{equation}
    \label{eqn:phi_t2_step_b}
    \norm{\nabla Q_j(x^{t}) - \nabla Q_j(x^{t-i})}\leq\mu\norm{x^t-x^{t-i}}.
\end{equation}
According to the update rule \eqref{eqn:update-straggler}, 
\begin{align}
    x^{t}&=x^{t-1}-\eta_{t-1}\sum_{h=0}^\tau\sum_{j\in T^{t-1;t-1-h}}\nabla Q_j(x^{t-1-h}) \nonumber \\
        &=x^{t-2}-\eta_{t-1}\sum_{h=0}^\tau\sum_{j\in T^{t-1;t-1-h}}\nabla Q_j(x^{t-1-h})-\eta_{t-2}\sum_{h=0}^\tau\sum_{j\in T^{t-2;t-2-h}}\nabla Q_j(x^{t-2-h}) \nonumber \\
        &=x^{t-2}-\sum_{k=1}^2\eta_{t-k}\sum_{h=0}^\tau\sum_{j\in T^{t-k;t-k-h}}\nabla Q_j(x^{t-k-h}) \nonumber \\
        &=\cdots \nonumber \\
        &=x^{t-i}-\sum_{k=1}^i\eta_{t-k}\sum_{h=0}^\tau\sum_{j\in T^{t-k;t-k-h}}\nabla Q_j(x^{t-k-h}).
\end{align}
Therefore,
\begin{align}
    \norm{x^t-x^{t-i}}=&\norm{\sum_{k=1}^i\eta_{t-k}\sum_{h=0}^\tau\sum_{j\in T^{t-k;t-k-h}}\nabla Q_j(x^{t-k-h})} \nonumber \\
        \leq&\sum_{k=1}^i\eta_{t-k}\norm{\sum_{h=0}^\tau\sum_{j\in T^{t-k;t-k-h}}\nabla Q_j(x^{t-k-h})}.
\end{align}
By \eqref{eqn:update-bnd-generalized}, let $\overline{G}=n(2n\mu\epsilon+\mu\Gamma)$. Also note that for all $t$, $\eta_{t}\geq\eta_{t+1}$. From above we have
\begin{equation}
    \norm{x^t-x^{t-i}}\leq\sum_{k=1}^i\eta_{t-k}\overline{G}\leq i\cdot \eta_{t-i}\overline{G}. 
\end{equation}
Combining \eqref{eqn:phi_t2_step_a}, \eqref{eqn:phi_t2_step_b} and above, we have
\begin{align}
    &\norm{\sum_{i=0}^\tau\sum_{j\in T^{t;t-i}}\left(\nabla Q_j(x^{t}) - \nabla Q_j(x^{t-i})\right)}\leq\sum_{i=0}^\tau\sum_{j\in T^{t;t-i}}\norm{\nabla Q_j(x^{t}) - \nabla Q_j(x^{t-i})} \nonumber \\
        \leq&\sum_{i=0}^\tau\sum_{j\in T^{t;t-i}}\mu i\cdot \eta_{t-i}\overline{G} \leq\overline{G}\cdot\sum_{i=0}^\tau\sum_{j\in T^{t;t-i}}\mu\tau\eta_{t-\tau}=\overline{G}\cdot\sum_{j\in T^{t}}\mu\tau\eta_{t-\tau} \nonumber \\
        \leq&\mu\tau\eta_{t-\tau}n\overline{G}.
\end{align}
Combining with \eqref{eqn:phi_t2}, from above we have
\begin{equation}
    \label{eqn:phi_t2_bnd}
    \phi_{t;2}\leq\mu\tau\eta_{t-\tau}n\overline{G}\norm{x^t-x^*}.
\end{equation}
In summary, $\phi_t=\phi_{t;1}-\phi_{t;2}$. 
% Combining \eqref{eqn:phi_t1_bnd} and \eqref{eqn:phi_t2_bnd}, we have
% \begin{align}
%     \phi_t\geq& n\gamma\norm{x^t-x^*}^2-r\norm{x^t-x^*}\left(2n\mu\epsilon+\mu\norm{x-x^*}\right)-\mu\tau\eta_{t-\tau}n\overline{G}\norm{x^t-x^*}.
% \end{align}

\textbf{Third}\footnote{Note that this part of the proof is similar to what we have in the proof of Lemma~\ref{lemma:bound} in Appendix~\ref{appdx:proof-lemma-bound}. Still, the full argument are presented with repeated contents to avoid confusion.}, we show that assuming that there exists some $\D^*\in[0,\max_{x\in\W}\norm{x-x^*})$ and $\xi>0$, such that $\phi_{t;1}\geq\xi$ when $\norm{x^t-x^*}\geq\D^*$, we have $\lim_{t\rightarrow\infty}\norm{x^t-x^*}\leq\D^*$.

Let $e^t$ denote $\norm{x^t-x^*}$. Define a scalar function $\psi$,
\begin{equation}
    \psi(y)=\left\{\begin{array}{ll}
        0, & y<\left(\D^*\right)^2, \\
        \left(y-\left(\D^*\right)^2\right)^2, & \textrm{otherwise}.
    \end{array}\right.
    \label{eqn:psi_def-gen}
\end{equation}
Let $\psi'(y)$ denote the derivative of $\psi$ at $y$. Then (cf. \cite{bottou1998online})
\begin{equation}
    \label{eqn:psi_bnd-gen}
    \psi(z)-\psi(y)\leq(z-y)\psi'(y)+(z-y)^2,~\forall y,z\in\mathbb{R}_{\geq0}.
\end{equation}
Note,
\begin{equation}
    \label{eqn:psi_prime-gen}
    \psi'(y)=\max\left\{0,2\left(y-\left(\D^*\right)^2\right)\right\}.
\end{equation}

Now, define
\begin{equation}
    \label{eqn:ht_def-gen}
    h_t\triangleq\psi(e_t^2).
\end{equation}
From \eqref{eqn:psi_bnd-gen} and \eqref{eqn:ht_def-gen},
\begin{equation}
    h_{t+1}-h_t=\psi\left(e_{t+1}^2\right)-\psi\left(e_t^2\right)\leq\left(e_{t+1}^2-e_t^2\right)\psi'\left(e_t^2\right)+\left(e_{t+1}^2-e_t^2\right)^2,~\forall t\in\mathbb{Z}_{\geq0}.
\end{equation}
From now on, we use $\psi_t'$ as the shorthand for $\psi'\left(e_t^2\right)$, i.e.,
\begin{equation}
    \label{eqn:def-psi-t-gen}
    \psi_t'\triangleq\psi'\left(e_t^2\right).
\end{equation}
From above, for all $t\geq0$,
\begin{equation}
    \label{eqn:ht_1-gen}
    h_{t+1}-h_t\leq\left(e_{t+1}^2-e_t^2\right)\psi'_t+\left(e_{t+1}^2-e_t^2\right)^2.%~\forall t\in\mathbb{Z}_{\geq0}.
\end{equation}

Recall the iterative process \eqref{eqn:update-straggler}. From now on, we use $\mathsf{StragAgg}[t]$ as a short hand for the output of the gradient aggregation rule at iteration $t$, i.e. 
\begin{equation}
    \label{eqn:def-stragg-t}
    \mathsf{StragAgg}[t]\triangleq\sum_{i=0}^\tau\sum_{j\in T^{t;t-i}}\nabla Q_j(x^{t-i}).
\end{equation}
Using the non-expansion property of Euclidean projection onto a closed convex set\footnote{$\norm{x-x^*}\geq\norm{[x]_\W-x^*},~\forall w\in\mathbb{R}^d$.}, 
\begin{equation}
    \norm{x^{t+1}-x^*}\leq\norm{x^t-\eta_t\mathsf{StragAgg}[t]-x^*}.%~\forall t\in\mathbb{Z}_{\geq0}.
\end{equation}
Taking square on both sides,
\begin{equation*}
    e_{t+1}^2\leq e_t^2-2\eta_t\iprod{x_t-x^*}{\mathsf{StragAgg}[t]}+\eta_t^2\norm{\mathsf{StragAgg}[t]}^2.%~\forall t\in\mathbb{Z}_{\geq0}.
\end{equation*}
Recall from \eqref{eqn:phi_t-gen} that $\iprod{x_t-x^*}{\mathsf{StragAgg}[t]}=\phi_t$, therefore (cf. \eqref{eqn:origin-of-phi-t}),
\begin{equation}
    \label{eqn:proj_bound-generalized}
    e_{t+1}^2\leq e_t^2-2\eta_t\phi_t+\eta_t^2\norm{\mathsf{StragAgg}[t]}^2,~\forall t\geq0.
\end{equation}

As $\psi'_t\geq0,~\forall t\in\mathbb{Z}_{\geq0}$, combining \eqref{eqn:ht_1-gen} and \eqref{eqn:proj_bound-generalized},
\begin{equation}
    \label{eqn:ht_2-gen}
    h_{t+1}-h_t\leq\left(-2\eta_t\phi_t+\eta_t^2\norm{\mathsf{StragAgg}[t]}^2\right)\psi'_t+\left(e_{t+1}^2-e_t^2\right)^2,~\forall t\geq0.
\end{equation}
Note that 
\begin{equation}
    \mnorm{e_{t+1}^2-e_t^2}=(e_{t+1}+e_t)\mnorm{e_{t+1}-e_t}.
\end{equation}
As $\W$ is assumed compact, there exists
\begin{equation}
    \Gamma=\max_{x\in\W}\norm{x-x^*}\leq\infty.
\end{equation}
Let $\Gamma>0$, since otherwise $\W=\{x^*\}$ only contains one point, and the problem becomes trivial. As $x^t\in\W$, $\forall t\geq0$,
\begin{equation}
    \label{eqn:e_t_bound-gen}
    e_t\leq\Gamma,
\end{equation}
which implies
\begin{equation}
    e_{t+1}+e_t\leq2\Gamma.
\end{equation}
Therefore,
\begin{equation}
    \mnorm{e_{t+1}^2-e_t^2}\leq2\Gamma\mnorm{e_{t+1}-e_t},~\forall t\geq0.
    \label{eqn:e2t-bound-1-gen}
\end{equation}
By triangle inequality,
\begin{equation}
    \mnorm{e_{t+1}-e_t}=\mnorm{\norm{x^{t+1}-x^*}-\norm{x^t-x^*}}\leq\norm{x^{t+1}-x^t}.
    \label{eqn:e2t-bound-2-gen}
\end{equation}
From \eqref{eqn:update-straggler} and the non-expansion property of Euclidean projection onto a closed convex set,
\begin{equation}
    \norm{x^{t+1}-x^t}=\norm{\left[x^t-\eta_t\mathsf{StragAgg}[t]\right]_\W-x^t}\leq\eta_t\norm{\mathsf{StragAgg}[t]}.
    \label{eqn:e2t-bound-3-gen}
\end{equation}
So from \eqref{eqn:e2t-bound-1-gen}, \eqref{eqn:e2t-bound-2-gen}, and \eqref{eqn:e2t-bound-3-gen},
\begin{align}
    &\mnorm{e_{t+1}^2-e_t^2}\leq2\eta_t\Gamma\norm{\mathsf{StragAgg}[t]}, \nonumber \\
    \Longrightarrow&\left(e_{t+1}^2-e_t^2\right)^2\leq4\eta_t^2\Gamma^2\norm{\mathsf{StragAgg}[t]}^2.
\end{align}
Substituting above in \eqref{eqn:ht_2-gen},
\begin{align}
    %\label{eqn:ht_3-gen}
    h_{t+1}-h_t\leq&\left(-2\eta_t\phi_t+\eta_t^2\norm{\mathsf{StragAgg}[t]}^2\right)\psi'_t+4\eta_t^2\Gamma^2\norm{\mathsf{StragAgg}[t]}^2, \nonumber \\
    =&-2\eta_t\phi_t\psi'_t+(\psi'_t+4\Gamma^2)\eta_t^2\norm{\mathsf{StragAgg}[t]}^2,~\forall t\geq0.
\end{align}
Recall that $\phi_t=\phi_{t;1}-\phi_{t;2}$. Also, $\eta_t\geq\eta_{t+1}$ for all $t$. Substitute for \eqref{eqn:e_t_bound-gen} and \eqref{eqn:phi_t2_bnd} in above, we obtain that
\begin{align}
    \label{eqn:ht_3-gen}
    h_{t+1}-h_t\leq&-2\eta_t(\phi_{t;1}-\phi_{t;2})\psi'_t+(\psi'_t+4\Gamma^2)\eta_t^2\norm{\mathsf{StragAgg}[t]}^2,~\forall t\geq0 \nonumber \\
    \leq&-2\eta_t\phi_{t;1}\psi'_t+2\eta_t(\mu\tau\eta_{t-\tau}n\overline{G}e_t)\psi'_t+(\psi'_t+4\Gamma^2)\eta_t^2\norm{\mathsf{StragAgg}[t]}^2,~\forall t\geq0 \nonumber \\
    \leq&-2\eta_t\phi_{t;1}\psi'_t+2\eta_t^2\mu\tau n\overline{G}\Gamma\psi'_t+(\psi'_t+4\Gamma^2)\eta_t^2\norm{\mathsf{StragAgg}[t]}^2,~\forall t\geq0. 
\end{align}

Let us assume for now that $\D^*\in[0,\max_{x\in\W}\norm{x-x^*})$, which indicates $\D^*<\Gamma$. Using \eqref{eqn:psi_prime-gen} and \eqref{eqn:e_t_bound-gen}, we have
\begin{equation}
    0\leq\psi'_t\leq2\left(e_t^2-\left(\D^*\right)^2\right)\leq2\left(\Gamma^2-\left(\D^*\right)^2\right)\leq2\Gamma^2.
    \label{eqn:psi_prime_t_bnd-gen}
\end{equation}
From \eqref{eqn:update-bnd-generalized} (recall that $\overline{G}=n(2n\mu\epsilon+\mu\Gamma)$), $\norm{\mathsf{StragAgg}[t]}\leq\overline{G}<\infty$ for all $t$. Substituting \eqref{eqn:psi_prime_t_bnd-gen} in \eqref{eqn:ht_3-gen},
\begin{align}
    \label{eqn:ht_4-gen}
    h_{t+1}-h_t\leq&-2\eta_t\phi_{t;1}\psi'_t+2\eta_t^2\mu\tau n\overline{G}\Gamma\left(2\Gamma^2\right)+\left(2\Gamma^2+4\Gamma^2\right)\eta_t^2\overline{G}^2 \nonumber \\
    =&-2\eta_t\phi_{t;1}\psi'_t+\eta_t^2\left(4\mu\tau n\Gamma^3\overline{G}+6\Gamma^2\overline{G}^2\right),~\forall t\geq0.
\end{align}

Now we use Lemma~\ref{lemma:converge} to show that $h_\infty=0$ as follows. For each iteration $t$, consider the following two cases:
\begin{description}
    \item[Case 1)] Suppose $e_t<\D^*$. In this case, $\psi'_t=0$. By Cauchy-Schwartz inequality,
        \begin{equation}
            \mnorm{\phi_{t;1}}=\mnorm{\iprod{x^t-x^*}{\mathsf{StragAgg}[t]}}\leq e_t\norm{\mathsf{StragAgg}[t]}.
        \end{equation}
        By \eqref{eqn:update-bnd-generalized} and \eqref{eqn:e_t_bound-gen}, this implies that
        \begin{equation}
            \mnorm{\phi_{t;1}}\leq\Gamma\overline{G}<\infty.
        \end{equation}
        Thus,
        \begin{equation}
            \label{eqn:phitpsit_1-gen}
            \phi_{t;1}\psi'_t=0.
        \end{equation}
    \item[Case 2)] Suppose $e_t\geq\D^*$. Therefore, there exists $\delta\geq0$, $e_t=\D^*+\delta$. From \eqref{eqn:psi_prime-gen}, we obtain that
    \begin{equation}
        \psi_t'=2\left(\left(\D^*+\delta\right)^2-\left(\D^*\right)^2\right)=2\delta\left(2\D^*+\delta\right).
    \end{equation}
    The statement of Lemma~\ref{lemma:bound} assumes that $\phi_t\geq\xi>0$ when $e_t\geq\D^*$, thus, 
        \begin{equation}
            \label{eqn:phitpsit_2-gen}
            \phi_{t;1}\psi'_t\geq2\delta\xi\left(2\D^*+\delta\right)>0.%,~\textrm{if}~e_t\geq\D^*+\delta.
        \end{equation}
\end{description}
From \eqref{eqn:phitpsit_1-gen} and \eqref{eqn:phitpsit_2-gen}, for both cases,
\begin{equation}
    \phi_{t;1}\psi'_t\geq0, ~\forall t\geq0.
    \label{eqn:phi_psi_bound-gen}
\end{equation}
Combining this with \eqref{eqn:ht_4-gen},
\begin{equation}
    h_{t+1}-h_t\leq\eta_t^2\left(4\mu\tau n\Gamma^3\overline{G}+6\Gamma^2\overline{G}^2\right).
\end{equation}
%Recall Lemma~\ref{lemma:converge}. 
From above we have
\begin{equation}
    \left(h_{t+1}-h_t\right)_+\leq\eta_t^2\left(4\mu\tau n\Gamma^3\overline{G}+6\Gamma^2\overline{G}^2\right).%,~\forall t\in\mathbb{Z}_{\geq0}.
\end{equation}
Since $\sum_{t=0}^\infty\eta_t^2<\infty$, $\Gamma,\overline{G}<\infty$,
\begin{equation}
    \sum_{t=0}^\infty\left(h_{t+1}-h_t\right)_+\leq\left(4\mu\tau n\Gamma^3\overline{G}+6\Gamma^2\overline{G}^2\right)\sum_{t=0}^\infty\eta_t^2<\infty.
\end{equation}
Then Lemma~\ref{lemma:converge} implies that by the definition of $h_t$, we have $h_t\geq0,~\forall t$, 
\begin{align}
    &h_t\xrightarrow[t\rightarrow\infty]{}h_\infty<\infty,~\textrm{and} \label{eqn:upper_bound_h_infty-gen}\\
    &\sum_{t=0}^\infty\left(h_{t+1}-h_t\right)_->-\infty.
\end{align}
Note that $h_\infty-h_0=\sum_{t=0}^\infty(h_{t+1}-h_t)$. Thus, from \eqref{eqn:ht_4-gen} we have 
\begin{equation}
    h_\infty-h_0\leq-2\sum_{t=0}^\infty\eta_t\phi_{t;1}\psi_t'+\left(4\mu\tau n\Gamma^3\overline{G}+6\Gamma^2\overline{G}^2\right)\sum_{t=0}^\infty\eta_t^2.
\end{equation}
By \eqref{eqn:ht_def} the definition of $h_t$, $h_t\geq0$ for all $t$. Therefore, from above we obtain
\begin{equation}
    2\sum_{t=0}^\infty\eta_t\phi_{t;1}\psi_t'\leq h_0-h_\infty+\left(4\mu\tau n\Gamma^3\overline{G}+6\Gamma^2\overline{G}^2\right)\sum_{t=0}^\infty\eta_t^2.
    \label{eqn:bound_2_sum-gen}
\end{equation}
% Thus,
% \begin{equation}
%     2\mnorm{\sum_{t=0}^\infty\eta_t\phi_t\psi_t'}\leq h_0+h_\infty+6\Gamma^2\M^2\sum_{t=0}^\infty\eta_t^2.
%     \label{eqn:bound_abs_sum}
% \end{equation}
% By assumption, $\sum_{t=0}^\infty\eta_t^2<\infty$. From \eqref{eqn:upper_bound_h_infty}, $h_\infty<\infty$. Substituting from \eqref{eqn:e_t_bound} that $e_t<\infty$ in \eqref{eqn:ht_def}, we obtain that $h_0=\psi\left(e_0^2\right)<\infty$. Also from \eqref{eqn:ht_def}, $h_t\geq0~(\forall t)$. Therefore, \eqref{eqn:bound_abs_sum} implies that
% \begin{equation}
%     2\mnorm{\sum_{t=0}^\infty\eta_t\phi_t\psi_t'}<\infty.
% \end{equation}
By assumption, $\sum_{t=0}^\infty\eta_t^2<\infty$. Substituting from \eqref{eqn:e_t_bound-gen} that $e_t<\infty$ in \eqref{eqn:ht_def-gen}, we obtain that 
\begin{equation}
    h_0=\psi\left(e_0^2\right)<\infty.
\end{equation}
Therefore, \eqref{eqn:bound_2_sum-gen} implies that
\begin{equation}
    2\sum_{t=0}^\infty\eta_t\phi_{t;1}\psi_t'\leq h_0+\left(4\mu\tau n\Gamma^3\overline{G}+6\Gamma^2\overline{G}^2\right)\sum_{t=0}^\infty\eta_t^2<\infty.
\end{equation}
Or simply,
\begin{equation}
    \sum_{t=0}^\infty\eta_t\phi_{t;1}\psi_t'<\infty.
    \label{eqn:upper_bound_etatphitpsit-gen}
\end{equation}

Finally, we reason below by contradiction that $h_\infty=0$. Note that for any $\zeta>0$, there exists a unique positive value $\beta$ such that $\zeta=2\beta\left(2\D^*+\sqrt{\beta}\right)^2$. Suppose that $h_\infty=2\beta\left(2\D^*+\sqrt{\beta}\right)^2$ for some positive value $\beta$. As the sequence $\{h_t\}_{t=0}^\infty$ converges to $h_\infty$ (see \eqref{eqn:upper_bound_h_infty-gen}), there exists some finite $\tau\in\Z_{\geq0}$ such that for all $t\geq\tau$, 
\begin{align}
    &\mnorm{h_t-h_\infty}\leq\beta\left(2\D^*+\sqrt{\beta}\right)^2 \\
    \Longrightarrow & h_t\geq h_\infty-\beta\left(2\D^*+\sqrt{\beta}\right)^2.
\end{align}
As $h_\infty=2\beta\left(2\D^*+\sqrt{\beta}\right)^2$, the above implies that
\begin{equation}
    h_t\geq \beta\left(2\D^*+\sqrt{\beta}\right)^2, \forall t\geq\tau.
    \label{eqn:ht_lower_bound-gen}
\end{equation}
Therefore (cf. \eqref{eqn:psi_def-gen} and \eqref{eqn:ht_def-gen}), for all $t\geq\tau$,
\begin{eqnarray*}
    \left(e_t^2-\left(\D^*\right)^2\right)^2\geq\beta\left(2\D^*+\sqrt{\beta}\right)^2, \textrm{ or} \\
    \mnorm{e_t^2-\left(\D^*\right)^2}\geq\sqrt{\beta}\left(2\D^*+\sqrt{\beta}\right).
\end{eqnarray*}
Thus, for each $t\geq\tau$, either
\begin{equation}
    e^2_t\geq\left(\D^*\right)^2+\sqrt{\beta}\left(2\D^*+\sqrt{\beta}\right)=\left(\D^*+\sqrt{\beta}\right)^2,
    \label{eqn:et_case_1-gen}
\end{equation}
or
\begin{equation}
    e^2_t\leq\left(\D^*\right)^2-\sqrt{\beta}\left(2\D^*+\sqrt{\beta}\right)<\left(\D^*\right)^2.
    \label{eqn:et_case_2-gen}
\end{equation}
If the latter, i.e., \eqref{eqn:et_case_2-gen} holds true for some $t'\geq\tau$, 
\begin{equation}
    h_{t'}=\psi\left(e_{t'}^2\right)=0,
\end{equation}
which contradicts \eqref{eqn:ht_lower_bound-gen}. Therefore, \eqref{eqn:ht_lower_bound-gen} implies \eqref{eqn:et_case_1-gen}.

From above we obtain that if $h_\infty=2\beta\left(2\D^*+\sqrt{\beta}\right)^2$, there exists $\tau<\infty$ such that for all $t\geq\tau$, 
\begin{equation}
    e_t\geq\D^*+\sqrt{\beta}.
\end{equation}
Thus, from \eqref{eqn:phitpsit_2-gen}, with $\delta=\sqrt{\beta}$, we obtain that %there exists a $\xi>0$, for $\epsilon=2\xi\sqrt{\beta}\left(2\D^*+\sqrt{\beta}\right)>0$ and $\delta=\sqrt{\beta}>0$ in \eqref{eqn:phitpsit_2},
\begin{equation}
    \phi_t\psi_t'\geq2\xi\sqrt{\beta}\left(2\D^*+\sqrt{\beta}\right), \forall t\geq\tau.
\end{equation}
Therefore,
\begin{equation}
    \sum_{t=\tau}^\infty\eta_t\phi_t\psi_t'\geq2\xi\sqrt{\beta}\left(2\D^*+\sqrt{\beta}\right)\sum_{t=\tau}^\infty\eta_t=\infty.
\end{equation}
This is a contradiction to \eqref{eqn:upper_bound_etatphitpsit-gen}. Therefore, $h_\infty=0$, and by \eqref{eqn:ht_def-gen}, the definition of $h_t$, 
\begin{equation}
    h_\infty=\lim_{t\rightarrow\infty}\psi\left(e_t^2\right)=0.
\end{equation}
Hence, by \eqref{eqn:psi_def-gen}, the definition of $\psi(\cdot)$, 
\begin{equation}
    \lim_{t\rightarrow\infty}\norm{x^t-x^*}\leq\D^*.
\end{equation}

\textbf{Finally}, recall the result in \eqref{eqn:phi_t1_final_final}, for arbitrary $\delta>0$,
\begin{equation}
    \tag{\ref{eqn:phi_t1_final_final}}
    \phi_{t;1}\geq\alpha n\gamma\delta\left(D+\delta\right)>0~\textrm{ when }~\norm{x^t-x^*}\geq\D+\delta,
\end{equation}
with $\displaystyle\alpha=1-\dfrac{r}{n}\cdot\dfrac{\mu}{\gamma}>0$ and $\D=\dfrac{2r\mu}{\alpha\gamma}\epsilon$. Combining with the \textbf{third} part of this proof, $\lim_{t\rightarrow\infty}\norm{x^t-x^*}\leq\D$.
\end{proof}

\vfill \newpage
\subsection{Proof of Theorems~\ref{thm:async-fault-toler} and the special case when using CGE gradient filter}

Theorem~\ref{thm:async-fault-toler} states the asymptotic convergence of Algorithm~\ref{alg} when solving \textbf{Problem D}, asynchronous Byzantine fault-tolerant optimization ($f,r\geq0$), using the following gradient aggregator:
\begin{align}
    &\textstyle\mathsf{GradAgg}\left(g_j^t|j\in S^t;n,f,r\right) 
    =\mathsf{GradFilter}\left(g_j^t|j\in S^t;n-r,f\right).
    \label{eqn:aggregation-rule-async-ft}
\end{align}
i.e., any gradient filter that satisfies the condition listed in the theorem. The result is established upon Assumption~\ref{assum:lipschitz} and the following assumption:

% \begin{assumption}
    \textbf{Assumption \ref{assum:strongly-convex-ft}.}
    \textit{For any set $S\subset\H$, we define the average cost function to be $Q_S(x)=({1}/{\mnorm{S}})\sum_{j\in S}Q_j(x)$. We assume that $Q_S(x)$ is $\gamma$-strongly convex for any $S$ subject to $\mnorm{S}\geq n-f$, i.e., $\forall x, x'\in\mathbb{R}^d$, 
    \begin{equation}
        \iprod{\nabla Q_S(x)-\nabla Q_S(x')}{x-x'}\geq\gamma\norm{x-x'}^2.
    \end{equation}}
% \end{assumption}

Note that within the convex compact set $\W$, $x_\H=\arg\min_{x\in\W}\sum_{j\in\H}Q_j(x)$ is the unique minimum of the aggregate cost functions of agents in $\H$, where $\H\subset[n]$ is a set of non-faulty agents with $\H=n-f$, and 
\begin{equation}
    g_j^t=\left\{\begin{array}{cl}
        \nabla Q_j(x^t), & \textrm{ if the agent is non-faulty,} \\
        \textrm{arbitrary vector}, & \textrm{ if the agent is faulty.} 
    \end{array}\right.
\end{equation}

\begin{mdframed}
% \begin{theorem}
    \textbf{Theorem \ref{thm:async-fault-toler}.}
    \textit{Suppose that Assumptions~\ref{assum:lipschitz} and \ref{assum:strongly-convex-ft} hold true, and the cost functions of all agents satisfies $(f,r;\epsilon)$-redundancy. Assume that $\eta_t$ %in \eqref{eqn:step-size} 
    satisfies $\sum_{t=0}^\infty=\infty$ and $\sum_{t=0}^\infty\eta_t^2<\infty$. Suppose that $\norm{\mathsf{GradFilter}\left(n-r,f;\,\left\{g_i^t\right\}_{i\in S^t}\right)}<\infty$ for all $t$. The proposed algorithm with aggregation rule \eqref{eqn:aggregation-rule-async-ft} satisfies the following: For some point $x_\H\in\W$, if there exists a real-valued constant $\D^*\in\left[0,\max_{x\in\W}\norm{x-x_\H}\right)$ and $\xi>0$ such that for each iteration $t$,
    \begin{equation*}
        \phi_t\triangleq\iprod{x^t-x_\H}{\mathsf{GradFilter}\left(n-r,f;\,\left\{g_i^t\right\}_{i\in S^t}\right)}\geq\xi \textrm{ when } \norm{x^t-x_\H}\geq\D^*,
    \end{equation*}
    we have $\lim_{t\rightarrow\infty}\norm{x^t-x_\H}\leq\D^*$.}
% \end{theorem}
\end{mdframed}
Note that by combining \eqref{eqn:aggregation-rule-async-ft} and Lemma~\ref{lemma:bound}, Theorem~\ref{thm:async-fault-toler} has already been proven. To illustrate the correctness of Theorem~\ref{thm:async-fault-toler} in practice, we present in the main paper a group of parameters when CGE gradient filter is in use:
\begin{gather*}
    \alpha = 1 - \frac{f-r}{n-r} + \frac{2\mu}{\gamma}\cdot\frac{f+r}{n-r}>0, \\
    \xi = \alpha m \gamma \delta \left( \frac{4\mu (f+r)\epsilon}{\alpha\gamma} + \delta \right) > 0, \\
    \D^* = \frac{4\mu (f+r)\epsilon}{\alpha\gamma}+\delta.
\end{gather*}

The gradient filter CGE \citep{liu2021approximate} can be defined as a function on $\R^{d\times m}\rightarrow\R^d$, with two hyperparameters $m$ and $f$. The server receives $m$ gradients $\left\{g_i^t\right\}_{i\in S^t}$ from $m$ agents in the set $S^t$ in iteration t. The server sorts the gradients as per their Euclidean norms (ties broken arbitrarily):
\begin{equation*}
    \norm{g_{i_1}^t}\leq \ldots \leq \norm{g_{i_{m-f}}^t} \leq \norm{g_{i_{m-f+1}}^t}\leq \ldots \leq \norm{g_{i_m}^t}.
\end{equation*}
That is, the gradient with the smallest norm, $g_{i_1}^t$, is received from agent $i_1$, and the gradient with the largest norm, $g_{i_m}^t$, is received from agent $i_m$, with $i_j\in S^t$ for all $j$. Then, the output of the CGE gradient-filter is the vector sum of the $m-f$ gradients with smallest $m-f$ Euclidean norms. Specifically,
\begin{align}
    \gf\left(m,f;\,\left\{g^t_i\right\}_{i\in S^t} \right) = \sum_{j=1}^{m-f}g_{i_j}^t. \label{eqn:cge_gf}
\end{align}

We now write this result in the full form as follows, and provide its proof. In this proof, we set $m=n-r$.
\begin{mdframed}
    \textbf{Theorem~\ref{thm:async-fault-toler}-CGE.}
    \textit{Suppose that Assumptions~\ref{assum:lipschitz} and \ref{assum:strongly-convex-ft} %, and \ref{assum:existence-ft} 
    hold true, and the cost functions of all agents satisfies $(f,r;\epsilon)$-redundancy. Assume that $\eta_t$ %in \eqref{eqn:step-size} 
    satisfies $\sum_{t=0}^\infty=\infty$ and $\sum_{t=0}^\infty\eta_t^2<\infty$. Suppose $\mathsf{GradFilter}$ in use is \emph{CGE}. If the following conditions hold true: 
    \begin{enumerate}[nosep, label=(\roman*)]
        \item $\norm{\mathsf{GradFilter}\left(n-r,f;\,\left\{g_i^t\right\}_{i\in S^t}\right)}<\infty$ for all $t$, and 
        \item if 
        \begin{equation}
            \label{eqn:alpha-cge}
            \alpha = 1 - \frac{f-r}{n-r} + \frac{2\mu}{\gamma}\cdot\frac{f+r}{n-r}>0,
        \end{equation}
        then for each set of $n-f$ non-faulty agents $\H$, for an arbitrary $\delta > 0$,
        \begin{align*}
            \phi_t \geq \alpha m \gamma \delta \left( \frac{4\mu (f+r)\epsilon}{\alpha\gamma} + \delta \right) > 0 \textrm{ when } \norm{x^t-x_{\H}}\geq\frac{4\mu (f+r)\epsilon}{\alpha\gamma}+\delta.
        \end{align*}
    \end{enumerate}
    Then for Algorithm~\ref{alg} with aggregation rule \eqref{eqn:aggregation-rule-async-ft} with \emph{CGE} gradient filter, we have
    $$\lim_{t\rightarrow\infty}\norm{x^t-x_\H}\leq\frac{4\mu (f+r)\epsilon}{\alpha\gamma}.$$}
\end{mdframed}

\begin{proof}
Throughout, we assume $f > 0$ to ignore the trivial case of $f = 0$. 

\textbf{First}, we show that $\norm{\gf\left(m,f;\,\left\{g^t_i\right\}_{i\in S^t} \right)}=\norm{\sum_{j=1}^{m-f}g_{i_j}^t} < \infty, ~ \forall t$. Consider a subset $S_1 \subset \H$ with $\mnorm{S_1} = m-2f$. From triangle inequality,
\begin{align*}
    \norm{\sum_{j\in S_1}\nabla Q_j(x)-\sum_{j\in S_1}\nabla Q_j(x_{\H})} \leq \sum_{j \in S_1}\norm{\nabla Q_j(x) - \nabla Q_j(x_{\H})}, \quad \forall x \in \R^d.
\end{align*}
Under Assumption~\ref{assum:lipschitz}, i.e., Lipschitz continuity of non-faulty gradients, for each non-faulty agent $j$, $\norm{\nabla Q_j(x) - \nabla Q_j(x_{\H})} \leq \mu \norm{x - x_{\H}}$. Substituting this above implies that 
\begin{equation}
    \norm{\sum_{j\in S_1}\nabla Q_j(x)-\sum_{j\in S_1}\nabla Q_j(x_{\H})} \leq \mnorm{S_1} \mu \, \norm{x-x_{\H}}.
    \label{eqn:lipschitz-distance}
\end{equation}
As $\mnorm{S_1} = m-2f$, the $(f,r;\epsilon)$-redundancy property defined in Definition~\ref{def:redundancy} implies that for all $x_1\in\arg\min_x\sum_{j\in S_1}Q_j(x)$,
\[\norm{x_1-x_{\H}} \leq \epsilon.\]
Substituting from above in~\eqref{eqn:lipschitz-distance} implies that, for all $x_1\in\arg\min_x\sum_{j\in S_1}Q_j(x)$,
\begin{equation}
    \norm{\sum_{j\in S_1}\nabla Q_j(x_1)-\sum_{j\in S_1}\nabla Q_j(x_{\H})} \leq \mnorm{S_1} \mu \, \norm{x_1-x_{\H}} \leq \mnorm{S_1} \mu \epsilon.
    \label{eqn:lipschitz-distance-2}
\end{equation}
For all $x_1\in\arg\min_x\sum_{j\in S_1}Q_j(x)$, $\nabla Q_j(x_1) = 0$. Thus,~\eqref{eqn:lipschitz-distance-2} implies that
\begin{equation}
    \norm{\sum_{j\in S_1}\nabla Q_j(x_{\H})} \leq \mnorm{S_1} \mu \epsilon.
    \label{eqn:lipschitz-distance-S1}
\end{equation}
Now, consider an arbitrary non-faulty agent $i \in \H\setminus S_1$. Let $S_2=S_1\cup\{i\}$. 
% Also suppose $i\in\H\backslash S_1$, and $S_2=S_1\cup\{i\}$. By Assumption \ref{assum:lipschitz}, there exists some $\mu$, such that %$(2f,\epsilon)$-redundancy, 
% For any $x_1\in\arg\min_x\sum_{j\in S_1}Q_j(x)$, by $(2f,\epsilon)$-redundancy, $\norm{x_1-x^*}\leq\epsilon$. Combining with \eqref{eqn:lipschitz-distance}
% \begin{equation}
%     \norm{\sum_{j\in S_1}\nabla Q_j(x^*)}=\norm{\sum_{j\in S_1}\nabla Q_j(x_1)-\sum_{j\in S_1}\nabla Q_j(x^*)}\leq\mu\epsilon.
% \end{equation}
Using similar arguments as above we obtain that under the $(f,r;\epsilon)$-redundancy property and Assumption~\ref{assum:lipschitz}, for all $x_2\in\arg\min_x\sum_{j\in S_2}Q_j(x)$,
\begin{align}
    \norm{\sum_{j\in S_2}\nabla Q_j(x_{\H})}=\norm{\sum_{j\in S_2}\nabla Q_j(x_2)-\sum_{j\in S_2}\nabla Q_j(x_{\H})}\leq\mnorm{S_2}\mu\epsilon.
\end{align}
Note that $\sum_{j\in S_2}\nabla Q_j(x)=\sum_{j\in S_1}\nabla Q_j(x)+\nabla Q_i(x)$. From triangle inequality,
\begin{equation}
    \norm{\nabla Q_i(x_{\H})}-\norm{\sum_{j\in S_1}\nabla Q_j(x_{\H})}\leq\norm{\sum_{j\in S_1}\nabla Q_j(x_{\H})+\nabla Q_i(x_{\H})}.
\end{equation}
Therefore, for each non-faulty agent $i\in\H$, 
\begin{align}
    \norm{\nabla Q_i(x_{\H})}&\leq\norm{\sum_{j\in S_1}\nabla Q_j(x_{\H})+\nabla Q_i(x_{\H})}+\norm{\sum_{j\in S_1}\nabla Q_j(x_{\H})} \leq\mnorm{S_2}\mu\epsilon+\mnorm{S_1}\mu\epsilon \nonumber\\
    & = (m-2f+1)\mu \epsilon + (m-2f) \mu \epsilon = (2m-4f+1)\mu\epsilon.
    \label{eqn:honest-norm-bound}
\end{align}
Now, for all $x$ and $i\in\H$, by Assumption \ref{assum:lipschitz},
\begin{equation*}
    \norm{\nabla Q_i(x)-\nabla Q_i(x_{\H})}\leq\mu\norm{x-x_{\H}}.
\end{equation*}
By triangle inequality,
\begin{equation*}
    \norm{\nabla Q_i(x)}\leq\norm{\nabla Q_i(x_{\H})}+\mu\norm{x-x_{\H}}.
\end{equation*}
Substituting from~\eqref{eqn:honest-norm-bound} above we obtain that
\begin{equation}
    \norm{\nabla Q_i(x)}\leq(2m-4f+1)\mu\epsilon+\mu\norm{x-x_{\H}}\leq2m\mu\epsilon+\mu\norm{x-x_{\H}}.
    \label{eqn:honest-bound-everywhere}
\end{equation}
% The last inequality in~\eqref{eqn:honest-bound-everywhere} follows from the fact that $f > 0$.\\
We use the above inequality~\eqref{eqn:honest-bound-everywhere} to show below that $\norm{\sum_{j=1}^{m-f}g_{i_j}^t}$ is bounded for all $t$. Recall that for each iteration $t$, 
\begin{equation*}
    \norm{g_{i_1}^t}\leq...\leq\norm{g_{i_{m-f}}^t}\leq\norm{g_{i_{m-f+1}}^t}\leq...\leq\norm{g_{i_m}^t}.
\end{equation*}
%With \eqref{eqn:update}, the construction of $\widehat{g_j^t}$, we have
%\begin{equation}
%    \norm{\widehat{g_j^t}}\leq\norm{g_{i_{n-f}}^t}, ~\forall j\in[n],~t\in\mathbb{Z}_{\geq0}.
%    \label{eqn:norm-upperbound}
%\end{equation}
As there are at most $f$ Byzantine agents, for each $t$ there exists $\sigma_t\in\H$ such that
\begin{equation}
    \norm{g_{i_{m-f}}^t}\leq\norm{g_{i_{\sigma_t}}^t}.
    \label{eqn:honest-bound}
\end{equation}
As $g_j^t=\nabla Q_j(x^t)$ for all $j\in\H$, from~\eqref{eqn:honest-bound} we obtain that
\begin{equation*}
    \norm{g_{i_j}^t}\leq\norm{\nabla Q_{\sigma_t}(x^t)}, \quad \forall j \in \{1, \ldots, m-f\}, ~ t.
\end{equation*}
Substituting from~\eqref{eqn:honest-bound-everywhere} above we obtain that for every $j \in \{1, \ldots, m-f\}$,
\begin{equation*}
    \norm{g_{i_j}^t}\leq\norm{g_{i_{m-f}}^t}\leq2m\mu\epsilon+\mu\norm{x^t-x_{\H}}.
\end{equation*}
Therefore, from triangle inequality,
\begin{equation}
    \norm{\sum_{j=1}^{m-f}g_{i_j}^t}\leq\sum_{j=1}^{m-f}\norm{g_{i_j}^t}\leq(m-f)\left(2m\mu\epsilon+\mu\norm{x^t-x_{\H}}\right). \label{eqn:filtered-upperbound}
\end{equation}
Recall that $x_{\H} \in \W$. Let $\Gamma = \max_{x \in \W} \norm{x - x_{\H}}$. As $\W$ is a compact set, $\Gamma < \infty$. Recall from the update rule~\eqref{eqn:update} that $x^t \in \W$ for all $t$. Thus, $\norm{x^t - x_{\H}} \leq \max_{x \in \W} \norm{x - x_{\H}} = \Gamma < \infty$. Substituting this in~\eqref{eqn:filtered-upperbound} implies that
\begin{equation}
    \norm{\sum_{j=1}^{m-f}g_{i_j}^t} \leq (m-f) \left( 2m \mu \epsilon + \mu \Gamma\right) < \infty. 
\end{equation}
Recall that in this particular case, $\sum_{j=1}^{m-f}g_{i_j}^t = \gf\left(m,f;\,\left\{g^t_i\right\}_{i\in S^t} \right)$ (see \eqref{eqn:cge_gf}). Therefore, from above we obtain that
\begin{align}
    \norm{\gf\left(m,f;\,\left\{g^t_i\right\}_{i\in S^t} \right)} < \infty, \quad \forall t. \label{eqn:cge_bnd_grd}
\end{align}

% Recall from the update rule~\eqref{eqn:update} that, for all $t \geq 0$,
% \begin{align*}
%     \norm{x^{t+1}-x^*}^2 &= \norm{\left[x^t - \eta_t\sum_{j=1}^{n-f}g_{i_j}^t\right]_{\W} - x^*}^2.
% \end{align*}
% Due to the well-known non-expansion property of projection onto a convex set (see~\cite{boyd2004convex}) we obtain that
% \begin{align*}
%     \norm{x^{t+1}-x^*}^2 &\leq \norm{x^t - \eta_t\sum_{j=1}^{n-f}g_{i_j}^t - x^*}^2 \\
%     &=\norm{x^t-x^*}^2 - 2\eta_t\iprod{x^t-x^*}{\sum_{j=1}^{n-f}g_{i_j}^t}+\eta_t^2\norm{\sum_{j=1}^{n-f}g_{i_j}^t}^2.
% \end{align*}
\textbf{Second}, we show that for an arbitrary $\delta > 0$ there exists $\xi > 0$ such that
\[\phi_t\triangleq\iprod{x^t-x_{\H}}{\sum_{j=1}^{m-f}g_{i_j}^t} \geq \xi ~ \text{ when } ~ \norm{x^t-x_{\H}} \geq \D \, \epsilon + \delta.\]
Consider an arbitrary iteration $t$. Note that, as $\mnorm{\H} = n-f$, there are at least $m-2f$ agents that are common to both sets $\H$ and $\{i_1,...,i_{m-f}\}$. We let $\H^t = \{i_1,...,i_{m-f}\} \cap \H$. The remaining set of agents $\B^t = \{i_1,...,i_{m-f}\} \setminus \H^t$ comprises of only faulty agents. Note that $\mnorm{\H^t} \geq m-2f $ and $\mnorm{\B^t} \leq f$. Therefore,
\begin{equation}
    \phi_t=\iprod{x^t-x_{\H}}{\sum_{j\in\H^t}g_j^t}+\iprod{x^t-x_{\H}}{\sum_{k\in\B^t}g_k^t}.
    \label{eqn:phi-t-two-parts}
\end{equation}
Consider the first term in the right-hand side of~\eqref{eqn:phi-t-two-parts}. Note that
\begin{align*}
    \iprod{x^t-x_{\H}}{\sum_{j\in\H^t}g_j^t}&=\iprod{x^t-x_{\H}}{\sum_{j\in\H^t}g_j^t+\sum_{j\in\H\backslash\H^t}g_j^t-\sum_{j\in\H\backslash\H^t}g_j^t} \\ 
    &=\iprod{x^t-x_{\H}}{\sum_{j\in\H}g_j^t}-\iprod{x^t-x_{\H}}{\sum_{j\in\H\backslash\H^t}g_j^t}.
\end{align*}
Recall that $g_j^t=\nabla Q_j(x^t)$, $\forall j\in\H$. Therefore,
\begin{equation}
    \iprod{x^t-x_{\H}}{\sum_{j\in\H^t}g_j^t}=\iprod{x^t-x_{\H}}{\sum_{j\in\H}\nabla Q_j(x^t)}-\iprod{x^t-x_{\H}}{\sum_{j\in\H\backslash\H^t}\nabla Q_j(x^t)}. \label{eqn:first_phi_1}
\end{equation}

\noindent Due to the strong convexity assumption (i.e., Assumption~\ref{assum:strongly-convex-ft}), for all $x, \, y \in \R^d$,
\begin{equation*}
    \iprod{x- y}{\nabla \sum_{j\in\H}Q_j(x)-\nabla\sum_{j\in\H} Q_j(y)} \geq \mnorm{\H}\, \gamma\norm{x-y}^2.
\end{equation*}
As $x_{\H}$ is minimum point of $\sum_{j \in \H}Q_j(x)$, $\nabla \sum_{j \in \H} Q_j(x_{\H}) = 0$. Thus, 
\begin{align}
    \iprod{x^t-x_{\H}}{\sum_{j\in\H}\nabla Q_j(x^t)} & = \iprod{x^t-x_{\H}}{\nabla\sum_{j\in\H}Q_j(x^t)-\nabla\sum_{j\in\H}Q_j(x_{\H})} \nonumber \\
    & \geq \mnorm{\H} \, \gamma\norm{x^t-x_{\H}}^2.
    \label{eqn:inner-prod-h}
\end{align}
Now, due to the Cauchy-Schwartz inequality, 
\begin{align}
    \iprod{x^t-x_{\H}}{\sum_{j\in\H\backslash\H^t}\nabla Q_j(x^t)}&=\sum_{j\in\H\backslash\H^t}\iprod{x^t-x_{\H}}{\nabla Q_j(x^t)} \nonumber\\
    &\leq\sum_{j\in\H\backslash\H^t}\norm{x^t-x_{\H}}\, \norm{\nabla Q_j(x^t)}.
    \label{eqn:inner-prod-h-ht}
\end{align}
Substituting from~\eqref{eqn:inner-prod-h} and~\eqref{eqn:inner-prod-h-ht} in~\eqref{eqn:first_phi_1} we obtain that
\begin{equation}
    \iprod{x^t-x_{\H}}{\sum_{j\in\H^t}g_j^t} \geq \gamma\mnorm{\H}\, \norm{x^t-x_{\H}}^2-\sum_{j\in\H\backslash\H^t}\norm{x^t-x_{\H}}\, \norm{\nabla Q_j(x^t)}.
    \label{eqn:phi-t-first-part}
\end{equation}

Next, we consider the second term in the right-hand side of~\eqref{eqn:phi-t-two-parts}. From the Cauchy-Schwartz inequality, 
\begin{equation*}
    \iprod{x^t-x_{\H}}{g_k^t}\geq-\norm{x^t-x_{\H}}\, \norm{g_k^t}.
\end{equation*}
Substituting from~\eqref{eqn:phi-t-first-part} and above in~\eqref{eqn:phi-t-two-parts} we obtain that
\begin{equation}
    \phi_t\geq\gamma\mnorm{\H}\, \norm{x^t-x_{\H}}^2-\sum_{j\in\H\backslash\H^t}\norm{x^t-x_{\H}}\, \norm{\nabla Q_j(x^t)}-\sum_{k\in\B^t}\norm{x^t-x_{\H}}\, \norm{g_k^t}.
    \label{eqn:phi-t-2}
\end{equation}

Recall that, due to the sorting of the gradients, for an arbitrary $k \in \B^t$ and an arbitrary $j \in \H\backslash\H^t$,
% for each $k \in \B^t$ there exists $j \in \H\backslash\H^t$ such that 
\begin{equation}
    \norm{g_k^t}\leq\norm{g_j^t}=\norm{\nabla Q_j(x^t)}. \label{eqn:k_B_j_H}
\end{equation}
Recall that $\B^t = \{i_1, \ldots, \, i_{m-f}\} \setminus \H^t$. Thus, $\mnorm{\B^t} = m-f-\mnorm{\H^t}$. Also, as $\mnorm{\H} = n-f$, $\mnorm{\H\backslash\H^t} = n- f - \mnorm{\H^t}$. That is, $\mnorm{\B^t} \leq \mnorm{\H\backslash\H^t}$. Therefore,~\eqref{eqn:k_B_j_H} implies that
\begin{align*}
    \sum_{k \in \B^t} \norm{g_k^t} \leq \sum_{j \in \H\backslash\H^t} \norm{\nabla Q_j(x^t)}.
\end{align*}
Substituting from above in~\eqref{eqn:phi-t-2}, we obtain that
\begin{align*}
    \phi_t&\geq\gamma\mnorm{\H}\, \norm{x^t-x_{\H}}^2-2\sum_{j\in\H\backslash\H^t}\norm{x^t-x_{\H}}\, \norm{\nabla Q_j(x^t)}.
\end{align*}
Substituting from \eqref{eqn:honest-bound-everywhere}, i.e., $\norm{\nabla Q_i(x)}\leq2m\mu\epsilon+\mu\norm{x-x_{\H}}$, above we obtain that
% . With \eqref{eqn:honest-bound-everywhere}, 
\begin{align*}
    \phi_t&\geq\gamma\mnorm{\H}\, \norm{x^t-x_{\H}}^2 - 2\mnorm{\H\backslash\H^t}\, \norm{x^t-x_{\H}}\, (2m\mu\epsilon+\mu\norm{x^t-x_{\H}}) \nonumber\\
    & \geq \left(\gamma\mnorm{\H}-2\mu\mnorm{\H\backslash\H^t}\right)\norm{x^t-x_{\H}}^2 - 4m\mu\epsilon \mnorm{\H\backslash\H^t}\, \norm{x^t-x_{\H}}.
\end{align*}
As $\mnorm{\H} = n-f$ and $\mnorm{\H\backslash\H^t}\leq f+r$, the above implies that
\begin{align}
    \begin{split}
    \label{eqn:phi-t-pre-final}
        \phi_t & \geq\left(\gamma(n-f)-2\mu (f+r)\right)\norm{x^t-x_{\H}}^2-4m\mu\epsilon (f+r) \norm{x^t-x_{\H}} \\
            & = \left(\gamma(n-f)-2\mu (f+r)\right)\norm{x^t-x_{\H}} \left(\norm{x^t-x_{\H}}-\dfrac{4m\mu\epsilon (f+r)}{\gamma(n-f)-2\mu (f+r)} \right) \\
            & = m \gamma \left( 1 - \frac{f-r}{m} + \frac{2\mu}{\gamma}\cdot\frac{f+r}{m}\right) \norm{x^t-x_{\H}}  \left(\norm{x^t-x_{\H}} - \frac{4 \mu (f+r) \epsilon}{\gamma\left( 1 - \frac{f-r}{m} + \frac{2\mu}{\gamma}\cdot\frac{f+r}{m}\right)} \right).
    \end{split}
\end{align}
Recall that we defined %from~\eqref{eqn:alpha-cge} that
\[\alpha = 1 - \frac{f-r}{m} + \frac{2\mu}{\gamma}\cdot\frac{f+r}{m}. \]
Substituting from above in~\eqref{eqn:phi-t-pre-final} we obtain that
% \[\D = \frac{4\mu f}{\alpha\gamma}\]
% and 
% $$\D = \dfrac{4n\mu f}{n\gamma\left(1-\dfrac{f}{n}\left(1+\dfrac{2\mu}{\gamma}\right)\right)}=\frac{4\mu f}{\alpha\gamma}.$$
% We have
\begin{align}
    \phi_t \geq \alpha m \gamma \norm{x^t-x_{\H}} \left(\norm{x^t-x_{\H}}- \frac{4\mu (f+r)\epsilon}{\alpha\gamma} \right).
    \label{eqn:phi-t-final}
\end{align}
As it is assumed that $\alpha > 0$,~\eqref{eqn:phi-t-final} implies that for an arbitrary $\delta > 0$,
\begin{align}
    \phi_t \geq \alpha m \gamma \delta \left( \frac{4\mu (f+r)\epsilon}{\alpha\gamma} + \delta \right) > 0 ~ \text{ when } \norm{x^t-x_{\H}}>\frac{4\mu (f+r)\epsilon}{\alpha\gamma}.
    \label{eqn:final-cge}
\end{align}
Combine \eqref{eqn:final-cge} with Theorem~\ref{thm:async-fault-toler}, hence, the proof.
% $\lim_{t\rightarrow\infty}\norm{x^t - x_{\H}} \not> \D \, \epsilon$. Hence, and therefore, $\lim_{t\rightarrow\infty}\norm{x^t - x_{\H}} \leq \D \, \epsilon$.
\end{proof}
\vfill \newpage
\section{Proofs of theorems in Section 4}

In this section, we present the detailed proof of the convergence rate results presented in Section 4 of the main paper, i.e., the three cases of Theorem~\ref{thm:cge}, one for each of \textbf{Problems BS}, \textbf{CS}, and \textbf{DS}. We state them separately, then proceed with the proofs.

\subsection{Preliminaries}

First, we revisit these definitions and notations from Section 4.

To compute a stochastic gradient in iteration $t$, each non-faulty agent $i$ samples $k$ i.i.d. data points $z_{i_1}^t,...,z_{i_k}^t$ from the distribution $\mathcal{D}_i$ and computes
\begin{equation}
    g_i^t=\frac{1}{k}\sum_{i=1}^k\nabla\ell(x^t, z_{i_j}^t),
    \label{eqn:def-g-i-t-cge}
\end{equation}
where $k$ is referred as the \textit{batch size} of the training procedure.

Suppose faulty agents in the system are \textit{fixed} during a certain execution. For each non-faulty agent $i$, let $\z_i^t=\left\{z_{i_1}^t,...,z_{i_k}^t\right\}$ denote the collection of $k$ i.i.d. data points sampled by agent $i$ at iteration $t$. Now for each agent $i$ and iteration $t$, we define a random variable 
\begin{equation}
    \zeta_i^t=\begin{cases}
        \z_i^t, & \textrm{ agent $i$ is non-faulty}, \\
        g_i^t, & \textrm{ agent $i$ is faulty}.
    \end{cases}
    \label{eqn:zeta-i-t}
\end{equation}
Recall that $g_i^t$ can be an arbitrary $d$-dimensional random variable for each Byzantine faulty agent $i$. For each iteration $t$, let $\zeta^t=\left\{\zeta_i^t,\,i=1,...,n\right\}$, and let $\E_t$ denote the expectation with repect to the collective random variables $\zeta^0,...,\zeta^t$, given the initial estimate $x^0$. Specifically, 
\begin{equation}
    \E_t(\cdot)=\E_{\zeta^0,...,\zeta^t}(\cdot), ~\forall t\geq0.
\end{equation}

Similar to Section~3, we make standard Assumptions~\ref{assum:lipschitz} and \ref{assum:strongly-convex} (or Assumption~\ref{assum:strongly-convex-ft} depending on the existence of Byzantine agents). We also need an extra assumption to bound the variance of stochastic gradients from all (non-faulty) agents. 

% \begin{assumption}
\textbf{Assumption \ref{assum:bound-grad}.}
\textit{For each (non-faulty) agent $i$, assume that the variance of $g_i^t$ is bounded. Specifically, there exists a finite real value $\sigma$ such that for all non-faulty agent $i$,
\begin{equation}
    \E_{\zeta_i^t}\norm{g_i^t-\E_{\zeta_i^t}\left(g_i^t\right)}^2\leq\sigma^2.
    \label{eqn:assum-bound-var-ft}
\end{equation}}
We make no assumption over Byzantine agents. 

Also, recall that we defined for the asynchronous case without Byzantine agents ($r\geq0, f=0$), by Assumptions~\ref{assum:lipschitz} and \ref{assum:strongly-convex}, there is only one minimum point
\begin{align}
    x^*=\arg\min_{x}\sum_{j\in[n]}Q_j(x)
\end{align}
for the aggregate cost functions of all agents. If Byzantine agents can exist ($f\geq0$), by Assumptions~\ref{assum:lipschitz} and \ref{assum:strongly-convex-ft}, there is only one minimum point
\begin{align}
    x_\H=\arg\min_{x}\sum_{j\in\H}Q_j(x)
\end{align}
for the aggregate cost functions of a set of non-faulty agents $\H$, where $\mnorm{\H}=n-f$. None trivially, we assume $x^*$ (or $x_\H$ if allowing Byzantine agents) is contained in the convex compact set $\W$.

\subsection{Observations and lemmas from the assumptions}

Before the proof, we first introduce and prove some results from the assumptions we made. 
From the definition of $\zeta^t$, for each non-faulty agent $i$ and deterministic real-valued function $\Psi$, 
\begin{equation}
    \E_{\zeta^t}\Psi\left(g_i^t\right)=\E_{\zeta_1^t,...,\zeta_n^t}\Psi\left(g_i^t\right).
\end{equation}
Also, for non-faulty each agent $i$, $\zeta_i^t=\z_i^t$. For a given current estimate $x^t$, the stochastic gradient $g_i^t$ is a function of data points $\z_i^t$ sampled by agent $i$. As non-faulty agents choose their data points independently in each iteration, from above we have that for each non-faulty agent $i$,
\begin{equation}
    \E_{\zeta_1^t,...,\zeta_n^t}\Psi\left(g_i^t\right)=\E_{\z_i^t}\Psi\left(g_i^t\right).
\end{equation}
Therefore, for each non-faulty agent $i$, 
\begin{equation}
    \E_{\zeta^t}\Psi\left(g_i^t\right)=\E_{\z_i^t}\Psi\left(g_i^t\right).
    \label{eqn:expectation-zeta-cge}
\end{equation}

Trivially, from \eqref{eqn:expectation-zeta-cge} we obtain that for an arbitrary non-faulty agent $i$ and iteration $t$, 
\begin{equation}
    \E_{\zeta^t}\left[g_i^t\right]=\E_{\z_i^t}\left[g_i^t\right].
    \label{eqn:def-expectation-zeta-g-cge}
\end{equation}
From the definition of a stochastic gradient \eqref{eqn:def-g-i-t-cge} and \eqref{eqn:def-expectation-zeta-g-cge} we have
\begin{equation}
    \E_{\zeta^t}\left[g_i^t\right]=\frac{1}{k}\cdot\E_{\z_i^t}\sum_{j=1}^k\left(\nabla\ell(x^t,z_{i_j}^t)\right),
\end{equation}
where the gradient of loss function $\ell(\cdot,\cdot)$ is with respect to its first argument $w$, the estimate of the machine learning model. Recall that $\z_i^t$ constitutes $k$ data points sampled i.i.d. from the distribution $\mathcal{D}_i$, therefore,
\begin{equation}
    \E_{\zeta^t}\left[g_i^t\right]=\frac{1}{k}\cdot\sum_{j=1}^k\E_{z_{i_j}^t\sim\mathcal{D}_i}\nabla\ell(x^t,z_{i_j}^t).
    \label{eqn:expectation-zeta-g-cge}
\end{equation}
Note that
\begin{equation}
    \nabla Q_i(x)=\E_{z\sim\mathcal{D}_i}\nabla \ell(w,z),~\forall w\in\R^d.
    \label{eqn:qi-expectation-cge}
\end{equation}
Substituting from \eqref{eqn:qi-expectation-cge} in \eqref{eqn:expectation-zeta-g-cge}, we obtain that for any non-faulty agent $i$,
\begin{equation}
    \E_{\zeta^t}\left[g_i^t\right]=\frac{1}{k}\sum_{j=1}^k\nabla Q_i(x^t)=\nabla Q_i(x^t).
    \label{eqn:expectation-g-i-t-cge}
\end{equation}

\subsubsection{Lemma~\ref{lemma:1-cge} on expectations related to gradients}
\begin{mdframed}
\begin{lemma}
    \label{lemma:1-cge}
    For any iteration $t$, if Assumption~\ref{assum:bound-grad} holds true, for each non-faulty agent $i$, the two inequalities hold true:
    \begin{equation}
        \E_{\zeta_i^t}\norm{g_i^t-\E_{\zeta_i^t}\left[g_i^t\right]}\leq\sigma.
        \label{eqn:assum-bound-grad-cge}
    \end{equation}
    \begin{equation}
        \E_{\zeta^t}\norm{g_i^t}^2\leq\sigma^2+\norm{\nabla Q_i(x^t)}^2.
        \label{eqn:lemma-1-cge}
    \end{equation}
\end{lemma}
\end{mdframed}
\begin{proof}
    For the \textbf{first} inequality: Note that $\left(\cdot\right)^2$ is a convex function. By Jensen's inequality, 
    \begin{equation}
        \left(\E_{\zeta_i^t}\norm{g_i^t-\E_{\zeta_i^t}\left[g_i^t\right]}\right)^2\leq\E_{\zeta_i^t}\norm{g_i^t-\E_{\zeta_i^t}\left[g_i^t\right]}^2.
    \end{equation}
    Substitute above in Assumption~\ref{assum:bound-grad}, we have
    \begin{equation}
        \left(\E_{\zeta_i^t}\norm{g_i^t-\E_{\zeta_i^t}\left[g_i^t\right]}\right)^2\leq\sigma^2.
    \end{equation}
    Since both sides are non-negative, taking square roots on both sides we have
    \begin{equation}
        \E_{\zeta_i^t}\norm{g_i^t-\E_{\zeta_i^t}\left[g_i^t\right]}\leq\sigma.
        \tag{\ref{eqn:assum-bound-grad-cge}}
    \end{equation}

    For the \textbf{second} inequality: Let $i$ be an arbitrary non-faulty agent. Using the definition of Euclidean norm, for any $t$,
    \begin{equation}
        \norm{g_i^t-\E_{\zeta^t}\left[g_i^t\right]}^2 = \norm{g_i^t}^2-2\iprod{g_i^t}{\E_{\zeta^t}\left[g_i^t\right]} + \norm{\E_{\zeta^t}\left[g_i^t\right]}^2.
    \end{equation}
    %As the expected value of a constant is the constant itself,
    Upon taking expectations on both sides, we obtain that 
    \begin{align}
        \E_{\zeta^t}\norm{g_i^t-\E_{\zeta^t}\left[g_i^t\right]}^2 &= \E_{\zeta^t}\norm{g_i^t}^2-\E_{\zeta^t}2\iprod{g_i^t}{\E_{\zeta^t}\left[g_i^t\right]} + \E_{\zeta^t}\norm{\E_{\zeta^t}\left[g_i^t\right]}^2 \nonumber \\\
        &= \E_{\zeta^t}\norm{g_i^t}^2-2\E_{\zeta^t}\left[g_i^t\right]\cdot\E_{\zeta^t}\left[g_i^t\right] + \norm{\E_{\zeta^t}\left[g_i^t\right]}^2 \nonumber \\
        &= \E_{\zeta^t}\norm{g_i^t}^2- \norm{\E_{\zeta^t}\left[g_i^t\right]}^2.
    \end{align}
    Note that from \eqref{eqn:expectation-zeta-cge}, $\E_{\zeta^t}\norm{g_i^t-\E_{\zeta^t}\left[g_i^t\right]}^2=\E_{\z^t}\norm{g_i^t-\E_{\z^t}\left[g_i^t\right]}^2$, therefore
    \begin{equation}
        \E_{\z^t}\norm{g_i^t-\E_{\z^t}\left[g_i^t\right]}^2= \E_{\zeta^t}\norm{g_i^t}^2- \norm{\E_{\z^t}\left[g_i^t\right]}^2.
        \label{eqn:expectation-difference-cge}
    \end{equation}
    Recall from \eqref{eqn:expectation-g-i-t-cge} that $\E_{\zeta_i^t}\left[g_i^t\right]=\nabla Q_i(x^t)$. Substituting this in \eqref{eqn:expectation-difference-cge}, 
    \begin{equation}
        \E_{\z^t}\norm{g_i^t-\E_{\z^t}\left[g_i^t\right]}^2= \E_{\zeta^t}\norm{g_i^t}^2-\norm{\nabla Q_i(x^t)}^2.
        \label{eqn:expectation-difference-cge-2}
    \end{equation}
    Combining Assumption~\ref{assum:bound-grad} and \eqref{eqn:expectation-zeta-cge}, we have $\E_{\z^t}\norm{g_i^t-\E_{\z^t}\left[g_i^t\right]}^2\leq\sigma^2$. Substituting this in \eqref{eqn:expectation-difference-cge-2}, we obtain that
    \begin{equation}
        \E_{\zeta^t}\norm{g_i^t}^2\leq\sigma^2+\norm{\nabla Q_i(x^t)}^2. \tag{\ref{eqn:lemma-1-cge}}
    \end{equation}
\end{proof}

\subsubsection{Lemmas on $\gamma$ and $\mu$}
Suppose there is no faulty agents in the system ($f=0$), we have the following result:
\begin{mdframed}
\begin{lemma}
    \label{lemma:gamma-mu}
    If Assumptions~\ref{assum:lipschitz} and \ref{assum:strongly-convex} hold true, 
    \begin{equation}
        \gamma \leq \mu.
    \end{equation}
\end{lemma}
\end{mdframed}

\begin{proof}
    Consider a set $S$ of at least $n-r$ agents. By Assumption~\ref{assum:strongly-convex}, for any $w\in\R^d$,
    \begin{align}
        \iprod{\sum_{j\in S}\nabla Q_j(x)-\sum_{j\in S}\nabla Q_j(x^*)}{x-x^*}\geq\mnorm{S}\gamma\norm{x-x^*}^2.
        \label{eqn:lemma-2-gamma}
    \end{align}
    By Cauchy-Schwartz inequality and triangle inequality,
    \begin{align}
        \iprod{\sum_{j\in S}\nabla Q_j(x)-\sum_{j\in S}\nabla Q_j(x^*)}{x-x^*}&\leq\norm{\sum_{j\in S}\nabla Q_j(x)-\sum_{j\in S}\nabla Q_j(x^*)}\norm{x-x^*} \nonumber \\
        &\leq \sum_{j\in S}\norm{\nabla Q_j(x) - \nabla Q_j(x^*)}\norm{x-x^*}.
        \label{eqn:lemma-2-mu-0}
    \end{align}
    By Assumption~\ref{assum:lipschitz}, for any agent $j$ and $w\in\R^d$,
    \begin{equation}
        \norm{\nabla Q_j(x) - \nabla Q_j(x^*)}\leq \mu\norm{x-x^*}.
    \end{equation}
    Therefore, 
    \begin{equation}
        \sum_{j\in S}\norm{\nabla Q_j(x) - \nabla Q_j(x^*)}\norm{x-x^*}\leq\mnorm{S}\mu\norm{x-x^*}^2.
        \label{eqn:lemma-2-mu}
    \end{equation}
    Combining \eqref{eqn:lemma-2-gamma}, \eqref{eqn:lemma-2-mu-0}, and \eqref{eqn:lemma-2-mu}, we have
    \begin{equation}
        \gamma \leq \mu.
    \end{equation}
\end{proof}

The same result can also be obtained when there are faulty agents in the system ($f\geq0$), only that a different assumption needs to be applied:
\begin{mdframed}
\begin{lemma}
    \label{lemma:gamma-mu-cge}
    If Assumptions~\ref{assum:lipschitz} and \ref{assum:strongly-convex-ft} hold true, 
    \begin{equation}
        \gamma \leq \mu.
    \end{equation}
\end{lemma}
\end{mdframed}
The proof follows the same steps as Lemma~\ref{lemma:gamma-mu-cge}.

\subsubsection{Lemma~\ref{lemma:converge-sgd} on expected convergence of estimates}

\begin{mdframed}
\begin{lemma}
    \label{lemma:converge-sgd}
    Consider the general iterative update rule \eqref{eqn:update}. Suppose $\M\geq0$ and $\rho\in[0,1)$ are two real values. If 
    \begin{align}
        \E_{\zeta^t}\norm{x^{t+1}-x_\H}^2\leq&\rho\norm{x^t-x_\H}^2 + \M.
        \label{eqn:expectation-w-t-square-sgd-cge-5}
    \end{align}
    we have
    \begin{align}
        \E_{t}\norm{x^{t+1}-x_\H}^2&\leq\rho^{t+1}\norm{x^0-x_\H}^2 + \left(\frac{1-\rho^{t+1}}{1-\rho}\right)\M.
        \label{eqn:expectation-bound-1-cge}
    \end{align}
\end{lemma}
\end{mdframed}

\begin{proof}
    We start from the condition 
    \begin{align}
        \E_{\zeta^t}\norm{x^{t+1}-x_\H}^2\leq&\rho\norm{x^t-x_\H}^2 + \M.
        \label{eqn:expectation-w-t-square-sgd-cge-5}
    \end{align}
    Specifically, since $\E_0=\E_{\zeta^0}$, for $t=0$, we have 
    \begin{align}
        \E_0\norm{x^{1}-x_\H}^2\leq&\rho\norm{x^0-x_\H}^2 + \M.
        \label{eqn:expectation-w-t-square-sgd-cge-6}
    \end{align}
    
    Recall that $x^t$ is a function of random variable $\zeta^{t-1}=\left\{\zeta_1^{t-1},...,\zeta_n^{t-1}\right\}$ given $x^{t-1}$. Recursively, we obtain that $x^t$ is a function of random variables $\zeta^{0},...,\zeta^{t-1}$, given the initial estimate $x^0$. Therefore, $\norm{x^{t+1}-x_\H}^2$ is a function of random variables $\zeta^{0},...,\zeta^{t}$, given the initial estimate $x^0$. For all $t>0$, let 
    \begin{align}
        \E_{\zeta^t|\zeta^0,...,\zeta^{t-1}}\norm{x^{t+1}-x_\H}^2
    \end{align}
    denote the conditional expectation of $\norm{x^{t+1}-x_\H}^2$ given the random variables $\zeta^0,...,\zeta^{t-1}$, and $x^0$. Thus,
    \begin{align}
        \E_{\zeta^t}\norm{x^{t+1}-x_\H}^2=\E_{\zeta^t|\zeta^0,...,\zeta^{t-1}}\norm{x^{t+1}-x_\H}^2.
        \label{eqn:condition-exp-norm-sgd-cge}
    \end{align}
    Substituting \eqref{eqn:condition-exp-norm-sgd-cge} in \eqref{eqn:expectation-w-t-square-sgd-cge-5}, we obtain that given $x^0$,
    \begin{equation}
        \E_{\zeta^t|\zeta^0,...,\zeta^{t-1}}\norm{x^{t+1}-x_\H}^2\leq\rho\norm{x^t-x_\H}^2 + \M.
        \label{eqn:expectation-w-t-square-sgd-cge-7}
    \end{equation}
    Note that by Bayes' rule, for all $t>0$,
    \begin{align}
        \E_{\zeta^0,...,\zeta^{t}}\norm{x^{t+1}-x_\H}^2=\E_{\zeta^0,...,\zeta^{t-1}}\left(\E_{\zeta^t|\zeta^0,...,\zeta^{t-1}}\norm{x^{t+1}-x_\H}^2\right).
    \end{align}
    Taking expectation $\E_{\zeta^0,...,\zeta^{t-1}}$ on both sides of \eqref{eqn:expectation-w-t-square-sgd-cge-7}, we obtain that given $x^0$,
    \begin{align}
        \E_{\zeta^0,...,\zeta^{t}}\norm{x^{t+1}-x_\H}^2&\leq\E_{\zeta^0,...,\zeta^{t-1}}\left(\rho\norm{x^t-x_\H}^2 + \M\right) \nonumber \\
        &\leq\rho\E_{\zeta^0,...,\zeta^{t-1}}\norm{x^t-x_\H}^2 + \M.
    \end{align}
    Recall that we defined the notation $\E_t$ as a shorthand for the joint expectation $\E_{\zeta^0,...,\zeta^{t}}$ given $x^0$ for all $t$. Therefore, with this notation,
    \begin{align}
        \E_{t}\norm{x^{t+1}-x_\H}^2&\leq\rho\E_{t-1}\norm{x^t-x_\H}^2 + \M.
        \label{eqn:expectation-w-t-square-recursive-sgd-cge}
    \end{align}
    
    By induction, we now prove the following bound
    \begin{align}
        \E_{t}\norm{x^{t+1}-x_\H}^2&\leq\rho^{t+1}\norm{x^0-x_\H}^2 + \left(\frac{1-\rho^{t+1}}{1-\rho}\right)\M.
        \tag{\ref{eqn:expectation-bound-1-cge}}
    \end{align}
    We have already shown that \eqref{eqn:expectation-bound-1-cge} is true when $t=0$ (see \eqref{eqn:expectation-w-t-square-sgd-cge-6}). Suppose \eqref{eqn:expectation-bound-1-cge} is true when $t=\tau-1$ where $\tau$ is an arbitrary integer with $\tau>1$, i.e.,
    \begin{align}
        \E_{\tau-1}\norm{x^{\tau}-x_\H}^2&\leq\rho^{\tau}\norm{x^0-x_\H}^2 + \left(\frac{1-\rho^{\tau}}{1-\rho}\right)\M.
        \label{eqn:expectation-bound-2-cge}
    \end{align}
    Combining \eqref{eqn:expectation-w-t-square-recursive-sgd-cge} with above, we have
    \begin{align}
        \E_{\tau}\norm{x^{\tau+1}-x_\H}^2&\leq\rho\E_{\tau-1}\norm{x^\tau-x_\H}^2 + \M \nonumber \\
        &\leq\rho\left(\rho^{\tau}\norm{x^0-x_\H}^2 + \left(\frac{1-\rho^{\tau}}{1-\rho}\right)\M\right)+\M \nonumber \\
        &\leq\rho^{\tau+1}\norm{x^0-x_\H}^2 + \left(\rho\left(\frac{1-\rho^{\tau}}{1-\rho}\right)+1\right)\M \nonumber \\
        &\leq\rho^{\tau+1}\norm{x^0-x_\H}^2 + \left(\frac{1-\rho^{\tau+1}}{1-\rho}\right)\M.
    \end{align}
    That is, \eqref{eqn:expectation-bound-1-cge} is also true when $t=\tau$. By induction, \eqref{eqn:expectation-bound-1-cge} holds true for all $t\geq0$.
\end{proof}

Note that Lemma~\ref{lemma:converge-sgd} also holds when replacing all $x_\H$ with $x^*$ if there is no faulty agents in the system.

In the following part of this section, we present the detailed proofs of Theorem 4 for three cases.

\vfill \newpage
\subsection{Proof of Theorem \ref{thm:cge} - Problem BS}

Recall that for \textbf{Problem BS}, there is up to $f$ Byzantine faulty agents, and 0 stragglers. Let us define the following parameters:
\begin{itemize}[nosep]
    \item The \textit{resilience margin}
        \begin{align}
            \alpha = 1-\frac{f}{n}\cdot\frac{\gamma+2\mu}{\gamma}.
            \label{eqn:def-alpha-ft}
        \end{align}
    \item The parameter that determines step size
        \begin{align}
            \overline{\eta} = \frac{2n\gamma\alpha}{(n-f)^2\mu^2}.
            \label{eqn:def-eta-bar-ft}
        \end{align}
\end{itemize}

\begin{mdframed}
    \textbf{Theorem~\ref{thm:cge}-BS.} 
    \textit{Consider Algorithm~\ref{alg} with stochastic gradients and the following gradient aggregator
    \begin{align*}
        \textstyle\mathsf{GradAgg}\left(g_j^t|j\in S^t;n,f,0\right)
        =\mathsf{GradFilter}\left(g_j^t|j\in[n];n,f\right)
        % \label{eqn:aggregation-rule-ft}
    \end{align*}
    using CGE gradient filter.
    Suppose Assumptions~\ref{assum:lipschitz}, \ref{assum:strongly-convex-ft}, and \ref{assum:bound-grad} hold true, the expected cost functions of non-faulty functions satisfy $(f,0;\epsilon)$-redundancy, $\alpha>0$ and step size in \eqref{eqn:update} $\eta_t=\eta>0$ for all $t$. Let 
    \begin{align}
        \M &= 4n\eta\mu\epsilon\left(2f+(n-f)^2\eta\mu\right)\Gamma + 4n^2(n-f)^2\eta^2\mu^2\epsilon^2 + 2f\eta\sigma\Gamma + (n-f)^2\eta^2\sigma^2.
        \label{eqn:def-m-ft}
    \end{align}
    If $\eta<\overline{\eta}$, the following holds true:
    \begin{itemize}[nosep]
        \item The value of a convergence rate parameter 
            \begin{align}
                \rho &= 1-2(n-f)\eta\gamma+4f\eta\mu+(n-f)^2\eta^2\mu^2.
                \label{eqn:def-rho-ft}
            \end{align}
            satisfies $0\leq\rho<1$, and 
        \item Given the initial estimate $x^0$ arbitrarily chosen from $\W$, for all $t\geq0$, \footnote{In the main paper $\W$ was mistyped as $\R^d$.}
            \begin{align}
                \E_{t}\norm{x^{t+1}-x_\H}^2&\leq\rho^{t+1}\norm{x^0-x_\H}^2 + \frac{1-\rho^{t+1}}{1-\rho}\M.
                \label{eqn:expectation-bound-1}
            \end{align}
    \end{itemize}}
\end{mdframed}

\begin{proof}
    \textbf{First}, we show a recursive bound over the expected value of $\norm{x^t-x^*}^2$. Following the \textbf{first} part of the proof of Theorem~\ref{thm:async-fault-toler}-CGE, we know that by Assumption~\ref{assum:lipschitz} and $(f,r;\epsilon)$-redundancy, for every non-faulty agent $i\in[n]$, we obtain \eqref{eqn:honest-bound-everywhere} that
    \begin{equation*}
        \norm{\nabla Q_i(x)}\leq(2m-4f+1)\mu\epsilon+\mu\norm{x-x_{\H}}\leq2m\mu\epsilon+\mu\norm{x-x_{\H}}.
    \end{equation*}
    This result stands when $r=0$, the current case.
    
    Let us denote $\g^t$ the output of CGE gradient-filter applied, i.e.
    \begin{equation}
        \g^t=\gf\left(g_j^t|j\in[n];n,f\right)=\sum_{l=1}^{n-f}g_{i_l}^t,
        \label{eqn:gothic-g-t-cge}
    \end{equation}
    where the gradients are sorted by their norms as 
    \begin{equation}
        \norm{g_{i_1}^t}\leq...\leq\norm{g_{i_{n-f-1}}^t}\leq\norm{g_{i_{n-f}}^t}\leq\norm{g_{i_{n-f+1}}^t}\leq...\leq\norm{g_{i_n}^t}.
    \end{equation}
    With $\eta_t=\eta$ and above, for each iteration $t$ we have
    \begin{equation}
        x^{t+1}=x^t-\eta\g^t.
        \label{eqn:update-gothic-g-t-cge}
    \end{equation}
    % From the definition of Euclidean norm,
    % By triangle inequality, 
    Subtracting $x_\H$ and taking norm on both sides, we have
    \begin{equation}
        \norm{x^{t+1}-x_\H}^2=\norm{x^t-x_\H}^2-2\eta\iprod{x^t-x_\H}{\g^t}+\eta^2\norm{\g^t}^2.
        \label{eqn:iterate-general}
    \end{equation}
    
    By triangle inequality,
    \begin{equation}
        \norm{\g^t}\leq\sum_{l=1}^{n-f}\norm{g_{i_l}^t}.
    \end{equation}
    Note that $\H$ denotes a set of $n-f$ non-faulty agents. We have
    \begin{equation}
        \sum_{l=1}^{n-f}\norm{g_{i_l}^t} \leq \sum_{j\in\H}\norm{g_{j}^t}.
    \end{equation}
    Combining the above two inequalities together, we have
    \begin{equation}
        \norm{\g^t}\leq \sum_{j\in\H}\norm{g_{j}^t}.
    \end{equation}
    Thus, by and AM-QM inequality (i.e., $\frac{1}{n}\sum_{j=1}^na_j\leq\sqrt{\frac{1}{n}\sum_{j=1}^na_j^2}$ for any $n$ positive real $a_j$'s), from the above we have 
    \begin{equation}
        \norm{\g^t}^2\leq\left(\sum_{j\in\H}\norm{g_j^t}\right)^2\leq\mnorm{\H}\sum_{j\in\H}\norm{g_j^t}^2=(n-f)\sum_{j\in\H}\norm{g_j^t}^2.
        \label{eqn:cge-square-bound}
    \end{equation}
    On the other hand, consider the term $\iprod{x^t-x_\H}{\g^t}$. let $\H^t=\{i_1,...,i_{n-f}\}\cap\H$, and let $\B^t=\{i_1,...,i_{n-f}\}\backslash\H^t$. Note that 
    \begin{equation}
        \mnorm{\H^t}\geq\mnorm{\H}-f = n-2f, ~\textrm{ and }~\mnorm{\B^t}\leq f.
    \end{equation}
    From \eqref{eqn:gothic-g-t-cge}, 
    \begin{equation}
        \g^t=\sum_{j\in\H^t}g_j^t+\sum_{j\in\B^t}g_j^t.
    \end{equation}
    Therefore, 
    \begin{equation}
        \iprod{x^t-x_\H}{\g^t} = \iprod{x^t-x_\H}{\sum_{j\in\H^t}g_j^t} + \iprod{x^t-x_\H}{\sum_{j\in\B^t}g_j^t}.
    \end{equation}
    We define 
    \begin{equation}
        % \phi_t=\sum_{j\in\H^t}\iprod{x^t-x_\H}{g_j^t} - f\norm{x^t-x_\H}\norm{g_{v_t}^t}.
        \phi_t=\iprod{x^t-x_\H}{\g^t}=\sum_{j\in\H^t}\iprod{x^t-x_\H}{g_j^t} + \sum_{j\in\B^t}\iprod{x^t-x_\H}{g_j^t}.
        \label{eqn:phi-t}
    \end{equation}
    % Note that $\iprod{x^t-x_\H}{\g^t}\geq\phi_t$. 
    Substituting above and \eqref{eqn:cge-square-bound} in \eqref{eqn:iterate-general}, we obtain that
    \begin{equation}
        \norm{x^{t+1}-x_\H}^2\leq\norm{x^t-x_\H}^2-2\eta\phi_t+(n-f)\eta^2\sum_{j\in\H}\norm{g_j^t}^2.
    \end{equation}
    Recall that $\zeta^t=\left\{\zeta_1^t,...,\zeta_n^t\right\}$, and $x^{t+1}$ is a function of the set of random variables $\zeta^t$. Also note that $\E_{\zeta^t}\norm{x^t-x_\H}^2=\norm{x^t-x_\H}^2$. Taking expectation $\E_{\zeta^t}$ on both sides, we have
    \begin{equation}
        \E_{\zeta^t}\norm{x^{t+1}-x_\H}^2\leq\norm{x^t-x_\H}^2-2\eta\E_{\zeta^t}\left[\phi_t\right]+(n-f)\eta^2\sum_{j\in\H}\E_{\zeta^t}\norm{g_j^t}^2. 
        \label{eqn:expectation-w-t-square}
    \end{equation}
    
    Consider $\E_{\zeta^t}\left[\phi_t\right]$. From \eqref{eqn:phi-t} the definition of $\phi_t$,
    \begin{align}
        \E_{\zeta^t}\left[\phi_t\right]& = \E_{\zeta^t}\sum_{j\in\H^t}\iprod{x^t-x_\H}{g_j^t} + \E_{\zeta^t}\sum_{j\in\B^t}\iprod{x^t-x_\H}{g_j^t} \nonumber \\
        & = \sum_{j\in\H^t}\iprod{x^t-x_\H}{\E_{\zeta^t}\left[g_j^t\right]} +\E_{\zeta^t}\sum_{j\in\B^t}\iprod{x^t-x_\H}{g_j^t}.
        \label{eqn:exp-phi-t}
    \end{align}
    Recall from \eqref{eqn:expectation-g-i-t-cge} that for any $j\in\H$, $\E_{\zeta^t}\left[g_j^t\right] = \nabla Q_j(x^t)$. The first term of \eqref{eqn:exp-phi-t} becomes
    \begin{align}
        \sum_{j\in\H^t}\iprod{x^t-x_\H}{\E_{\zeta^t}\left[g_j^t\right]} &= \sum_{j\in\H^t}\iprod{x^t-x_\H}{\nabla Q_j(x^t)} \nonumber \\
        &= \iprod{x^t-x_\H}{\sum_{j\in\H}\nabla Q_j(x^t)} - \iprod{x^t-x_\H}{\sum_{j\in\H\backslash\H^t}\nabla Q_j(x^t)}
    \end{align}
    By Assumption~\ref{assum:strongly-convex-ft}, also the fact that $\nabla\sum_{j\in\H}Q_j(x_\H)=0$,
    \begin{align}
        \iprod{x^t-x_\H}{\sum_{j\in\H}\nabla Q_j(x^t)} = \iprod{x^t-x_\H}{\sum_{j\in\H}\nabla Q_j(x^t)-\sum_{j\in\H}\nabla Q_j(x_\H)}\geq \mnorm{\H}\gamma\norm{x^t-x_\H}^2.
        \label{eqn:exp-phit-part1A}
    \end{align}
    By Cauchy-Schwartz inequality,
    \begin{equation}
        \iprod{x^t-x_\H}{\sum_{j\in\H\backslash\H^t}\nabla Q_j(x^t)}= \sum_{j\in\H\backslash\H^t}\iprod{x^t-x_\H}{\nabla Q_j(x^t)} \leq \sum_{j\in\H\backslash\H^t}\norm{x^t-x_\H}\norm{\nabla Q_j(x^t)}.
        \label{eqn:exp-phit-part1B}
    \end{equation}
    Also by Cauchy-Schwartz inequality, for any $j\in[n]$, 
    \begin{equation}
        \iprod{x^t-x_\H}{g_j^t}\geq -\norm{x^t-x_\H}\norm{g_j^t}.
    \end{equation}
    Recall the sorting of vectors $\{g_j^t\}_{j=1}^n$. For an arbitrary $j\in\B^t$ and an arbitrary $j'\in\H\backslash\H^t$, 
    \begin{equation}
        \norm{g_j^t}\leq\norm{g_{j'}^t}.
        \label{eqn:bound-B-t}
    \end{equation}
    Recall that $\B^t=\left\{i_1,...,i_{n-f}\right\}\backslash\H^t$. Thus, $\mnorm{\B^t}=n-f-\mnorm{\H^t}$. Also, as $\mnorm{\H}=n-f$, $\mnorm{\H\backslash\H^t}=n-f-\mnorm{\H^t}$. That is, $\mnorm{\B^t}=\mnorm{\H\backslash\H^t}$. Therefore,  we have
    \begin{align}
       \sum_{j\in\B^t}\iprod{x^t-x_\H}{g_j^t}&\geq-\sum_{j\in\B^t}\norm{x^t-x_\H}\norm{g_j^t}\geq-\sum_{j\in\H\backslash\H^t}\norm{x^t-x_\H}\norm{g_j^t}
       \label{eqn:B-t}
    \end{align}
    Taking expectation on both sides, for the second term of \eqref{eqn:exp-phi-t} we have
    \begin{align}
       \E_{\zeta^t}\sum_{j\in\B^t}\iprod{x^t-x_\H}{g_j^t}\geq\E_{\zeta^t}\left[-\sum_{j\in\H\backslash\H^t}\norm{x^t-x_\H}\norm{g_j^t}\right]=-\sum_{j\in\H\backslash\H^t}\norm{x^t-x_\H}\E_{\zeta^t}\norm{g_j^t}.
       \label{eqn:exp-B-t}
    \end{align}
    Recall \eqref{eqn:assum-bound-grad-cge} from Lemma~\ref{lemma:1-cge} that for any non-faulty agent $i$, 
    \begin{equation*}
        \E_{\zeta_i^t}\norm{g_i^t-\E_{\zeta_i^t}\left[g_i^t\right]}\leq\sigma.
    \end{equation*}
    Note that by triangle inequality,
    \begin{equation}
        \norm{g_i^t-\E_{\zeta_i^t}\left[g_i^t\right]}\geq\norm{g_i^t}-\norm{\E_{\zeta_i^t}\left[g_i^t\right]}.
    \end{equation}
    Taking expectation on both sides, we obtain that
    \begin{equation}
        \E_{\zeta_i^t}\norm{g_i^t-\E_{\zeta_i^t}\left[g_i^t\right]}\geq\E_{\zeta_i^t}\norm{g_i^t}-\norm{\E_{\zeta_i^t}\left[g_i^t\right]}.
    \end{equation}
    Combining above and \eqref{eqn:assum-bound-grad-cge}, we have
    \begin{equation}
        \E_{\zeta_i^t}\norm{g_i^t}\leq\sigma+\norm{\E_{\zeta_i^t}\left[g_i^t\right]}=\sigma+\norm{\nabla Q_i\left(x^t\right)}.
        \label{eqn:norm-expectation-bound}
    \end{equation}
    Substituting \eqref{eqn:norm-expectation-bound} in \eqref{eqn:exp-B-t}, we obtain that %\textcolor{red}{(check this)}
    \begin{align}
        \E_{\zeta^t}\sum_{j\in\B^t}\iprod{x^t-x_\H}{g_j^t}&\geq-\sum_{j\in\H\backslash\H^t}\norm{x^t-x_\H}\E_{\zeta^t}\norm{g_j^t} \nonumber \\
        &\geq-\sum_{j\in\H\backslash\H^t}\norm{x^t-x_\H}\left(\sigma+\norm{\nabla Q_j(x^t)}\right) \nonumber \\
        &=-\sigma\mnorm{\H\backslash\H^t}\norm{x^t-x_\H}-\sum_{j\in\H\backslash\H^t}\norm{x^t-x_\H}\norm{\nabla Q_j(x^t)}.
        \label{eqn:exp-phit-part2}
    \end{align}
    Combining \eqref{eqn:exp-phi-t}, \eqref{eqn:exp-phit-part1A}, \eqref{eqn:exp-phit-part1B}, and \eqref{eqn:exp-phit-part2}, we have
    \begin{align}
        \E_{\zeta^t}\left[\phi_t\right] &\geq \mnorm{\H}\gamma\norm{x^t-x_\H}^2-2\sum_{j\in\H\backslash\H^t}\norm{x^t-x_\H}\norm{\nabla Q_j(x^t)}-\sigma\mnorm{\H\backslash\H^t}\norm{x^t-x_\H}. 
    \end{align}
    Substituting \eqref{eqn:honest-bound-everywhere} from the beginning of this proof in above, we have
    \begin{align}
        \E_{\zeta^t}\left[\phi_t\right]&\geq \mnorm{\H}\gamma\norm{x^t-x_\H}^2-2\mnorm{\H\backslash\H^t}\norm{x^t-x_\H}\left(2n\mu\epsilon + \mu\norm{x^t-x_\H}\right)-\sigma\mnorm{\H\backslash\H^t}\norm{x^t-x_\H}.
    \end{align}
    As $\mnorm{\H}=n-f$ and $\mnorm{\H\backslash\H^t}\leq f$, the above implies that
    \begin{align}
        \E_{\zeta^t}\left[\phi_t\right]&\geq (n-f)\gamma\norm{x^t-x_\H}^2-2f\norm{x^t-x_\H}\left(2n\mu\epsilon + \mu\norm{x^t-x_\H}\right)-f\sigma\norm{x^t-x_\H} \nonumber \\
        &\geq \left((n-f)\gamma - 2f\mu\right)\norm{x^t-x_\H}^2 - (4nf\mu\epsilon+f\sigma)\norm{x^t-x_\H}.
        \label{eqn:expectation-second-term}
    \end{align}
    
    Consider the term $\sum_{j\in\H}\E_{\zeta^t}\norm{g_j^t}^2$. Recall from Lemma~\ref{lemma:1-cge} that 
    \begin{equation*}
        \E_{\zeta^t}\norm{g_j^t}^2\leq\sigma^2+\norm{\nabla Q_j(x^t)}^2. 
    \end{equation*}
    Substituting \eqref{eqn:honest-bound-everywhere} in above, we have
    \begin{align}
        \E_{\zeta^t}\norm{g_j^t}^2\leq \sigma^2+\norm{\nabla Q_j(x^t)}^2= \sigma^2+\left(2n\mu\epsilon + \mu\norm{x^t-x_\H}\right)^2.
    \end{align}
    Therefore, as $\mnorm{\H}=n-f$,
    \begin{align}
        \sum_{j\in\H}\E_{\zeta^t}\norm{g_j^t}^2&\leq \mnorm{\H}\left(\sigma^2+\left(2n\mu\epsilon + \mu\norm{x^t-x_\H}\right)^2\right) \nonumber \\
        &= (n-f)\left(\sigma^2+\left(2n\mu\epsilon + \mu\norm{x^t-x_\H}\right)^2\right).
        \label{eqn:expectation-third-term}
    \end{align}
    
    Substituting \eqref{eqn:expectation-second-term} and \eqref{eqn:expectation-third-term} in \eqref{eqn:expectation-w-t-square}, we have
    \begin{align}
        &\E_{\zeta^t}\norm{x^{t+1}-x_\H}^2\leq\norm{x^t-x_\H}^2-2\eta\E_{\zeta^t}\left[\phi_t\right]+(n-f)\eta^2\sum_{j\in\H}\E_{\zeta^t}\norm{g_j^t}^2 \nonumber \\
        \leq&\norm{x^t-x_\H}^2-2\eta\left(\left((n-f)\gamma - 2f\mu\right)\norm{x^t-x_\H}^2 - (4nf\mu\epsilon+f\sigma)\norm{x^t-x_\H}\right) \nonumber \\
            &\qquad\qquad +(n-f)^2\eta^2\left(2n\mu\epsilon + \mu\norm{x^t-x_\H}\right)^2 + (n-f)^2\eta^2\sigma^2 \nonumber \\
        =&\left(1-2(n-f)\eta\gamma+4f\eta\mu+(n-f)^2\eta^2\mu^2\right)\norm{x^t-x_\H}^2 +4n^2(n-f)^2\eta^2\mu^2\epsilon^2 \nonumber \\
        &\qquad\qquad+\left(4n\eta\mu\epsilon\left(2f+(n-f)^2\eta\mu\right)+ 2f\eta\sigma\right)\norm{x^t-x_\H} + (n-f)^2\eta^2\sigma^2.
        \label{eqn:expectation-w-t-square-2}
    \end{align}
    Let $B=4n\eta\mu\epsilon\left(2f+(n-f)^2\eta\mu\right)$, $C=4n^2(n-f)^2\eta^2\mu^2\epsilon^2$, $D=2f\eta\sigma$, $E=(n-f)^2\eta^2\sigma^2$. Recall that $x^t\in\W$ for all $t$, where $\W$ is a convex compact set. There exists a $\Gamma=\max_{x\in\W}\norm{x-x^*}<\infty$, such that $\norm{x^t-x^*}\leq\Gamma$ for all $t$. Thus, from \eqref{eqn:expectation-w-t-square-2} we obtain
    \begin{align}
        \label{eqn:expectation-w-t-square-3}
        \E_{\zeta^t}\norm{x^{t+1}-x^*}^2&\leq \left(1-2(n-f)\eta\gamma+4f\eta\mu+(n-f)^2\eta^2\mu^2\right)\norm{x^t-x^*}^2 \nonumber \\
        &\qquad+B\Gamma+C+D\Gamma+E.
    \end{align}
    Recall \eqref{eqn:def-rho-ft} the definition of $\rho$, we have
    \begin{align}
        \label{eqn:expectation-w-t-square-4}
        \E_{\zeta^t}\norm{x^{t+1}-x^*}^2&\leq \rho\norm{x^t-x^*}^2+B\Gamma+C+D\Gamma+E.
    \end{align}
    Note that $B,C,D\geq0$. 
    
    \textbf{Second}, we show $0\leq\rho<1$. 
    Recall that in \eqref{eqn:def-alpha-ft} we defined 
    \begin{equation}
        \alpha=1-\cfrac{f}{n}\cdot\cfrac{\gamma+2\mu}{\gamma}. \tag{\ref{eqn:def-alpha}}
    \end{equation} 
    We have
    \begin{equation}
        (n-f)\gamma-2f\mu = n\gamma\alpha.
    \end{equation}
    So $\rho$ can be written as
    \begin{align}
        \rho=1-2n\eta\gamma\alpha+(n-f)^2\eta^2\mu^2 = 1-(n-f)^2\mu^2\eta\left(\frac{2n\gamma\alpha}{(n-f)^2\mu^2}-\eta\right).
        \label{eqn:rho-sgd-cge-2}
    \end{align}
    Recall \eqref{eqn:def-eta-bar-ft} that $\overline{\eta}=\cfrac{2n\gamma\alpha}{(n-f)^2\mu^2}$. From above we obtain that
    \begin{align}
        \rho = 1-(n-f)^2\mu^2\eta(\overline{\eta}-\eta).
    \end{align}
    Note that 
    \begin{align}
        \eta(\overline{\eta}-\eta)=\left(\frac{\overline{\eta}}{2}\right)^2-\left(\eta-\frac{\overline{\eta}}{2}\right)^2.
    \end{align}
    Therefore,
    \begin{align}
        \rho &= 1-(n-f)^2\mu^2\left[\left(\frac{\overline{\eta}}{2}\right)^2-\left(\eta-\frac{\overline{\eta}}{2}\right)^2\right] \nonumber \\
        &= (n-f)^2\mu^2\left(\eta-\frac{\overline{\eta}}{2}\right)^2 +  1-(n-f)^2\mu^2\left(\frac{\overline{\eta}}{2}\right)^2.
    \end{align}
    Since $\eta\in(0,\overline{\eta})$, the minimum value of $\rho$ can be obtained when $\eta=\overline{\eta}/2$,
    \begin{align}
        \min_\eta\rho=1-(n-f)^2\mu^2\left(\frac{\overline{\eta}}{2}\right)^2.
    \end{align}
    On the other hand, since $\eta\in(0,\overline{\eta})$, from \eqref{eqn:rho-sgd-cge-2}, $\rho<1$. Thus,
    \begin{align}
        1-(n-f)^2\mu^2\left(\frac{\overline{\eta}}{2}\right)^2\leq\rho<1.
    \end{align}
    Substituting \eqref{eqn:def-eta-bar-ft} in above implies that $A\in\left[1-\left(\cfrac{n\gamma\alpha}{(n-f)\mu}\right)^2,1\right)$. Note that since $\alpha>0$, we have $(n-f)\gamma-2f\mu>0$. Thus,
    \begin{equation}
        ((n-f)\gamma-2f\mu)^2<(n-f)^2\gamma^2.
        \label{eqn:gamma-mu-related-ineq-sgd-cge-1}
    \end{equation}
    Recall from Lemma~\ref{lemma:gamma-mu-cge} that $\gamma\leq\mu$, and that $(n-f)\gamma-2f\mu=n\gamma\alpha$, we have $\left(\cfrac{n\gamma\alpha}{(n-f)\mu}\right)^2<\cfrac{\gamma}{\mu}\leq1$.
    Therefore, $\rho\in[0,1)$. 
    
    \textbf{Third}, we show the convergence property. Let $F=B\Gamma + C$,  $G=D\Gamma + E$. Note that $F(\epsilon)=\O(\epsilon^2)$ is a function of $\epsilon$, and $G(\sigma)=\O(\sigma^2)$ is a function of $\sigma$.
    
    Recall that in \eqref{eqn:def-m-ft} we defined
    \begin{equation*}
        \M = 4n\eta\mu\epsilon\left(2f+(n-f)^2\eta\mu\right)\Gamma + 4n^2(n-f)^2\eta^2\mu^2\epsilon^2 + 2f\eta\sigma\Gamma + (n-f)^2\eta^2\sigma^2 = F+G. 
    \end{equation*}
    From \eqref{eqn:expectation-w-t-square-4}, definition of $\rho$ in \eqref{eqn:def-rho-ft}, and definition of $\M$ above, we have
    \begin{align}
        \E_{\zeta^t}\norm{x^{t+1}-x_\H}^2\leq&\rho\norm{x^t-x_\H}^2 + \M.
        \label{eqn:expectation-w-t-square-5}
    \end{align}
    By Lemma~\ref{lemma:converge-sgd}, since $\M\geq0$ and $\rho\in[0,1)$, we have
    \begin{align*}
        \E_{t}\norm{x^{t+1}-x_\H}^2&\leq\rho^{t+1}\norm{x^0-x_\H}^2 + \left(\frac{1-\rho^{t+1}}{1-\rho}\right)\M. \qedhere
    \end{align*}
\end{proof}

\vfill \newpage
\subsection{Proof of Theorem~\ref{thm:cge} - Problem CS}

Recall that for \textbf{Problem CS}, there is 0 Byzantine agents, and up to $r$ stragglers. Let us define the following parameters:
\begin{itemize}[nosep]
    \item The \textit{resilience margin}
    \begin{align}
        \alpha = 1-\frac{r}{n}\cdot\frac{\mu}{\gamma},
        \label{eqn:def-alpha}
    \end{align}
    \item The parameter that determines step size
    \begin{align}
        \overline{\eta} = \frac{2n\gamma\alpha}{(n-r)^2\mu^2},
        \label{eqn:def-eta-bar}
    \end{align}
\end{itemize}

\begin{mdframed}
    \textbf{Theorem~\ref{thm:cge}-CS.}
    \textit{Consider Algorithm~\ref{alg} with stochastic updates and the following gradient aggregator \eqref{eqn:aggregation-rule}
    \begin{align*}
        \textstyle\mathsf{GradAgg}\left(g_j^t|j\in S^t;n,f,r\right)=\sum_{j\in{S^t}}\nabla g_j^t,
    \end{align*}
    where $g_j^t$ is the stochastic gradient of agent $j$ at iteration $i$.
    Suppose Assumptions~\ref{assum:lipschitz}, \ref{assum:strongly-convex}, and \ref{assum:bound-grad} hold true, the expected cost functions of the agents in the system satisfy $(r,\epsilon)$-redundancy, $\alpha>0$ and step size in \eqref{eqn:update} $\eta_t=\eta>0$ for all $t$. Let
    \begin{align}
        \M &=  4n\eta\mu\epsilon\left(r + (n-r)^2\eta\mu\right)\Gamma + 4n^2(n-r)^2\eta^2\mu^2\epsilon^2 + (n-r)^2\eta^2\sigma^2.
        \label{eqn:def-m}
    \end{align}
    If $\eta<\overline{\eta}$, the following holds true:
    \begin{itemize}
        \item The value of 
        \begin{align}
            \rho = 1-2(n\gamma-r\mu)\eta + (n-r)^2\eta^2\mu^2,
            \label{eqn:def-rho}
        \end{align}
        satisfies $0<\rho<1$, and 
        \item Given the initial estimate $x^0$ arbitrarily chosen from $\W$, for all $t\geq0$,
        \begin{align}
            \E_{t}\norm{x^{t+1}-x^*}^2&\leq\rho^{t+1}\norm{x^0-x^*}^2 + \left(\frac{1-\rho^{t+1}}{1-\rho}\right)\M.
            \label{eqn:expectation-bound-1}
        \end{align}
    \end{itemize}}
\end{mdframed}

\begin{proof}
    \textbf{First}, we show a recursive bound over the expected value of $\norm{x^t-x^*}^2$. Following the \textbf{first} part of the proof of Theorem~\ref{thm:approx}, we know that by Assumption~\ref{assum:lipschitz} and $(0,r;\epsilon)$-redundancy, for all $i\in[n]$, we obtain \eqref{eqn:apprx_gradient_bnd_2} that
    \begin{equation*}
        \norm{\nabla Q_i(x)}\leq2n\mu\epsilon+\mu\norm{x-x^*}.
    \end{equation*}
    
    Define $\g^t=\sum_{j\in S^t}g_j^t$. Recall our iterative update \eqref{eqn:update}. Using the non-expansion property of Euclidean projection onto a closed convex set\footnote{$\norm{x-x^*}\geq\norm{[x]_\W-x^*},~\forall w\in\mathbb{R}^d$.}, with $\eta_t=\eta$ for all $t$, we have
    \begin{equation}
        \label{eqn:apprx-one-step-sgd}
        \norm{x^{t+1}-x^*}\leq\norm{x^t-\eta\g^t-x^*}.%~\forall t\in\mathbb{Z}_{\geq0}.
    \end{equation}
    Taking square on both sides, we have
    \begin{align}
        \label{eqn:rate-sgd-0}
        \norm{x^{t+1}-x^*}^2\leq&\norm{x^t-x^*}^2-2\eta\iprod{x^t-x^*}{\g^t}+\eta^2\norm{\g^t}^2
    \end{align}
    By triangle inequality and AM-QM inequality (i.e., $\frac{1}{n}\sum_{j=1}^nx_j\leq\sqrt{\frac{1}{n}\sum_{j=1}^nx_j^2}$ for any $n$ positive real $x_j$'s),
    \begin{align}
        \norm{\g^t}^2 \leq \left(\sum_{j\in S^t}\norm{g_j^t}\right)^2\leq\mnorm{S^t}\sum_{j\in S^t}\norm{g_j^t}^2=(n-r)\sum_{j\in S^t}\norm{g_j^t}^2.
    \end{align}
    Substitute above in \eqref{eqn:rate-sgd-0}, we have
    \begin{align}
        \label{eqn:rate-sgd-1}
        \norm{x^{t+1}-x^*}^2\leq&\norm{x^t-x^*}^2-2\eta\iprod{x^t-x^*}{\g^t}+(n-r)\eta^2\sum_{j\in S^t}\norm{g_j^t}^2.
    \end{align}
    We define
    \begin{align}
        \phi_t=\iprod{x^t-x^*}{\g^t} = \sum_{j\in S^t}\iprod{x^t-x^*}{g_j^t}.
    \end{align}
    Substituting above in \eqref{eqn:rate-sgd-1}, we have
    \begin{align}
        \label{eqn:rate-sgd-2}
        \norm{x^{t+1}-x^*}^2\leq&\norm{x^t-x^*}^2-2\eta\phi_t+(n-r)\eta^2\sum_{j\in S^t}\norm{g_j^t}^2.
    \end{align}
    Recall that $\zeta^t=\{\zeta_1^t,...,\zeta_n^t\}$, and $x^{t+1}$ is a function of the set of random variables $\zeta^t$. Also note that $\E_{\zeta^t}\norm{x^t-x^*}^2=\norm{x^t-x^*}^2$. Taking expectation $\E_{\zeta^t}$ on both sides, we have
    \begin{align}
        \label{eqn:rate-sgd-3}
        \E_{\zeta^t}\norm{x^{t+1}-x^*}^2\leq&\norm{x^t-x^*}^2-2\eta\E_{\zeta^t}\left[\phi_t\right]+(n-r)\eta^2\sum_{j\in S^t}\E_{\zeta^t}\norm{g_j^t}^2.
    \end{align}
    
    Consider $\E_{\zeta^t}\left[\phi_t\right]$. With the definition of $\g^t$ we have
    \begin{align}
        \label{eqn:phi_t_bnd-sgd}
        \E_{\zeta^t}\left[\phi_t\right] = \E_{\zeta^t}\iprod{x^t-x^*}{\g^t} = \E_{\zeta^t}\iprod{x^t-x^*}{\sum_{j\in S^t}g_j^t}.
    \end{align}
    Recall from \eqref{eqn:expectation-g-i-t-cge} that for any $j\in[n]$, $\E_{\zeta^t}\left[g_j^t\right]=\nabla Q_j(x^t)$. Substituting this in \eqref{eqn:phi_t_bnd-sgd}, we have
    \begin{align}
        \label{eqn:phi_t_bnd-sgd-1}
        \E_{\zeta^t}\left[\phi_t\right] &= \E_{\zeta^t}\iprod{x^t-x^*}{\sum_{j\in S^t}g_j^t} = \iprod{x^t-x^*}{\E_{\zeta^t}\sum_{j\in S^t}g_j^t} \nonumber \\
        &= \iprod{x^t-x^*}{\sum_{j\in S^t}\E_{\zeta^t}\left[g_j^t\right]} = \iprod{x^t-x^*}{\sum_{j\in S^t}\nabla Q_j(x^t)} \nonumber \\
        & = \iprod{x^t-x^*}{\sum_{j\in S^t}\nabla Q_j(x^t) + \sum_{j\in [n]\backslash S^t}\nabla Q_j(x^t)-\sum_{j\in[n]\backslash S^t}\nabla Q_j(x^t)} \nonumber \\
        & = \iprod{x^t-x^*}{\sum_{j\in [n]}\nabla Q_j(x^t)} - \iprod{x^t-x^*}{\sum_{j\in[n]\backslash S^t}\nabla Q_j(x^t)}.
    \end{align}
    For the first term in \eqref{eqn:phi_t_bnd-sgd-1}, recall that $x^*$ is the minimum of  $\sum_{j\in[n]}Q_j(x)$, i.e., $\sum_{j\in[n]}\nabla Q_j(x^*)=0$. By Assumption~\ref{assum:strongly-convex}, we have
    \begin{align}
        \label{eqn:phi_t_bnd-sgd-2-1}
        \iprod{x^t-x^*}{\sum_{j\in [n]}\nabla Q_j(x^t)} = \iprod{x^t-x^*}{\sum_{j\in[n]}\nabla Q_j(x^t)-\sum_{j\in[n]}\nabla Q_j(x^*)} \geq n\gamma\norm{x^t-x^*}^2.
    \end{align}
    For the second term in \eqref{eqn:phi_t_bnd-sgd-1}, by Cauchy-Schwartz inequality,
    \begin{align}
        \iprod{x^t-x^*}{\sum_{j\in[n]\backslash S^t}\nabla Q_j(x^t)} &\geq - \sum_{j\in[n]\backslash S^t}\norm{x^t-x^*}\norm{\nabla Q_j(x^t)}.
    \end{align}
    Note that $\mnorm{[n]\backslash S^t}=r$. Substituting \eqref{eqn:apprx_gradient_bnd_2} in above, 
    \begin{align}
        \label{eqn:phi_t_bnd-sgd-2-2}
        \iprod{x^t-x^*}{\sum_{j\in[n]\backslash S^t}\nabla Q_j(x^t)} &\geq - r\norm{x^t-x^*}(2n\mu\epsilon +\mu\norm{x^t-x^*}).
    \end{align}
    Substituting \eqref{eqn:phi_t_bnd-sgd-2-1} and \eqref{eqn:phi_t_bnd-sgd-2-2} in \eqref{eqn:phi_t_bnd-sgd-1}, we have
    \begin{align}
        \label{eqn:phi_t_bnd-sgd-3}
        \E_{\zeta^t}\left[\phi_t\right] & = \iprod{x^t-x^*}{\sum_{j\in [n]}\nabla Q_j(x^t)} - \iprod{x^t-x^*}{\sum_{j\in[n]\backslash S^t}\nabla Q_j(x^t)} \nonumber \\
        &\geq (n\gamma - r\mu)\norm{x^t-x^*}^2 -2nr\mu\epsilon\norm{x^t-x^*}.
    \end{align}
    
    Now consider $\sum_{j\in S^t}\E_{\zeta^t}\norm{g_j^t}^2$. Recall \eqref{eqn:lemma-1-cge} from Lemma~\ref{lemma:1-cge} that
    \begin{equation*}
        \E_{\zeta^t}\norm{g_i^t}^2\leq\sigma^2+\norm{\nabla Q_i(x^t)}^2.
    \end{equation*}
    Combining above and \eqref{eqn:apprx_gradient_bnd_2}, we have
    \begin{equation}
        \label{eqn:expectation-second-term-sgd}
        \sum_{j\in S^t}\E_{\zeta^t}\norm{g_j^t}^2\leq\mnorm{S^t}\left(\sigma^2+\left(2n\mu\epsilon+\mu\norm{x^t-x^*}\right)^2\right).
    \end{equation}
    
    Note that $\mnorm{S^t}=n-r$. Substituting \eqref{eqn:phi_t_bnd-sgd-3} and \eqref{eqn:expectation-second-term-sgd} in \eqref{eqn:rate-sgd-3},
    \begin{align}
        \label{eqn:rate-sgd-4}
        \E_{\zeta^t}\norm{x^{t+1}-x^*}^2&\leq\norm{x^t-x^*}^2 - 2\eta\left((n\gamma - r\mu)\norm{x^t-x^*}^2 -2nr\mu\epsilon\norm{x^t-x^*}\right) \nonumber \\
        &\qquad+(n-r)^2\eta^2\left(\sigma^2+\left(2n\mu\epsilon+\mu\norm{x^t-x^*}\right)^2\right) \nonumber \\
        &\leq (1-2(n\gamma-r\mu)\eta + (n-r)^2\eta^2\mu^2)\norm{x^t-x^*}^2 \nonumber \\
        &\qquad+4n\eta\mu\epsilon\left(r + (n-r)^2\eta\mu\right)\norm{x^t-x^*} \nonumber \\
        &\qquad+(n-r)^2\eta^2(\sigma^2+4n^2\mu^2\epsilon^2).
    \end{align}
    
    Let $B=4n\eta\mu\epsilon\left(r + (n-r)^2\eta\mu\right)$, $C=4n^2(n-r)^2\eta^2\mu^2\epsilon^2$, and $D=(n-r)^2\eta^2\sigma^2$. Recall that $x^t\in\W$ for all $t$, where $\W$ is a convex compact set. There exists a $\Gamma=\max_{x\in\W}\norm{x-x^*}<\infty$, such that $\norm{x^t-x^*}\leq\Gamma$ for all $t$. Thus, from \eqref{eqn:rate-sgd-4} we obtain
    \begin{align}
        \label{eqn:rate-sgd-5}
        \E_{\zeta^t}\norm{x^{t+1}-x^*}^2&\leq (1-2(n\gamma-r\mu)\eta + (n-r)^2\eta^2\mu^2)\norm{x^t-x^*}^2+B\Gamma+C+D.
    \end{align}
    Recall \eqref{eqn:def-rho} the definition of $\rho$, we have
    \begin{align}
        \label{eqn:rate-sgd-6}
        \E_{\zeta^t}\norm{x^{t+1}-x^*}^2&\leq \rho\norm{x^t-x^*}^2+B\Gamma+C+D.
    \end{align}
    Note that $B,C,D\geq0$. 

    \textbf{Second}, we show $0\leq\rho<1$. Recall that in \eqref{eqn:def-alpha} we defined 
    \begin{align*}
        \alpha = 1-\frac{r}{n}\cdot\frac{\mu}{\gamma}.
    \end{align*}
    We have
    \begin{align}
    n\gamma-r\mu=n\gamma\alpha. 
    \end{align}
    So $\rho$ can be written as 
    \begin{align}
        \rho = 1-2n\eta\gamma\alpha + (n-r)^2\eta^2\mu^2 = 1- (n-r)^2\mu^2\eta\left(\frac{2n\gamma\alpha}{(n-r)^2\mu^2}-\eta\right).
    \end{align}
    Recall \eqref{eqn:def-eta-bar} that $\overline{\eta}=\cfrac{2n\gamma\alpha}{(n-r)^2\mu^2}$, from above we obtain that
    \begin{align}
        \rho = 1-(n-r)^2\mu^2\eta(\overline{\eta}-\eta).
        \label{eqn:rho-sgd-2}
    \end{align}
    Note that 
    \begin{align}
        \eta(\overline{\eta}-\eta)=\left(\frac{\overline{\eta}}{2}\right)^2-\left(\eta-\frac{\overline{\eta}}{2}\right)^2.
    \end{align}
    Therefore,
    \begin{align}
        \rho &= 1-(n-r)^2\mu^2\left[\left(\frac{\overline{\eta}}{2}\right)^2-\left(\eta-\frac{\overline{\eta}}{2}\right)^2\right] \nonumber \\
        &= (n-r)^2\mu^2\left(\eta-\frac{\overline{\eta}}{2}\right)^2 +  1-(n-r)^2\mu^2\left(\frac{\overline{\eta}}{2}\right)^2.
    \end{align}
    Since $\eta\in(0,\overline{\eta})$, the minimum value of $\rho$ can be obtained when $\eta=\overline{\eta}/2$,
    \begin{align}
        \min_\eta\rho=1-(n-r)^2\mu^2\left(\frac{\overline{\eta}}{2}\right)^2.
    \end{align}
    On the other hand, since $\eta\in(0,\overline{\eta})$, from \eqref{eqn:rho-sgd-2}, $\rho<1$. Thus,
    \begin{align}
        1-(n-r)^2\mu^2\left(\frac{\overline{\eta}}{2}\right)^2\leq\rho<1.
    \end{align}
    Substituting \eqref{eqn:def-eta-bar} in above implies that $A\in\left[1-\left(\cfrac{n\gamma\alpha}{(n-r)\mu}\right)^2,1\right)$. Note that since $\alpha>0$, we have $n\gamma-r\mu>0$. Thus, with $\gamma\leq\mu$ (cf. Lemma~\ref{lemma:gamma-mu}),
    \begin{align}
        (n\gamma-r\mu)^2\leq (n-r)^2\mu^2.
    \end{align}
    Recall that $n\gamma-r\mu=n\gamma\alpha$, we have $\left(\cfrac{n\gamma\alpha}{(n-r)\mu}\right)^2\leq1$. Therefore, $\rho\in[0,1)$.

    \textbf{Third}, we show the convergence property. Let $E=B\Gamma+C$. Note that $E(\epsilon)=\O(\epsilon^2)$ is a function of $\epsilon$, and $D(\sigma)=\O(\sigma^2)$ is a function of $\sigma$.
    
    Recall that in \eqref{eqn:def-m} we defined
    \begin{align*}
        \M = 4n\eta\mu\epsilon\left(r + (n-r)^2\eta\mu\right)\Gamma + 4n^2(n-r)^2\eta^2\mu^2\epsilon^2 + (n-r)^2\eta^2\sigma^2 = E+D.
    \end{align*}
    From \eqref{eqn:rate-sgd-6}, definition of $\rho$ in \eqref{eqn:def-rho}, and definition of $\M$ above, we have
    \begin{align}
        \label{eqn:rate-sgd-7}
        \E_{\zeta^t}\norm{x^{t+1}-x^*}^2&\leq \rho\norm{x^t-x^*}^2 + \M.
    \end{align}
    By Lemma~\ref{lemma:converge-sgd}, since $\M\geq0$ and $\rho\in[0,1)$, we have
    \begin{align*}
        \E_{t}\norm{x^{t+1}-x_\H}^2&\leq\rho^{t+1}\norm{x^0-x_\H}^2 + \left(\frac{1-\rho^{t+1}}{1-\rho}\right)\M. \qedhere
    \end{align*}
\end{proof}

\vfill \newpage
\subsection{Proof of Theorem~\ref{thm:cge} - Problem DS}

Recall that for \textbf{Problem DS}, there is up to $f$ Byzantine agents, and up to $r$ stragglers. Let us define the following parameters:
\begin{itemize}[nosep]
    \item The \textit{resilience margin}
    \begin{align}
        \alpha=1-\frac{f-r}{n-r}+\frac{f+r}{n-r}\cdot\frac{2\mu}{\gamma}.
        \label{eqn:def-alpha-cge}
    \end{align}
    \item The parameter that determines step size
    \begin{align}
        \overline{\eta} = \cfrac{2(n-r)\gamma\alpha}{(n-r-f)^2\mu^2}.
        \label{eqn:def-eta-bar-cge}
    \end{align}
\end{itemize}

\begin{mdframed}
    \textbf{Theorem~\ref{thm:cge}-DS.}
    \textit{Consider Algorithm~\ref{alg} with stochastic updates and the following gradient aggregator
    \begin{align*}
        &\textstyle\mathsf{GradAgg}\left(g_j^t|j\in S^t;n,f,r\right) 
        =\mathsf{GradFilter}\left(g_j^t|j\in S^t;n-r,f\right).
        % \label{eqn:aggregation-rule-async-ft}
    \end{align*}
    using CGE gradient filter. 
    Suppose Assumptions~\ref{assum:lipschitz}, \ref{assum:strongly-convex-ft}, and \ref{assum:bound-grad} hold true, the expected cost functions of the agents in the system satisfy $(r,\epsilon)$-redundancy, $\alpha>0$, step size in \eqref{eqn:update} $\eta_t=\eta>0$ for all $t$, and $n\geq 2f+r/2$. Let
    \begin{align}
        \M =& 4(n-r)\eta\mu\epsilon\left(2(f+r)+(n-r-f)^2\eta\mu\right)\Gamma + 4(n-r)^2(n-r-f)^2\eta^2\mu^2\epsilon^2 \nonumber \\
        &\quad +2(f+r)\eta\sigma\Gamma + (n-r-f)^2\eta^2\sigma^2.
        \label{eqn:def-m-cge}
    \end{align}
    If $\eta<\overline{\eta}$, the following holds true:
    \begin{itemize}
        \item The value of 
        \begin{align}
            \rho = 1-2(n-f)\eta\gamma+4(f+r)\eta\mu+(n-r-f)^2\eta^2\mu^2,
            \label{eqn:def-rho-cge}
        \end{align}
        satisfies $0<\rho<1$, and 
        \item Given the initial estimate $x^0$ arbitrarily chosen from $\R^d$, for all $t\geq0$,
        \begin{align}
            \E_{t}\norm{x^{t+1}-x_\H}^2&\leq\rho^{t+1}\norm{x^0-x_\H}^2 + \left(\frac{1-\rho^{t+1}}{1-\rho}\right)\M.
            \label{eqn:expectation-bound-1-cge}
        \end{align}
    \end{itemize}}
\end{mdframed}

\begin{proof}
    \textbf{First}, we show a recursive bound over the expected value of $\norm{x^t-x_\H}^2$. Following the \textbf{first} part of the proof of Theorem~\ref{thm:async-fault-toler}-CGE, we know that by Assumption~\ref{assum:lipschitz} and $(f,r;\epsilon)$-redundancy, for every non-faulty agent $i\in[n]$, we obtain \eqref{eqn:honest-bound-everywhere} that
    \begin{equation}
        \norm{\nabla Q_i(x)}\leq(2m-4f+1)\mu\epsilon+\mu\norm{x-x_{\H}}\leq2m\mu\epsilon+\mu\norm{x-x_{\H}}.
    \end{equation}
    
    Let us denote $\g^t$ the output the gradient-filter applied, i.e.
    \begin{equation}
        \g^t=\gf\left(g_j^t|j\in S^t;m,f\right). 
        \label{eqn:gothic-g-t-async-cge}
    \end{equation}
    With $\eta_t=\eta$ and above, for each iteration $t$ we have
    \begin{equation}
        x^{t+1}=x^t-\eta\g^t.
        \label{eqn:update-gothic-g-t-async-cge}
    \end{equation}
    % From the definition of Euclidean norm,
    % By triangle inequality, 
    Subtracting $x_\H$ and taking norm on both sides, we have
    \begin{equation}
        \norm{x^{t+1}-x_\H}^2=\norm{x^t-x_\H}^2-2\eta\iprod{x^t-x_\H}{\g^t}+\eta^2\norm{\g^t}^2.
        \label{eqn:iterate-general-cge}
    \end{equation}
    
    Consider the CGE gradient-filter, where 
    \begin{equation}
        \g^t=\sum_{l=1}^{m-f}g_{i_l}^t,
        \label{eqn:cge-filter}
    \end{equation}
    where the gradients are sorted by their norms as 
    \begin{equation}
        \norm{g_{i_1}^t}\leq...\leq\norm{g_{i_{m-f-1}}^t}\leq\norm{g_{i_{m-f}}^t}\leq\norm{g_{i_{m-f+1}}^t}\leq...\leq\norm{g_{i_m}^t}.
    \end{equation}
    By triangle inequality, 
    \begin{equation}
        \norm{\g^t}\leq\sum_{l=1}^{m-f}\norm{g_{i_l}^t}.
    \end{equation}
    As there are at most $f$ Byzantine agents, for each iteration $t$ there exists a $\sigma_t\in\H$ such that 
    \begin{align}
        \norm{g_{i_{m-f}}^t}\leq \norm{g_{i_{\sigma_t}}^t}.
    \end{align}
    Combining the above two inequalities together, we have
    \begin{equation}
        \norm{\g^t}\leq \sum_{l=1}^{m-f}\norm{g_{i_{\sigma_t}}^t}=(m-f)\norm{g_{i_{\sigma_t}}^t}.
    \end{equation}
    Thus, 
    \begin{equation}
        \norm{\g^t}^2\leq(m-f)^2\norm{g_{i_{\sigma_t}}^t}^2.
        \label{eqn:cge-square-bound-cge}
    \end{equation}
    On the other hand, consider the term $\iprod{x^t-x_\H}{\g^t}$. let $\H^t=\{i_1,...,i_{m-f}\}\cap\H$, and let $\B^t=\{i_1,...,i_{m-f}\}\backslash\H^t$. Note that 
    \begin{equation}
        \mnorm{\H^t}\geq\mnorm{\H}-f = m-2f, ~\textrm{ and }~\mnorm{\B^t}\leq f.
    \end{equation}
    From \eqref{eqn:cge-filter}, 
    \begin{equation}
        \g^t=\sum_{j\in\H^t}g_j^t+\sum_{j\in\B^t}g_j^t.
    \end{equation}
    Therefore, 
    \begin{equation}
        \iprod{x^t-x_\H}{\g^t} = \iprod{x^t-x_\H}{\sum_{j\in\H^t}g_j^t} + \iprod{x^t-x_\H}{\sum_{j\in\B^t}g_j^t}.
    \end{equation}
    We define 
    \begin{equation}
        % \phi_t=\sum_{j\in\H^t}\iprod{x^t-x_\H}{g_j^t} - f\norm{x^t-x_\H}\norm{g_{v_t}^t}.
        \phi_t=\iprod{x^t-x_\H}{\g^t}=\sum_{j\in\H^t}\iprod{x^t-x_\H}{g_j^t} + \sum_{j\in\B^t}\iprod{x^t-x_\H}{g_j^t}.
        \label{eqn:phi-t-sgd-cge}
    \end{equation}
    % Note that $\iprod{x^t-x_\H}{\g^t}\geq\phi_t$. 
    Substituting above and \eqref{eqn:cge-square-bound-cge} in \eqref{eqn:iterate-general-cge}, we obtain that
    \begin{equation}
        \norm{x^{t+1}-x_\H}^2\leq\norm{x^t-x_\H}^2-2\eta\phi_t+(m-f)^2\eta^2\norm{g_{i_{\sigma_t}}^t}^2.
    \end{equation}
    Recall that $\zeta^t=\left\{\zeta_1^t,...,\zeta_n^t\right\}$, and $x^{t+1}$ is a function of the set of random variables $\zeta^t$. Also note that $\E_{\zeta^t}\norm{x^t-x_\H}^2=\norm{x^t-x_\H}^2$. Taking expectation $\E_{\zeta^t}$ on both sides, we have
    \begin{equation}
        \E_{\zeta^t}\norm{x^{t+1}-x_\H}^2\leq\norm{x^t-x_\H}^2-2\eta\E_{\zeta^t}\left[\phi_t\right]+(m-f)^2\eta^2\E_{\zeta^t}\norm{g_{i_{\sigma_t}}^t}^2. 
        \label{eqn:expectation-w-t-square-cge}
    \end{equation}
    
    From \eqref{eqn:phi-t-sgd-cge} the definition of $\phi_t$,
    \begin{align}
        \E_{\zeta^t}\left[\phi_t\right]& = \E_{\zeta^t}\sum_{j\in\H^t}\iprod{x^t-x_\H}{g_j^t} + \E_{\zeta^t}\sum_{j\in\B^t}\iprod{x^t-x_\H}{g_j^t} \nonumber \\
        & = \sum_{j\in\H^t}\iprod{x^t-x_\H}{\E_{\zeta^t}\left[g_j^t\right]} +\E_{\zeta^t}\sum_{j\in\B^t}\iprod{x^t-x_\H}{g_j^t}.
        \label{eqn:exp-phi-t-sgd-cge}
    \end{align}
    Recall from \eqref{eqn:expectation-g-i-t-cge} that for any $j\in\H$, $\E_{\zeta^t}\left[g_j^t\right] = \nabla Q_j(x^t)$. The first term of \eqref{eqn:exp-phi-t-sgd-cge} becomes
    \begin{align}
        \sum_{j\in\H^t}\iprod{x^t-x_\H}{\E_{\zeta^t}\left[g_j^t\right]} &= \sum_{j\in\H^t}\iprod{x^t-x_\H}{\nabla Q_j(x^t)} \nonumber \\
        &= \iprod{x^t-x_\H}{\sum_{j\in\H}\nabla Q_j(x^t)} - \iprod{x^t-x_\H}{\sum_{j\in\H\backslash\H^t}\nabla Q_j(x^t)}
    \end{align}
    By Assumption~\ref{assum:strongly-convex-ft}, also the fact that $\nabla\sum_{j\in\H}Q_j(x_\H)=0$,
    \begin{align}
        \iprod{x^t-x_\H}{\sum_{j\in\H}\nabla Q_j(x^t)} = \iprod{x^t-x_\H}{\sum_{j\in\H}\nabla Q_j(x^t)-\sum_{j\in\H}\nabla Q_j(x_\H)}\geq \mnorm{\H}\gamma\norm{x^t-x_\H}^2.
        \label{eqn:exp-phit-part1A-sgd-cge}
    \end{align}
    By Cauchy-Schwartz inequality,
    \begin{equation}
        \iprod{x^t-x_\H}{\sum_{j\in\H\backslash\H^t}\nabla Q_j(x^t)}= \sum_{j\in\H\backslash\H^t}\iprod{x^t-x_\H}{\nabla Q_j(x^t)} \leq \sum_{j\in\H\backslash\H^t}\norm{x^t-x_\H}\norm{\nabla Q_j(x^t)}.
        \label{eqn:exp-phit-part1B-sgd-cge}
    \end{equation}
    Also by Cauchy-Schwartz inequality, for any $j\in[n]$, 
    \begin{equation}
        \iprod{x^t-x_\H}{g_j^t}\geq -\norm{x^t-x_\H}\norm{g_j^t}.
    \end{equation}
    Recall the sorting of vectors $\{g_j^t\}_{j=1}^n$. For an arbitrary $j\in\B^t$ and an arbitrary $j'\in\H\backslash\H^t$, 
    \begin{equation}
        \norm{g_j^t}\leq\norm{g_{j'}^t}.
        \label{eqn:bound-B-t}
    \end{equation}
    Recall that $\B^t=\left\{i_1,...,i_{n-f}\right\}\backslash\H^t$. Thus, $\mnorm{\B^t}=m-f-\mnorm{\H^t}$. Also, as $\mnorm{\H}=n-f$, $\mnorm{\H\backslash\H^t}=n-f-\mnorm{\H^t}$. That is, $\mnorm{\B^t}\leq\mnorm{\H\backslash\H^t}$. Therefore,  we have
    \begin{align}
       \sum_{j\in\B^t}\iprod{x^t-x_\H}{g_j^t}&\geq-\sum_{j\in\B^t}\norm{x^t-x_\H}\norm{g_j^t}\geq-\sum_{j\in\H\backslash\H^t}\norm{x^t-x_\H}\norm{g_j^t}.
       \label{eqn:B-t-sgd-cge}
    \end{align}
    Taking expectation on both sides, for the second term of \eqref{eqn:exp-phi-t-sgd-cge} we have
    \begin{align}
       \E_{\zeta^t}\sum_{j\in\B^t}\iprod{x^t-x_\H}{g_j^t}\geq\E_{\zeta^t}\left[-\sum_{j\in\H\backslash\H^t}\norm{x^t-x_\H}\norm{g_j^t}\right]=-\sum_{j\in\H\backslash\H^t}\norm{x^t-x_\H}\E_{\zeta^t}\norm{g_j^t}.
       \label{eqn:exp-B-t-sgd-cge}
    \end{align}
    Recall \eqref{eqn:assum-bound-grad-cge} from Lemma~\ref{lemma:1-cge} that for any non-faulty agent $i$, 
    \begin{equation*}
        \E_{\zeta_i^t}\norm{g_i^t-\E_{\zeta_i^t}\left[g_i^t\right]}\leq\sigma.
    \end{equation*}
    Note that by triangle inequality,
    \begin{equation}
        \norm{g_i^t-\E_{\zeta_i^t}\left[g_i^t\right]}\geq\norm{g_i^t}-\norm{\E_{\zeta_i^t}\left[g_i^t\right]}.
    \end{equation}
    Taking expectation on both sides, we obtain that
    \begin{equation}
        \E_{\zeta_i^t}\norm{g_i^t-\E_{\zeta_i^t}\left[g_i^t\right]}\geq\E_{\zeta_i^t}\norm{g_i^t}-\norm{\E_{\zeta_i^t}\left[g_i^t\right]}.
    \end{equation}
    Combining above and \eqref{eqn:assum-bound-grad-cge}, we have
    \begin{equation}
        \E_{\zeta_i^t}\norm{g_i^t}\leq\sigma+\norm{\E_{\zeta_i^t}\left[g_i^t\right]}=\sigma+\norm{\nabla Q_i\left(x^t\right)}.
        \label{eqn:norm-expectation-bound-cge}
    \end{equation}
    Substituting \eqref{eqn:norm-expectation-bound-cge} in \eqref{eqn:exp-B-t-sgd-cge}, we obtain that %\textcolor{red}{(check this)}
    \begin{align}
        \E_{\zeta^t}\sum_{j\in\B^t}\iprod{x^t-x_\H}{g_j^t}&\geq-\sum_{j\in\H\backslash\H^t}\norm{x^t-x_\H}\E_{\zeta^t}\norm{g_j^t} \nonumber \\
        &\geq-\sum_{j\in\H\backslash\H^t}\norm{x^t-x_\H}\left(\sigma+\norm{\nabla Q_j(x^t)}\right) \nonumber \\
        &=-\sigma\mnorm{\H\backslash\H^t}\norm{x^t-x_\H}-\sum_{j\in\H\backslash\H^t}\norm{x^t-x_\H}\norm{\nabla Q_j(x^t)}.
        \label{eqn:exp-phit-part2-sgd-cge}
    \end{align}
    Combining \eqref{eqn:exp-phi-t-sgd-cge}, \eqref{eqn:exp-phit-part1A-sgd-cge}, \eqref{eqn:exp-phit-part1B-sgd-cge}, and \eqref{eqn:exp-phit-part2-sgd-cge}, we have
    \begin{align}
        \E_{\zeta^t}\left[\phi_t\right] &\geq \mnorm{\H}\gamma\norm{x^t-x_\H}^2-2\sum_{j\in\H\backslash\H^t}\norm{x^t-x_\H}\norm{\nabla Q_j(x^t)}-\sigma\mnorm{\H\backslash\H^t}\norm{x^t-x_\H}. 
    \end{align}
    Substituting \eqref{eqn:honest-bound-everywhere} in above, we have
    \begin{align}
        \E_{\zeta^t}\left[\phi_t\right]&\geq \mnorm{\H}\gamma\norm{x^t-x_\H}^2-2\mnorm{\H\backslash\H^t}\norm{x^t-x_\H}\left(2m\mu\epsilon + \mu\norm{x^t-x_\H}\right)-\sigma\mnorm{\H\backslash\H^t}\norm{x^t-x_\H}.
    \end{align}
    As $\mnorm{\H}=n-f$, $\mnorm{H^t}\geq m-2f$, and $\mnorm{\H\backslash\H^t}=n-f-\mnorm{H^t}\leq f+r$, the above implies that
    \begin{align}
        \E_{\zeta^t}\left[\phi_t\right]&\geq (n-f)\gamma\norm{x^t-x_\H}^2-2(f+r)\norm{x^t-x_\H}\left(2m\mu\epsilon + \mu\norm{x^t-x_\H}\right)-(f+r)\sigma\norm{x^t-x_\H} \nonumber \\
        &\geq \left((n-f)\gamma - 2(f+r)\mu\right)\norm{x^t-x_\H}^2 - (4m\mu\epsilon+\sigma)(f+r)\norm{x^t-x_\H}.
        \label{eqn:expectation-second-term-sgd-cge}
    \end{align}
    
    Now consider the third term $(m-f)^2\eta^2\E_{\zeta^t}\norm{g_{i_{\sigma_t}}^t}^2$ in \eqref{eqn:expectation-w-t-square-cge}. Recall from Lemma~\ref{lemma:1-cge} that for any non-faulty agent $j$,
    \begin{equation}
        \E_{\zeta^t}\norm{g_j^t}^2\leq\sigma^2+\norm{\nabla Q_j(x^t)}^2. \tag{\ref{eqn:lemma-1}}
    \end{equation}
    Substituting \eqref{eqn:honest-bound-everywhere} in above, we have
    \begin{align}
        \E_{\zeta^t}\norm{g_j^t}^2\leq \sigma^2+\norm{\nabla Q_j(x^t)}^2= \sigma^2+\left(2m\mu\epsilon + \mu\norm{x^t-x_\H}\right)^2.
    \end{align}
    Therefore, 
    \begin{align}
        (m-f)^2\eta^2\E_{\zeta^t}\norm{g_{i_{\sigma_t}}^t}^2&\leq (m-f)^2\eta^2\left(\sigma^2+\left(2m\mu\epsilon + \mu\norm{x^t-x_\H}\right)^2\right).
        \label{eqn:expectation-third-term-sgd-cge}
    \end{align}
    
    Substituting \eqref{eqn:expectation-second-term-sgd-cge} and \eqref{eqn:expectation-third-term-sgd-cge} in \eqref{eqn:expectation-w-t-square-cge}, we have
    \begin{align}
        &\E_{\zeta^t}\norm{x^{t+1}-x_\H}^2\leq\norm{x^t-x_\H}^2-2\eta\E_{\zeta^t}\left[\phi_t\right]+(m-f)^2\eta^2\E_{\zeta^t}\norm{g_{i_{\sigma_t}}^t}^2 \nonumber \\
        \leq&\norm{x^t-x_\H}^2-2\eta\left(\left((n-f)\gamma - 2(f+r)\mu\right)\norm{x^t-x_\H}^2 - (4m\mu\epsilon+\sigma)(f+r)\norm{x^t-x_\H}\right) \nonumber \\
            &\qquad\qquad +(m-f)^2\eta^2\left(2m\mu\epsilon + \mu\norm{x^t-x_\H}\right)^2 + (m-f)^2\eta^2\sigma^2 \nonumber \\
        =&\left(1-2(n-f)\eta\gamma+4(f+r)\eta\mu+(m-f)^2\eta^2\mu^2\right)\norm{x^t-x_\H}^2 \nonumber \\
        &\qquad\qquad+\left(4m\eta\mu\epsilon\left(2(f+r)+(m-f)^2\eta\mu\right)+2(f+r)\eta\sigma\right)\norm{x^t-x_\H} \nonumber \\
        &\qquad\qquad+4m^2(m-f)^2\eta^2\mu^2\epsilon^2+ (m-f)^2\eta^2\sigma^2.
        \label{eqn:expectation-w-t-square-sgd-cge-2}
    \end{align}
    Let $B=4m\eta\mu\epsilon\left(2(f+r)+(m-f)^2\eta\mu\right)$, $C=4m^2(m-f)^2\eta^2\mu^2\epsilon^2$, $D=2(f+r)\eta\sigma$, and $E=(m-f)^2\eta^2\sigma^2$. Recall that $x^t\in\W$ for all $t$, where $\W$ is a convex compact set. There exists a $\Gamma=\max_{x\in\W}\norm{x-x^*}<\infty$, such that $\norm{x^t-x^*}\leq\Gamma$ for all $t$. Thus, from \eqref{eqn:expectation-w-t-square-sgd-cge-2} we obtain
    \begin{align}
        \label{eqn:expectation-w-t-square-sgd-cge-3}
        \E_{\zeta^t}\norm{x^{t+1}-x^*}^2&\leq \left(1-2(n-f)\eta\gamma+4(f+r)\eta\mu+(m-f)^2\eta^2\mu^2\right)\norm{x^t-x^*}^2 \nonumber \\
        &\qquad+B\Gamma+C+D\Gamma+E.
    \end{align}
    Recall \eqref{eqn:def-rho-cge} the definition of $\rho$, we have
    \begin{align}
        \label{eqn:expectation-w-t-square-sgd-cge-4}
        \E_{\zeta^t}\norm{x^{t+1}-x^*}^2&\leq \rho\norm{x^t-x^*}^2+B\Gamma+C+D\Gamma+E.
    \end{align}
    Note that $B,C,D\geq0$. 

    \textbf{Second}, we show $0\leq\rho<1$. Recall that in \eqref{eqn:def-alpha-cge} we defined 
    \begin{equation*}
        \alpha=1-\frac{f-r}{m}+\frac{f+r}{m}\cdot\frac{2\mu}{\gamma}.
    \end{equation*} 
    We have
    \begin{equation}
        (n-f)\gamma-2(f+r)\mu = m\gamma\alpha.
    \end{equation}
    So $\rho$ can be written as
    \begin{align}
        \rho=1-2m\eta\gamma\alpha+(m-f)^2\eta^2\mu^2 = 1- (m-f)^2\mu^2\eta\left(\frac{2m\gamma\alpha}{(m-f)^2\mu^2}-\eta\right).
        \label{eqn:rho-sgd-cge-2}
    \end{align}
    Recall \eqref{eqn:def-eta-bar-cge} that $\overline{\eta}=\cfrac{2m\gamma\alpha}{(m-f)^2\mu^2}$. From above we obtain that
    \begin{align}
        \rho = 1-(m-f)^2\mu^2\eta(\overline{\eta}-\eta).
    \end{align}
    Note that 
    \begin{align}
        \eta(\overline{\eta}-\eta)=\left(\frac{\overline{\eta}}{2}\right)^2-\left(\eta-\frac{\overline{\eta}}{2}\right)^2.
    \end{align}
    Therefore,
    \begin{align}
        \rho &= 1-(m-f)^2\mu^2\left[\left(\frac{\overline{\eta}}{2}\right)^2-\left(\eta-\frac{\overline{\eta}}{2}\right)^2\right] \nonumber \\
        &= (m-f)^2\mu^2\left(\eta-\frac{\overline{\eta}}{2}\right)^2 +  1-(m-f)^2\mu^2\left(\frac{\overline{\eta}}{2}\right)^2.
    \end{align}
    Since $\eta\in(0,\overline{\eta})$, the minimum value of $\rho$ can be obtained when $\eta=\overline{\eta}/2$,
    \begin{align}
        \min_\eta\rho=1-(m-f)^2\mu^2\left(\frac{\overline{\eta}}{2}\right)^2.
    \end{align}
    On the other hand, since $\eta\in(0,\overline{\eta})$, from \eqref{eqn:rho-sgd-cge-2}, $\rho<1$. Thus,
    \begin{align}
        1-(m-f)^2\mu^2\left(\frac{\overline{\eta}}{2}\right)^2\leq\rho<1.
    \end{align}
    Substituting \eqref{eqn:def-eta-bar-cge} in above implies that $A\in\left[1-\left(\cfrac{m\gamma\alpha}{(m-f)\mu}\right)^2,1\right)$. Note that since $\alpha>0$, we have $(n-f)\gamma-2(f+r)\mu>0$. Recall from Lemma~\ref{lemma:gamma-mu-cge} that $\gamma\leq\mu$, thus
    \begin{equation}
        ((n-f)\gamma-2(f+r)\mu)^2\leq((n-f)-2(f+r))^2\mu^2.
        \label{eqn:gamma-mu-related-ineq-sgd-cge-1}
    \end{equation}
    Recall that $(n-f)\gamma-2(f+r)\mu=m\gamma\alpha$, we have 
    \begin{align}
        1-\left(\cfrac{m\gamma\alpha}{(m-f)\mu}\right)^2\geq1-\frac{((n-f)-2(f+r))^2}{(m-f)^2}=\frac{(2f+r)(2n-4f-3r)}{(m-f)^2}.
    \end{align}
    Note that we assume that $n\geq 2f+r/2$, thus the right hand side of the above is non-negative. 
    Therefore, $\rho\in[0,1)$. 
    
    \textbf{Third}, we show the convergence property. Let $F=B\Gamma+C$, $G=D\Gamma+E$. Note that $F(\epsilon)=\O(\epsilon^2)$ is a function of $\epsilon$, and $G(\sigma)=\O(\sigma^2)$ is a function of $\sigma$.

    Recall that in \eqref{eqn:def-m-cge} we defined
    \begin{align*}
        \M &= 4m\eta\mu\epsilon\left(2(f+r)+(m-f)^2\eta\mu\right)\Gamma + 4m^2(m-f)^2\eta^2\mu^2\epsilon^2 + 2\eta(f+r)\sigma\Gamma + (m-f)^2\eta^2\sigma^2 = F+G.
    \end{align*}
    From \eqref{eqn:expectation-w-t-square-sgd-cge-4}, definition of $\rho$ in \eqref{eqn:def-rho-cge}, and definition of $\M$ above, we have
    \begin{align}
        \E_{\zeta^t}\norm{x^{t+1}-x_\H}^2\leq&\rho\norm{x^t-x_\H}^2 + \M.
        \label{eqn:expectation-w-t-square-sgd-cge-5}
    \end{align}
    By Lemma~\ref{lemma:converge-sgd}, since $\M\geq0$ and $\rho\in[0,1)$, we have
    \begin{align*}
        \E_{t}\norm{x^{t+1}-x_\H}^2&\leq\rho^{t+1}\norm{x^0-x_\H}^2 + \left(\frac{1-\rho^{t+1}}{1-\rho}\right)\M. \qedhere
    \end{align*}
\end{proof}

\end{document}